\documentclass[reprint,amsmath,amssymb,aps,superscriptaddress]{revtex4-2}

\usepackage{graphicx}
\usepackage{dcolumn}
\usepackage{bm}
\usepackage{hyperref}
\usepackage{color}
\usepackage{xcolor}
\usepackage{physics}
\usepackage{overpic}

\begin{document}

\preprint{APS/123-QED}

\title{Dynamic stimulated emission for deterministic addition and subtraction of propagating photons}

\author{Haoyuan Luo}
\email{haoyuan.luo@sydney.edu.au}
\affiliation{
Centre for Engineered Quantum Systems, School of Physics, The University of Sydney, NSW 2006, Australia
}
\affiliation{Sydney Quantum Academy, Sydney, NSW, Australia}
\affiliation{%
 Institute for Photonics and Optical Sciences (IPOS), School of Physics, The University of Sydney, NSW 2006, Australia
}

\author{Parth S. Shah}
\affiliation{Moore Laboratory of Engineering, California Institute of Technology, Pasadena, California 91125, USA}
\affiliation{Institute for Quantum Information and Matter, California Institute of Technology, Pasadena, California 91125, USA}

\author{Frank Yang}
\affiliation{Moore Laboratory of Engineering, California Institute of Technology, Pasadena, California 91125, USA}
\affiliation{Institute for Quantum Information and Matter, California Institute of Technology, Pasadena, California 91125, USA}

\author{Mohammad Mirhosseini}
\affiliation{Moore Laboratory of Engineering, California Institute of Technology, Pasadena, California 91125, USA}
\affiliation{Institute for Quantum Information and Matter, California Institute of Technology, Pasadena, California 91125, USA}

\author{Sahand Mahmoodian}%
\email{sahand.mahmoodian@sydney.edu.au}
\affiliation{
Centre for Engineered Quantum Systems, School of Physics, The University of Sydney, NSW 2006, Australia
}
\affiliation{%
 Institute for Photonics and Optical Sciences (IPOS), School of Physics, The University of Sydney, NSW 2006, Australia
}%

\begin{abstract}
Photon subtraction and addition are essential non-Gaussian processes in quantum optics, where conventional methods using linear optics and number-resolving detection often suffer from low success probability. Here, we introduce the concept of \textit{dynamic stimulated emission}, whereby a quantum emitter undergoes stimulated emission with a time-dependent coupling. Despite interacting with a multimode photon field, we show that, for both two- and three-level emitters, their dynamics can mimic that of a single-mode Jaynes-Cummings model. This feature can be used to deterministically add or subtract a photon while maintaining a single propagating optical mode. We provide semi-analytic solutions to this problem for Fock states, enabling deterministic and unconditional photon addition and subtraction with fidelity ${\cal F}>0.996$. Our semi-analytic solutions are provided for both dynamically coupled two-level systems and for three-level systems whose dynamical coupling is controlled by a coherent laser drive. Moving beyond individual Fock states, we further showcase the ability to subtract and add single photons to photon-number superposition states. We show that Schr{\"o}dinger cat states can be prepared from squeezed vacuum input via cascaded subtraction or cascaded addition. Finally, we show that our photon-addition process can be used to add a photon to any squeezed and displaced state with high success probability and fidelity ${\cal F}>0.99$, thereby potentially converting quantum emitters from single-photon sources to sources of single-photon-added Gaussian states without the need for inline squeezing. Our protocols provide a path towards integrating quantum emitters to construct efficient sources of single-mode non-Gaussian light beyond single photons.
\end{abstract}

\maketitle

\section{Introduction}

Non-Gaussian quantum states of photons are key ingredients for unlocking quantum advantage in continuous-variable quantum computation \cite{Bartlett2002PRL,Mari2012PRL,Ralph2003PRA,Mirrahimi2014NJP,Knill2001Nature,Kok2007RevMP}, sensing \cite{Munro2002PRA,Gilchrist2004JOB}, and communication \cite{Braunstein2005RevMP,Kok2007RevMP,Brask2010PRL}. Exotic non-Gaussian states such as large-photon-number Schr\"{o}dinger cat states and the Gottesman-Kitaev-Preskill (GKP) state are particularly desired for their error-correcting properties \cite{Grimsmo2020PRX,Gottesman2001PRA,Grimsmo2021PRXQ}. While the preparation of these non-Gaussian states has matured in superconducting \cite{Ofek2016Nature,Grimsmo2021PRXQ} and trapped ion \cite{Fluhmann2019Nature, Matsos2024PRL} platforms, they remain elusive in optical systems. Conventional approaches typically rely on feeding squeezed states into linear optical circuits with photon-number-resolving detection acting both as a heralding signal and as the only non-Gaussian operation. These protocols typically have low success probabilities and can be very sensitive to photon loss \cite{Dakna1997PRA,Ourjoumtsev2006science,Ourjoumtsev2007PRL,Xanadu2025Nat,Takase2021PRA,Takase2024PRA,Hanamura2025}.

The two-level transition of a quantum emitter enables the generation of non-Gaussian states of light. For example, when strongly coupled to a single mode, quantum emitters are excellent single-photon sources \cite{Tomm2021NNANO, Lodahl2015RMP, Aharonovich2016NPho, Senellart2017NNano, Thomas2022Nature, Ding2025NPho}. However, beyond single-photon generation, few schemes for using quantum emitters for single-mode non-Gaussian state generation have been proposed. Apart from experimental challenges, this is largely because multi-photon emission processes such as superradiance \cite{Dicke1954PR, Yudson1985JETP, Paulisch2019PRA} produce states of photons featuring entanglement among many temporal modes \cite{Valente2012NJP,Gonzalez2015PRL,Gonzalez2017PRL, Fischer2018OSA, Kiilerich2019PRL}. This contrasts with most quantum information protocols which require photons to be in a single spatiotemporal mode. Extending the scope of quantum emitters to generate single-mode multi-photon states would position them as strong candidates for introducing non-Gaussian operations in optical systems, thereby having a significant impact on the efficiency of generating non-Gaussian states. 

In this work, we use quantum emitters to deterministically perform photon addition and subtraction while preserving a single-mode. These operations are fundamental for bosonic quantum information processing and useful for generating non-Gaussian states of light. Explicitly, photon addition and subtraction are transitions in the Fock basis, i.e., $\ket{n-1}\leftrightarrow\ket{n}$. While it is straightforward to perform these operations in the single-mode Jaynes-Cummings model, where photon addition and subtraction are realised by the signature Rabi oscillations, it is a challenging problem for propagating photons because scattering a single-mode multi-photon state off a quantum emitter generally results in a multimode state \cite{Mahmoodian2020PRX,Ralph2015PRL,Yang2022PRL}. Here, we emphasize the difference between a multimode state and a single-mode state. A general $n$-photon multimode state of a one-dimensional photon field can be expressed as
\begin{equation}
\label{general photon state}
    \frac{1}{\sqrt{n!}}\int_{\mathbb{R}^n}d^n\mathbf{x}~\alpha(\mathbf{x})\hat{\mathbf{a}}^\dagger(\mathbf{x})\ket{0},
\end{equation}
where $\ket{0}$ is the vacuum, $d^n\mathbf{x}=dx_1dx_2\dots dx_n$, $\hat{\mathbf{a}}^\dagger(\mathbf{x})=\hat{a}^\dagger(x_1)\hat{a}^\dagger(x_2)\dots\hat{a}^\dagger(x_n)$ are the photon creation operators and $\alpha(\mathbf{x})$ is the $n$-photon symmetric position wavefunction. In the special case when the wavefunction is separable, $\alpha(\mathbf{x})=f(x_1)\dots f(x_n)$, the state is single-moded with mode function, $f(x)$, and it is referred to as an $n$-photon Fock state. This can also be generalised to superpositions of Fock states. Lund \textit{et al.} presented a potential solution to the problem by scattering an $n$-photon Fock state with a passive two-level system (TLS) \cite{Lund2024PRL}. However, the proposed solution still produced a two-mode state whose mode functions share overlapping support, therefore an experimentally challenging quantum pulse gate \cite{Serino2023PRXQ} must be used to extract a single-mode output state. 

In a seminal work, Cirac, Zoller, Kimble, and Mabuchi showed that driven three-level quantum emitters provide an effective dynamic coupling that can be used to emit and absorb single photons with arbitrary mode shapes \cite{Cirac1997PRL}. This process can be considered as the single-excitation special case of photon addition and subtraction where the respective initial and final state of the photon field is the vacuum state. We note that the single-mode condition is automatically satisfied in their protocol since a single-photon state is always single-moded. Extending from the single-excitation scenario, here we consider multiple excitations. Scattering a multi-photon state can induce stimulated emission in a quantum emitter, which greatly complicates the interaction as the quantum emitter can simultaneously emit and absorb photons. This complicated scattering process is difficult to solve and most importantly, it generally leads multimode fragmentation.

\begin{figure}[!t]
    \centering
    \includegraphics[width=0.75\linewidth]{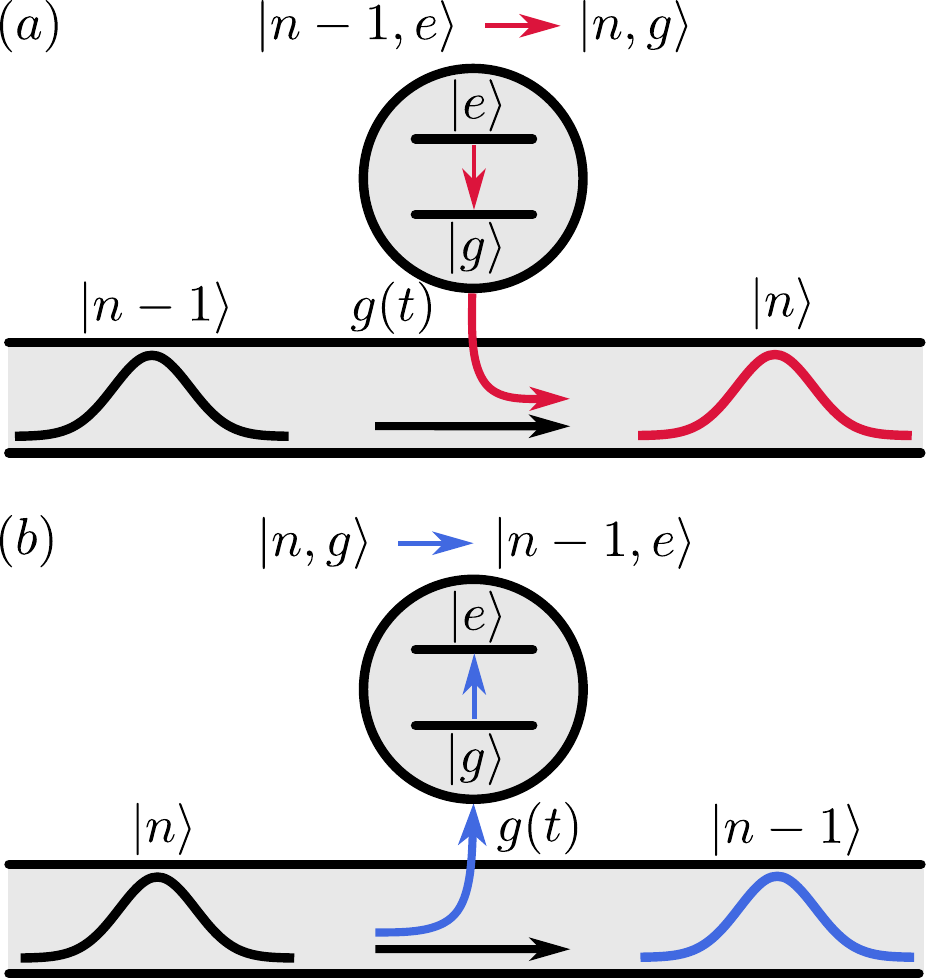}
    \caption{(a) Schematic of dynamic stimulated emission for the addition of a single photon to produce an $n$-photon Fock state in a one-dimensional waveguide QED system. 
    When a $(n-1)$-photon Fock state pulse is incident on a two-level system (TLS) prepared in the excited state, a carefully chosen dynamic coupling strength $g(t)$ enables the TLS to emit a photon such that output is a $n$-photon Fock state. (b) Conversely, the schematic of the reverse process (subtraction). The TLS is excited by an $n$-photon Fock state. The dynamic coupling strengths, $g(t)$ are distinct for each addition and subtraction process and they depend on the excitation number and the input mode shape; explicit expressions are given by Eq.~\eqref{n photon add g} and Eq.~\eqref{n photon sub g}.}
    \label{fig:TLS_schematic}
\end{figure}

Until now, it is unclear whether performing photon addition and subtraction with $n$ excitations while maintaining a single-mode output was possible. In this work, we demonstrate that under a carefully chosen dynamic coupling strength, $g(t)$, a quantum emitter is able to facilitate deterministic and single-mode-preserving photon addition and subtraction when the incoming Fock state has arbitrary number of excitations, see Fig.~\ref{fig:TLS_schematic}. When stimulated emission occurs while the quantum emitter is dynamically coupled, we refer to it as \textit{dynamic stimulated emission}. We demonstrate that under dynamic stimulated emission, the dynamics of a quantum emitter and a multimode quantum field can mimic that of the single-mode Jaynes-Cummings model. A direction consequence of establishing this connection is to use a quantum emitter for deterministic photon addition and subtraction.

We show that the unconditional fidelities for both the addition and subtraction processes using dynamic stimulated emission with a TLS are $\mathcal{F}>0.996$. Moreover, in optical experiment settings, an effective TLS with time-dependent coupling can be realised using a far-detuned three-level system (3LS) with the dynamic coupling controlled by a coherent laser drive. We show that the unconditional fidelities when using 3LSs for addition and subtraction of single photons are $\mathcal{F}>0.996$. Moving to more general superpositions of Fock states, we implement cascaded photon addition and subtraction on squeezed vacuum to prepare high-fidelity Schr\"{o}dinger cat states. Finally, we show that dynamic stimulated emission has the potential to transform single-photon sources -- which add a single photon to vacuum -- to sources of photon-added Gaussian states, where a photon is added to a squeezed displaced state.  These states can be generated with success probabilities greater than $0.54$ and fidelities $\mathcal{F}>0.99$. 

This article is organized as follows, Sec.~\ref{single excitation} reviews the dynamic-coupling solution for the addition and subtraction of a single photon using a TLS. In Sec.~\ref{two-excitation} we derive the two-excitation solution and present the general solution to an arbitrary number of excitations in Sec.~\ref{n-excitation}. In Sec.~\ref{TLS results} we discuss the results of photon addition and subtraction of up to five excitations. In Sec.~\ref{connection to the JC model} we draw the connection between waveguide QED and the single-mode Jaynes-Cummings model. In Sec.~\ref{cascaded addition and subtraction} we cascade up to five TLSs to successively perform photon addition/subtraction to generate transitions between Fock states. Moving beyond TLS, in Sec.~\ref{three-level system}, we present our solution for using 3LSs for performing photon addition and subtraction of Fock states and the results are discussed in Sec.~\ref{3LS results}. In Sec.~\ref{cat states} we cascade TLSs to produce Schr\"{o}dinger cat states and in Sec.~\ref{photon-added gaussian state} we use dynamic stimulated emission to generate photon-added Gaussian states. We provide concluding remarks in Sec.~\ref{conclusion}.

\section{Dynamic stimulated emission and absorption using two-level systems}
\label{two-level system}

The problem of single-photon addition and subtraction has been studied extensively and has been solved using Heisenberg's equations \cite{Cirac1997PRL,Gheri1998} and quantum stochastic differential equations \cite{Gough2015EPJ,Nurdin2016IEEE}. It was shown that, with a dynamic coupling, a quantum emitter can be used to absorb or emit a single photon with any reasonable temporal mode shape. These processes constitute dynamic emission and absorption of a single photon, where respectively, a photon is added to vacuum or a photon is absorbed, leaving only vacuum. Here, we extend the photon addition and subtraction processes to an arbitrary number of excitations, i.e., we extend dynamic emission and absorption to dynamic stimulated emission and absorption. The ideal photon subtraction process for a Fock state is illustrated in Fig.~\hyperref[fig:TLS_schematic]{1(a)}. This shows a propagating $n$-photon Fock state in a given temporal mode interacting with a TLS subject to the tailored dynamic coupling rate $g(t)$. The TLS perfectly absorbs one photon and transits to the excited state while the transmitted $(n-1)$-photon state remains in a single temporal mode. The output temporal mode may be identical to the input mode, but the key requirement is that it remains in a single mode. Since photon subtraction is deterministic and the input and output states are Fock states, the reverse process is the ideal photon addition process, illustrated by Fig.~\hyperref[fig:TLS_schematic]{1(b)}. 

Here, we provide a semi-analytic solution for the dynamical coupling required to implement these ideal addition and subtraction processes for arbitrary input pulse shapes. Our approach can be summarized as follows: start with the most general ansatz for each $n$-excitation subspace, solve the Schr\"{o}dinger equation with the known input state and the desired output conditions, then explicitly find the required dynamic coupling strength, $g(t)$. Although the coupling, $g(t)$ for the single-excitation subtraction and addition processes can be solved exactly, it is intractable with an arbitrary number of excitations. Therefore, we perform an effective linearisation of the solution to the Schr\"{o}dinger equation before solving for the couplings. The linearised solution has a form similar to the single-excitation solution and serves as a starting point for a variational ansatz.  This provides a semi-analytic form that achieves near-unit fidelity for arbitrary excitation number. 

In this section, we introduce the model before revisiting the single-excitation solution. We follow this by solving the Schr\"{o}dinger equation for the two-excitation subtraction process and then introduce the linearisation step before solving for the dynamic coupling strength explicitly. Using induction, we obtain the couplings for the $n$-excitation subtraction processes. We show that, under time and parity reversal, the couplings for the $n$-excitation addition processes can be obtained from the subtraction couplings. Lastly, we showcase the results using numerical simulations.

\subsection{Model and background}
We begin by considering a two-level system (TLS) coupled with a time-dependent coupling strength to a one-dimensional continuum of right-propagating photons under the dipole, rotating-wave, and Markov approximations. After renormalizing the frequency of the TLS, assuming it to be positioned at $x=0$, and setting the group velocity to $v_g=\hbar=1$, the Hamiltonian reads, 
\begin{equation}
\hat{H}(t)=-i\int_{\mathbb{R}} dx \, \hat{a}^\dagger(x) \partial_x \hat{a}(x)+g(t)\hat{\sigma}_+\hat{a}(0)+\textrm{h.c.}.
\end{equation} 
Here, $\hat{a}(x)$ ($\hat{a}^\dagger(x)$) is the photon annihilation (creation) operator at position $x$, $\hat{\sigma}_-$ ($\hat{\sigma}_+$) is the Pauli lowering (raising) operator of the TLS, and $\textrm{h.c.}$ indicates the Hermitian conjugate of the preceding term. The first term in the Hamiltonian generates propagation of the photons to the right, while the second and third terms describe the interaction between the TLS and the photon field. The unidirectional coupling is the signature of chiral waveguide QED (quantum electrodynamics) systems \cite{Lodahl2017Nature,Sollner2015NNA, Yudson1985JETP, Mahmoodian2020PRX,Suarez2025PRXQ,Joshi2023PRX}. In the interaction picture with respect to the free-propagation term, the wavefunction evolution is described by the Hamiltonian,
\begin{equation}
    \hat{H}_I(t)=g(t)\hat{\sigma}_+\hat{a}(-t)+g^*(t)\hat{a}^\dagger(-t)\hat{\sigma}_-\label{hamiltonian}.
\end{equation}
In this co-moving frame and with local interactions, one can interpret the TLS as a boundary that is moving to the left from $x=0$ to $x=-t$, while interacting with the photon field with coupling strength $g(t)$.

\subsection{Single excitation}
\label{single excitation}
Let us review the derivation of the dynamic coupling strength $g(t)$ by solving the Schr\"{o}dinger equation such that the TLS either absorbs or emits a single photon in a given temporal mode. Since the total number of excitation is conserved, the dynamics are entirely contained in the single-excitation subspace which is captured by the interaction-picture ansatz,
\begin{equation}
\ket{\psi(t)}=\int_\mathbb{R}dx~\alpha(x,t)\hat{a}^\dagger(x)\ket{0,g}+\beta(t)\ket{0,e}.
\end{equation}
Here, $\ket{0,i}=\ket{0}\otimes\ket{i}$ is the product state of the vacuum state of the one-dimensional continuum and the ground ($i=g$) or excited ($i=e$) state of the TLS, $\alpha(x,t)$ and $\beta(t)$ are the corresponding wavefunction amplitudes. At all times, they satisfy the normalisation condition, $\int_\mathbb{R}dx|\alpha(x,t)|^2+|\beta(t)|^2=1$. From the Schr\"{o}dinger equation, the equations of motion are,
\begin{align}
    &i\partial_t\alpha(x,t)=\delta(x+t)g^*(t)\beta(t),\label{one photon d alpha}\\
    &i\partial_t\beta(t)=g(t)\alpha(-t,t).\label{one photon d beta}
\end{align}
We highlight that the $-t$ argument arises as it corresponds to the $x$ coordinate of the TLS in the interaction picture at time $t$. Before solving these differential equations, let us adapt the language of input--output theory by labeling the single-photon wavefunction $\alpha(x,t)$ in two different regions corresponding to before and after interacting with the TLS,
\begin{equation}
    \alpha(x,t)=\left\{
  \begin{array}{rcr}
    \alpha_0(x,t), &  & x\leq-t \\
    \alpha_1(x,t), &  & x\geq -t \\
  \end{array}
\right.
\end{equation}
since the TLS is moving at position $-t$, then $\alpha_0(x,t)$ and $\alpha_1(x,t)$ are the wavefunctions to the left and right of the TLS. When evaluated at $x=-t$, $\alpha_0(-t,t)$ and $\alpha_1(-t,t)$ are the input and output wavefunctions, respectively. Since Eq.~\eqref{one photon d alpha} contains a $\delta$-function, we must impose a regularisation condition, $\alpha(-t,t)=[\alpha_0(-t,t)+\alpha_1(-t,t)]/2$, for the single-photon wavefunction evaluated at the position of the TLS \cite{Yudson1985JETP, Shen2007PRA}.

Integrating Eq.~\eqref{one photon d alpha} in a region away from the position of the photon gives $\alpha(x,t)=\alpha(x,t')$ for $-x\notin[t',t]$. Physically, this means that the wavefunction is constant in the absence of interaction. Alternatively, in the original picture, this implies that the photon freely propagates before and after the scattering process. This allows us to relate the photon input wavefunction to the initial state by $\alpha_0(-t,t)=\alpha_0(-t,0)$. On the other hand, integrating over the position of the photon produces the boundary condition or equivalently, the single-excitation input--output relation,
\begin{equation}
    \alpha_1(-t,t)-\alpha_0(-t,t)=-ig^*(t)\beta(t).\label{one photon bc}
\end{equation}
With the regularisation condition and the boundary condition, Eq.~\eqref{one photon d beta} becomes a differential equation for the wavefunction of the excited state population, $\partial_t\beta(t)=-|g(t)|^2\beta(t)/2-ig(t)\alpha_0(-t,0)$. The solution to this is
\begin{equation} \label{eq:singleBetaSol}
    \beta(t)=e^{-\frac{G(0,t)}{2}}\left[\beta(0)-i\int_0^t dt'~g(t')e^{\frac{G(0,t')}{2}}\alpha_0(-t',0)\right],
\end{equation}
where $G(0,t)=\int_0^tdt'~|g(t')|^2$. In the above equation for $\beta(t)$, the exponential factor outside the square brackets is responsible for the decay of the TLS, while the terms on the inside are the initial excited state amplitude and the source term arising from the incoming photon. 

To solve for subtraction and addition of a single photon we must substitute Eq.~\eqref{eq:singleBetaSol} into the boundary condition Eq.~\eqref{one photon bc}, then set the appropriate initial conditions and the desired boundary conditions. For single-photon emission i.e., addition to vacuum, the goal is to release the excitation from the TLS as a single photon of a specific temporal mode, $f_{\rm out}(-t)$. Therefore, the initial and boundary conditions are $\beta(0)=1$, $\alpha(x,0)=0$ and $\alpha_0(-t,t)=0$, $\alpha_1(-t,t)=f_{\rm out}(-t)$. All together with Eq.~\eqref{one photon bc}, we have the single-photon addition boundary condition, $f_{\rm out}(-t)=-ig^*(t)e^{-G(0,t)/2}$. This equation can be inverted by integrating its squared modulus, $|f_{\rm out}(-t)|^2=-\partial_te^{-G(0,t)}$, and the explicit time-dependent coupling strength reads \cite{Gough2015EPJ},
\begin{equation}
\label{one photon add g}
    g_{{\rm add},1}^*(t)=\frac{if_{\rm out}(-t)}{\sqrt{\int_t^\infty dt'~|f_{\rm out}(-t')|^2}}.
\end{equation}
Here, the subscript denotes that it is a tailored dynamic coupling for the single-excitation addition process. Conversely, the process of single-photon absorption is when a photon in the temporal mode $f_{\rm in}(-t)$ is completely absorbed and excites the TLS which remains in $\ket{e}$. Here, the initial and boundary conditions are $\beta(0)=0$, $\alpha_0(x,0)=f_{\rm in}(x)$, $\alpha_1(x,0)=0$ and $\alpha_0(-t,t)=f_{\rm in}(-t)$, $\alpha_1(-t,t)=0$ i.e., we require no photon to be transmitted. Substituting these conditions into Eq.~\eqref{one photon bc}, we obtain the single-photon subtraction boundary condition, $f_{\rm in}(-t)=g^*(t)e^{-G(0,t)/2}S_{1/2}(0,t)$, where
\begin{equation}
\label{source term}
    S_v(a,b)=\int_a^b dt'~g(t')e^{vG(a,t')}f_{\rm in}(-t'),
\end{equation}
is the source term to excite the TLS in the time interval $[a,b]$ with $v\in\mathbb{C}$. The physical interpretation behind the product of $S_{1/2}(0,t)$ and the decay term $g^*(t)e^{-G(0,t)/2}$ is straightforward: the TLS is being excited by a photon in the time interval $[0,t]$, but at the same time the excitation could also decay. The explicit time-dependent coupling strength is given by (see Appendix~\ref{appendix inversion}) \cite{Nurdin2016IEEE},
\begin{equation}
\label{one photon sub g}
    g_{{\rm sub},1}^*(t)=\frac{f_{\rm in}(-t)}{\sqrt{\int_0^tdt'|f_{\rm in}(-t')|^2}}.
\end{equation}

\subsection{Two excitations}
\label{two-excitation}
Now that we have reviewed the single-excitation case, in this subsection, we generalise the subtraction and addition processes to two excitations. Ideal two-excitation addition occurs when the TLS is initially in the excited state and undergoes dynamic stimulated emission to emit a photon to the incoming single-photon Fock state with unit probability. The output two-photon state must then be a Fock state with a separable wavefunction.  Conversely, ideal two-excitation subtraction occurs when the TLS absorbs a photon from the input two-photon Fock state with unit probability. Notice that if the subtraction process has unit absorption probability, the output state is automatically a single-photon Fock state. However, the same cannot be said about the addition process, as the TLS could emit a photon with unit probability, but the output wavefunction is not guaranteed to be single-moded. Moreover, we emphasize that the output mode does not necessarily have to be the same as the input mode. Solving for the dynamics required for ideal addition and subtraction in the two-excitation counterpart is significantly more difficult than for one excitation due to the nonlinearity of the TLS. Hence, we only explicitly obtain the time-dependent coupling strength after performing a linear approximation. Nevertheless, as we show in Sec.~\ref{TLS results}, these approximate solutions provide near-unit fidelities.

By the conservation of the total number of excitations, the dynamics are fully captured by the two-excitation ansatz,
\begin{align}
    \ket{\psi(t)}=&\frac{1}{\sqrt{2}}\int_{\mathbb{R}^2} dx_1dx_2 ~ \alpha(x_1,x_2,t)\hat{a}^\dagger(x_1)\hat{a}^\dagger(x_2)\ket{0,g}\nonumber\\
    &+\int_\mathbb{R} dx ~ \beta(x,t)\hat{a}^\dagger(x)\ket{0,e}.
\end{align}
where $\alpha(x_1,x_2,t)$ is the two-photon symmetric wavefunction with the TLS in the ground state, and similarly $\beta(x,t)$ is the single-photon wavefunction with the TLS in the excited state. These wavefunctions are normalised by $\int_{\mathbb{R}^2}dx_1dx_2~|\alpha(x_1,x_2,t)|^2+\int_\mathbb{R}dx~|\beta(x,t)|^2=1$. From the Schr\"{o}dinger equation, the equations of motion are
\begin{align}
    i\partial_t\alpha(x_1,x_2,t)&=\frac{g^*(t)}{\sqrt{2}}[\beta(x_1,t)\delta(x_2+t)\label{two photon d alpha}\\
    &+ \beta(x_2,t)\delta(x_1+t)], \nonumber\\
    i\partial_t\beta(x,t)&=\sqrt{2}g(t)\alpha(x,-t,t).\label{two photon d beta}
\end{align}
Restricting the domain to $x_1\geq x_2$ impose a position ordering on the two photons, rendering the two-photon wavefunction unsymmetrised. Importantly, this implies that the scattering process becomes \textit{sequential}, allowing us to solve for the wavefunctions in distinct regions defined by the photons' coordinates relative to the TLS. Specifically, we label the wavefunctions in different regions as follows,
\begin{align}
    \alpha_{\rm unsym}(x_1,x_2,t)=&\left\{
  \begin{array}{rcr}
    \alpha_0(x_1,x_2,t), &  & -t\geq x_1\geq x_2 \\
    \alpha_1(x_1,x_2,t), &  & x_1\geq -t\geq x_2 \\
    \alpha_2(x_1,x_2,t), &  & x_1\geq x_2 \geq -t
  \end{array}
\right.,\\
\beta(x,t)=&\left\{
  \begin{array}{rcr}
    \beta_1(x,t), &  & x\leq -t \\
    \beta_2(x,t), &  & x\geq -t \\
  \end{array}
\right..
\end{align}
The interpretation of each labeling is analogous to that of the input--output wavefunctions of the single-excitation problem. In the interaction picture, the TLS is moving to the left at position $-t$, then the subscripts of the two-photon wavefunctions in different regions indicate the number of photons to the right of the TLS, for example, $\alpha_1(x_1,x_2,t)$ is the two-photon wavefunction with the first photon at position $x_1$ to the right of the TLS and the second photon at position $x_2$ to the left. Meanwhile, the subscripts of the single-photon wavefunction label which photon has been absorbed by the TLS, for example, $\beta_1(x,t)$ is the single-photon wavefunction with the first photon absorbed while the second photon is at position $x$ that is to the left of the TLS. 

Again, we must impose regularisation conditions for the two-photon wavefunction when either one of the photons is evaluated at the position of the TLS: the unsymmetrical two-photon wavefunction is the average of the wavefunctions in neighbouring regions \cite{Yudson1985JETP, Shen2007PRA}, i.e., $\alpha_{\rm unsym}(-t,x,t)=[\alpha_0(-t,x,t)+\alpha_1(-t,x,t)]/2$ and similarly, $\alpha_{\rm unsym}(x,-t,t)=[\alpha_1(x,-t,t)+\alpha_2(x,-t,t)]/2$. Furthermore, the two-photon wavefunction is constant in the absence of interaction i.e., if $-x_1,-x_2\notin [t',t]$, then $\alpha_{\rm unsym}(x_1,x_2,t)=\alpha_{\rm unsym}(x_1,x_2,t')$. On the other hand, the effects of the interaction on the wavefunctions can be seen by integrating Eq.~\eqref{two photon d alpha} over the position of the photons; which produces the boundary conditions relating the neighboring regions of the two-photon wavefunctions and the single-photon wavefunctions
\begin{align}
    &\alpha_1(-t,x_2,t)-\alpha_0(-t,x_2,t)=\frac{-ig^*(t)}{\sqrt{2}}\beta_1(x_2,t), \label{two photon beta1 bc}\\
    &\alpha_2(x_1,-t,t)-\alpha_1(x_1,-t,t)=\frac{-ig^*(t)}{\sqrt{2}}\beta_2(x_1,t).\label{two photon beta2 bc}
\end{align}
Moreover, since Eq.~\eqref{two photon d beta} does not contain a $\delta$-function, the single-photon wavefunctions in different regions match on the boundary, i.e., $\beta_1(-t,t)=\beta_2(-t,t)$. 

We proceed by solving for the unknown wavefunctions in the sequence of the photons' coordinates. Specifically, we first solve for the single-photon wavefunction, $\beta_1(x_2,t)$. With the boundary condition in Eq.~\eqref{two photon beta1 bc}, we relate the two-photon wavefunction, $\alpha_1(x_1,x_2,t)$ to the initial conditions. Repeating the process, we solve for the single-photon wavefunction in the next region, $\beta_2(x_1,t)$, and with Eq.~\eqref{two photon beta2 bc}, we can express the output two-photon wavefunction, $\alpha_2(x_1,x_2,t)$ in terms of the initial conditions. 

By combining the regularisation and the boundary conditions, Eq.~\eqref{two photon d beta} produces differential equations for the single-photon wavefunctions, $\partial_t\beta_1(x_2,t)=-|g(t)|^2\beta_1(x_2,t)/2-i\sqrt{2}g(t)\alpha_0(-t,x_2,0)$ and $\partial_t\beta_2(x_1,t)=-|g(t)|^2\beta_2(x_1,t)/2-i\sqrt{2}g(t)\alpha_1(x_1,-t,-x_1)$. Depending on the initial condition, the $\beta_1(x_2,t)$ equation describes the dynamics of the arrival or the emission of the first photon in the domain, $t\in[0,-x_1]$. While the $\beta_2(x_1,t)$ equation describes the dynamics when a photon has scattered past the TLS and the second photon is incoming, with the corresponding domain, $t\in[-x_1,-x_2]$. Notably, we have taken advantage of the free propagation of photons by substituting $\alpha_0(-t,x_2,t)=\alpha_0(-t,x_2,0)$ and $\alpha_1(x_1,-t,t)=\alpha_1(x_1,-t,-x_1)$ into the single-photon wavefunction differential equations. Explicitly, the solution for $\beta_1(x_2,t)$ is
\begin{align}
    \beta_1(x_2,t)
    &=e^{-G(0,t)/2}\bigg[\beta_1(x_2,0)\nonumber\\
    &-i\sqrt{2}\int_0^tdt'g(t')e^{G(0,t')/2}\alpha_0(-t',x_2,0)\bigg].\label{two photon beta1 sol}
\end{align}
The dynamics of the TLS are similar to that of the single-excitation case, where the first term inside the square brackets of Eq.~\eqref{two photon beta1 sol} corresponds to the decay of any initial excitation of the TLS and constitutes the emission of the first photon at position $-t\geq x_2$. The second term is responsible for exciting the TLS, but only from the first photon. Together with the $\beta_1(x_2,t)$ solution and the boundary condition in Eq.~\eqref{two photon beta1 bc}, we can completely express $\alpha_1(x_1,x_2,t)$ in terms of the initial conditions. We are now ready to solve the interaction in the next region. Matching the single-photon wavefunctions at the boundary, the solution to the $\beta_2(x_1,t)$ differential equation is,
\begin{align}
    \beta_2&(x_1,t)=e^{-G(-x_1,t)/2}\bigg[\beta_1(x_1,-x_1)\nonumber\\
    &-i\sqrt{2}\int_{-x}^tdt'g(t')e^{G(-x_1,t')/2}\alpha_1(x_1,-t',-x_1)\bigg].\label{two photon beta2 sol}
\end{align}
The first term in the square brackets corresponds to the decay of the leftover excitation in the TLS from the previous region, but the emitted photon is instead the second photon as the first photon is to the right of the TLS at position $x_1$. At the same time, the TLS has a chance to be excited by the incoming second photon, this is indicated by the having $\alpha_1(x,-t,-x)$ in the source term. With the boundary condition given by Eq.~\eqref{two photon beta2 bc}, we can express the output two-photon wavefunction, $\alpha_2(x_1,x_2,t)$ in terms of the initial conditions. We have therefore solved for the dynamics of the system in terms of the initial conditions and the dynamic coupling coefficient $g(t)$.

We now wish to proceed to determine the dynamic stimulated emission process that enables photon addition and its reverse process that allows for photon subtraction. This means finding the appropriate dynamical coupling $g(t)$ that enables addition or subtraction. We proceed by first solving the two-excitation subtraction process. The corresponding initial state is,
\begin{equation}
    \ket{\psi(t=0)}=\frac{1}{\sqrt{2}}\left[\int_\mathbb{R} dx~f_{\rm in}(x)\hat{a}^\dagger(x)\right]^2\ket{0,g},
\end{equation}
with the normalised input mode, $\int_\mathbb{R} dx~|f_{\rm in}(x)|^2=1$. This initial state corresponds to the initial conditions, $\beta(x,0)=0$ and $\alpha_0(x_1,x_2,0)=f_{\rm in}(x_1)f_{\rm in}(x_2)$. For the ideal two-excitation subtraction process, a sufficient condition on the boundary condition is $\alpha_2(x_1,x_2,t)=0$; hence, Eq.~\eqref{two photon beta2 bc} becomes $\alpha_1(x_1,-t,t)=ig^*(t)\beta_2(x_1,t)/\sqrt{2}$. Expressed in terms of the initial condition, the boundary condition reads 
\begin{widetext}
\begin{align}
    f_{\rm in}(-t)B(x_1)
    =g^*(t)\left[f_{\rm in}(x_1)e^{-G(0,t)/2}S_{1/2}(0,-x_1)+B(x_1)e^{-G(-x_1,t)/2}S_{1/2}(-x_1,t)\right]. \label{output relation}
\end{align}
\end{widetext}
Here, $B(x)=f_{\rm in}(x)-g^*(-x)e^{-G(0,-x)/2}S_{1/2}(0,-x)$, where $S_{1/2}(0,-x)$ is defined by Eq.~\eqref{source term}. Notice that, $B(x)=0$ is the boundary condition required for the single-excitation subtraction process, indicating that its solution would not satisfy Eq.~\eqref{output relation} and fail in two excitations. The ramification is that no dynamic coupling exists such that a TLS is excited by a superposition of single- and two-photon Fock state with unit probability. An intuitive explanation is that the interaction in the single-excitation subspace is linear, and the nonlinearity of the TLS only comes into effect when the excitation number is greater than one. Hence, the single- and two-excitation dynamics are dramatically different, consequently, they do not share a common dynamic coupling for exciting the TLS.

The nonlinear response of the TLS can be examined by pinpointing the contributing terms in Eq.~\eqref{output relation}. The left-hand side of Eq.~\eqref{output relation} is the two-photon wavefunction with a photon on either side of the TLS; there are two contributing terms, namely, the free propagation term and the interaction term. The former can be seen by the product of the input temporal modes evaluated at different positions, $f_{\rm in}(-t)f_{\rm in}(x)$. The latter occurs when the TLS absorbs the first photon in the time interval $[0,-x_1]$ and decays to emit a photon just to the right of the TLS at position $x_1$, this is indicated by the product of the decay and the source term, $e^{-G(0,-x_1)/2}S_{1/2}(0,-x_1)$. 

Meanwhile, the right-hand side of Eq.~\eqref{output relation} is the single-photon wavefunction with a photon to the right of the excited TLS; it consists of a superposition of three different contributing processes. In the order of summation, 
\begin{enumerate}
    \item The first process occurs when the TLS absorbs the first photon in the time interval $[0,-x_1]$, as indicated by the source term, $S_{1/2}(0,-x_1)$. But instead of decaying, the TLS remains in the excited state in the time interval $[0,t]$. The excited TLS allows the second photon to freely propagate past the TLS, as suggested by having the temporal mode evaluated at $x_1$. Subsequently, the position ordering of the photons is swapped.
    \item The second process, hidden in $B(x_1)$ on the right-hand side of Eq.~\eqref{output relation}, occurs when the first photon freely propagates and passes the TLS at position $x_1$. The TLS is then excited by the second photon in the time interval $[-x_1,t]$, as can be seen by the source term $S_{1/2}(-x_1,t)$.
    \item The third process, also hidden in $B(x_1)$, occurs when the TLS absorbs the first photon between the time interval $[0,-x_1]$ then emits a photon at position $x_1$ and starts to absorb the second photon in the interval $[-x_1,t]$, as indicated by the product, $e^{-G(0,-x_1)/2}S_{1/2}(0,-x_1)e^{-G(-x_1,t)/2}S_{1/2}(-x_1,t)$.
\end{enumerate}

Finding the time-dependent coupling strength $g(t)$ that satisfies Eq.~\eqref{output relation} would produce the ideal two-excitation subtraction process. However, Eq.~\eqref{output relation} involves $g(t)$ evaluated at different times and integrated between different time intervals, hence it is difficult to find an explicit expression of the coupling strength. Instead, we make a linear approximation that eliminates the second and the third processes of the right-hand side of Eq.~\eqref{output relation} and only the first process remains, which corresponds to the TLS absorbing the first photon while the second photon freely propagates past the TLS. We note that the position ordering of the photons is merely a consequence of working with the unsymmetrised wavefunction, in actuality, the TLS could be excited by any of the two photons. Then this linear approximation can be understood as the TLS only interacts with one photon from the two-photon field. Moreover, the linear approximation corresponds to setting $x_1=-t$ in Eq.~\eqref{output relation} and it becomes the linearised two-excitation subtraction boundary condition, $f_{\rm in}(-t)=2g^*(t)e^{-G(0,t)/2}S_{1/2}(0,t)$, which differs from the boundary condition for the single-photon subtraction process by a multiplicative factor of 2. 

Before finding the time-dependent coupling strength that satisfies the linearised boundary condition, we present a more general boundary condition that appears in the $n$-photon subtraction process in Sec.~\ref{n-excitation} and the three-level system photon subtraction in Sec.~\ref{three-level system}. Together, the general boundary condition and its solution are (see Appendix~\ref{appendix inversion})
\begin{align}
    &f_{\rm in}(-t)=ug^*(t)\int_0^t dt'~g(t')e^{v G(t,t')}f_{\rm in}(-t'),\label{general bc}\\
    &g^*(t)=\frac{f_{\rm in}(-t)}{\sqrt{2\Re(u-v)F(t)^{\frac{u-v}{\Re(u-v)}}}},\label{general g(t)}
\end{align}
where $u,v\in\mathbb{C}$, $\Re(\cdot)$ denotes the real part of a complex number and $F(t)=\int_0^tdt'~|f_{\rm in}(-t')|^2$. From Eq.~\eqref{general g(t)}, it is clear that $\Re(u-v)\neq0$. 

Setting $u=2$ and $v=1/2$ in Eq.~\eqref{general bc} and Eq.~\eqref{general g(t)}, we recover the linearised two-excitation subtraction boundary condition and obtain its solution. To check the validity of the linearised solution, we derive the output single-photon wavefunction from Eq.~\eqref{two photon beta2 sol} by taking the infinite-time limit, $\lim_{t\rightarrow\infty}\beta_2(x,t)=-i\sqrt{\frac{3}{8}}\left[1+F(-x)^{2/3}\right]f_{\rm in}(x)$. Integrating the squared modulus of the single-photon wavefunction yields the probability of a successful subtraction, $\int_\mathbb{R} dx~|\lim_{t\rightarrow\infty}\beta_2(x,t)|^2=69/70\approx0.986$. It is remarkable that the solution obtained from solving the approximate boundary condition can produce such a high probability.

\begin{figure*}[ht!]
\centering
\begin{overpic}[width=0.85\textwidth]{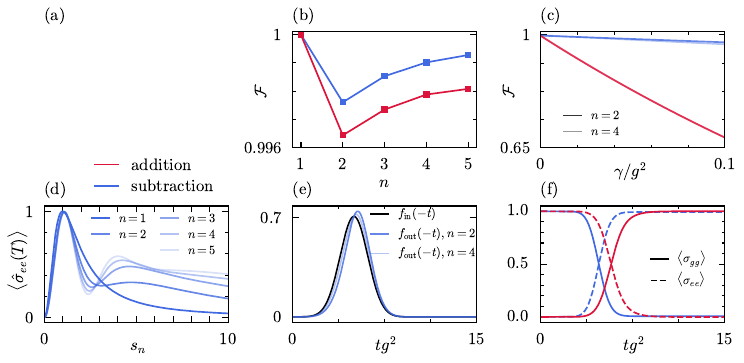}
  \put(5, 30){\includegraphics[width=0.2\textwidth]{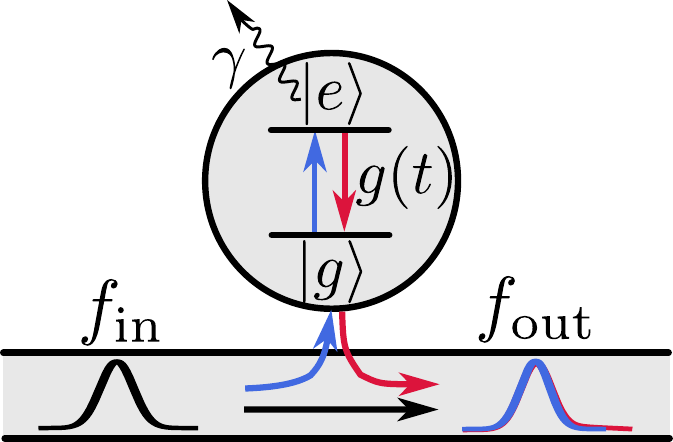}}
\end{overpic}
    \caption{Simulation results for the addition and subtraction processes, labeled by colours blue and red, respectively. The schematics of both processes are depicted in (a), where $f_{\rm in}$ is the input temporal mode and $f_{\rm out}$ the output mode. The excitation in the TLS of dynamic stimulated emission follows the red arrows, the reverse process where a photon is absorbed by the TLS follows the blue arrows. Results in all sub-figures (b-f) are simulated with a Gaussian input temporal mode of width $\sigma g^2=2/\sqrt{\pi}$, where $g^2=|g_{{\rm sub},1}(t_0)|^2$ is the reference coupling rate of the system (see main text). (b) The lossless unconditional fidelities for the subtraction and addition processes for up to five excitations and (c)  in the presence of loss for two- and four-excitations. The reduced addition fidelities in (c) are due to the TLS's idling time (see main text). (d) At final time $T$, the excited state population for the subtraction process versus the variational parameter $s_n$ for up to five excitations. (e) The input Gaussian temporal mode and the output temporal modes for the two- and four-excitation subtraction processes. (f) The ground (solid) and excited (dashed) state population dynamics for the two-excitation subtraction and addition processes.}
    \label{fig:TLS_sub_add}
\end{figure*}

We now make a critical assumption allowing us to construct a variational ansatz: we assume that, if we were to include the nonlinear terms in the solution, they would have a largely similar time-dependence thereby only changing the amplitude of $g(t)$ and preserving its shape. We therefore multiply our linearised solution with a variational parameter, $s_2\in \mathbb{R}$, here the subscript denotes the excitation number. In terms of Eq.~\eqref{general bc}, this variation corresponds to solving a slightly different boundary condition. Explicitly, the dynamic coupling strength ansatz for the two-excitation subtraction process is,
\begin{equation}
\label{two photon sub g}
    g_{\rm sub,2}^*(t)=\frac{s_2f_{\rm in}(-t)}{\sqrt{3\int_0^tdt'~|f_{\rm in}(-t')|^2}}.
\end{equation}
After optimising towards unit probability with a Gaussian input temporal mode, the value of the optimised variational parameter is $s_2=1.072$, which is \textit{invariant} under changes to the width of the Gaussian. The probability or the fidelity with the optimised parameter for the two-excitation subtraction process is $\mathcal{F}>0.996$, the result is presented in Fig.~\ref{fig:TLS_sub_add} and further discussed in Sec.~\ref{TLS results}.

We now move to the two-excitation addition problem. Here, the initial state consists of a single photon in the waveguide and the TLS excited, $\ket{\psi(t=0)}=\int_\mathbb{R}dx~f_{\rm in}(x)\hat{a}^\dagger(x)\ket{0,e}$. The goal is for the TLS to undergo a dynamic stimulated emission process such that the two-photon output state is a Fock state with unit probability. Solving for the two-excitation addition process in the same way as the subtraction process would relate the initial conditions to the output two-photon wavefunction, $\alpha_2(x_1,x_2,t)$. However, this is unwanted because the only constraint on the output two-photon wavefunction is for it to be separable, and we do not wish to impose any further constraints such as specifying the output temporal mode. 

Instead, we wish to derive the dynamic coupling for the addition process by applying reversal transformations to the subtraction process. Assuming the two-excitation subtraction process has unit probability, then the two-excitation addition process is the time-reversed process \cite{Lund2024PRL}. However, a time-reversal transformation is not enough to bring the Hamiltonian back into the form of Eq.~\eqref{hamiltonian} since the direction of travel is also flipped. Therefore, in the interaction picture, we must perform a time-reversal transformation, a translation to reposition the TLS back to the starting position, followed by a parity inversion such that the TLS still travels to the left. Moreover, the initial and final states of the subtraction process under these transformations are mapped to the final and initial states of the addition process, respectively. Hence, in order to reproduce the subtraction in reverse, we must prepare the initial state of the addition process according to the final state of the subtraction process. This would lead to a coupling strength that depends on the output state of the addition process, which is unwanted. To get around this issue, we make a single-mode approximation such that the temporal mode of the single-photon wavefunction of the two-excitation subtraction process is the same as the input temporal mode, $f_{\rm in}(-t)$, of the two-photon Fock state. This approximation is valid since their overlap is near-unity, $|\sqrt{70/69}\int_\mathbb{R} dx~f_{\rm in}^*(x)\lim_{t\rightarrow\infty}\beta_2(x,t)|^2=112/115\approx 0.974$. All together, we perform the three transformations with the single-mode approximation and obtain the dynamic coupling strength for the two-excitation addition process that only depends on the initial state (see Appendix~\ref{appendix reversal}),
\begin{equation}
\label{two photon add g}
    g_{\rm add,2}^*(t)=\frac{a_2f_{\rm in}(-t)}{\sqrt{3\int_t^\infty dt'~|f_{\rm in}(-t')|^2}}.
\end{equation}
Here, we have also included a variational parameter that takes the optimised value, $a_2=0.816$, and $f_{\rm in}(-t)$ is the temporal mode of the initial input single-photon Fock state. The fidelity of the two-excitation addition process with the optimised dynamic coupling is $\mathcal{F}>0.996$, the result is presented in Fig.~\ref{fig:TLS_sub_add} and further discussed in Sec.~\ref{TLS results}.

\subsection{$n$ excitations}
\label{n-excitation}

In this subsection, we generalise our solution for two-excitation subtraction and addition processes to $n$-excitation processes as illustrated by the schematic of Fig.~\ref{fig:TLS_schematic}. The ideal $n$-excitation subtraction process is when the TLS absorbs a photon from an incoming $n$-photon Fock state with unit probability, and the outgoing state is an $(n-1)$-photon Fock state. Conversely, the ideal $n$-excitation addition process is the reverse, the initially excited TLS emits a photon to an incoming $(n-1)$-photon Fock state with unit probability, and the outgoing state is a $n$-photon Fock state. The strategy to obtain the time-dependent coupling strength is similar to that of the two-excitation processes: we confine the dynamics to the $n$-excitation subspace and derive the equations of motion from the Schr\"{o}dinger equation. However, solving for the unsymmetrised wavefunctions from region to region is intractable as the number of terms grows quickly. Nonetheless, making the linear approximation (the TLS only interacts with one photon in total) at appropriate steps can greatly reduce the number of terms (see Appendix~\ref{appendix n-excitation}).

We solve the $n$-excitation subtraction process with the an initial $n$-photon Fock state in the waveguide and the TLS in the ground state, i.e., $\ket{\psi(t=0)}=\left[\int_\mathbb{R} dxf_{\rm in}(x)\hat{a}^\dagger(x)\right]^n\ket{0,g}/\sqrt{n!}$. The output boundary condition for $n$-excitation subtraction is $f_{\rm in}(-t)=ng^*(t)e^{-G(0,t)/2}S_{1/2}(0,t)$ (see Appendix~\ref{appendix n-excitation} for a derivation). Notice that this is in the form of the general boundary condition in Eq.~\eqref{general bc} and its solution is given by Eq.~\eqref{general g(t)} with $u=n$ and $v=1/2$. Similar to the treatment of the linearised two-excitation subtraction solution, we multiply the solution by a variational parameter, $s_n\in \mathbb{R}$, that we later optimise numerically. Explicitly, the time-dependent coupling strength for the $n$-excitation subtraction process is 
\begin{equation}
\label{n photon sub g}
    g_{{\rm sub},n}^*(t)=\frac{s_nf_{\rm in}(-t)}{\sqrt{(2n-1)\int_0^tdt'~|f_{\rm in}(-t')|^2}}.
\end{equation}

On the other hand, the initial state of the $n$-excitation addition process consists of an $(n-1)$-photon Fock state in the waveguide with the TLS excited, i.e., $\ket{\psi(t=0)}=[\int_{\mathbb{R}}dx~f_{\rm in}(x)\hat{a}^\dagger(x)]^{n-1}\ket{0,e}/\sqrt{(n-1)!}$. Applying the time-reversal, translation and parity inversion transformations to the Hamiltonian in Eq.~\eqref{hamiltonian} in the same way as described in Sec.~\ref{two-excitation}, we obtain the time-dependent coupling strength for the $n$-excitation addition process (see Appendix~\ref{appendix reversal}). Recall that we approximated the initial temporal mode and the output temporal mode of the subtraction processes to be the same. In fact, this approximation becomes more accurate with increasing $n$, as the overlap between the two increases with the excitation number (overlap $\approx0.99$). Explicitly, the coupling for addition reads
\begin{equation}
\label{n photon add g}
    g_{{\rm add},n}^*(t)=\frac{a_nf_{\rm in}(-t)}{\sqrt{(2n-1)\int_t^\infty dt'~|f_{\rm in}(-t')|^2}}.
\end{equation}
Here, $a_n\in\mathbb{R}$ is the variational parameter for the $n$-excitation addition process. The optimised parameters for subtraction and addition for up to five excitations are given in Table~\ref{tab:TLS variational parameters}. We emphasize that these optimised parameters also produce high fidelity for other mode shapes. Recall that the coupling strength for the single-excitation addition process is given by Eq.~\eqref{one photon add g}, which corresponds to setting $f_{\rm in}(-t)\rightarrow if_{\rm out}(-t)$ in Eq.~\eqref{n photon add g}. From this perspective, the single-excitation process is unique; the initial state is fixed, and the final state can be easily read off from the coupling strength.
\begin{table}[!t]
    \centering
    \begin{tabular}{|c|c|c|}
        \hline
        $n$ & $s_n$  & $a_n$ \\
        \hline
         1 & 1.000 & 1.000 \\
         \hline
         2 & 1.072 & 0.816\\
         \hline
         3 & 1.088 & 0.883\\
         \hline
         4 & 1.095 & 0.926\\
         \hline
         5 & 1.098 & 0.960\\ 
         \hline
    \end{tabular}
    \caption{Numerically optimised parameters for subtraction ($s_n$) and addition ($a_n$) processes for up to five excitations.}
    \label{tab:TLS variational parameters}
\end{table}

\subsection{Results and discussions}
\label{TLS results}

In this subsection, we present numerical results and discussions for subtraction and addition processes for up to five excitations. We first introduce the setup and the numerical techniques. Followed by showcasing the fidelities with and without losses. We show that the output temporal modes for two- and four-excitation subtraction processes are well approximated by the input temporal mode.

Since the couplings considered in this work are chiral, we model the dynamics using the quantum pulses formalism in Refs.~\cite{Combes2017APX,Kiilerich2019PRL}. Specifically, we set up an input virtual cavity in the upstream to inject the input state, followed by the TLS in the middle and a collection virtual cavity in the downstream to analyse the output state. We emphasize that these virtual cavities are absent in an experiment, and they are only introduced for the numerical analysis. To analyse a multimoded state, we factorise the first-order correlation function into a sum of orthogonal modes, $G^{(1)}(t,t')=\expval{\hat{a}^\dagger(-t)\hat{a}(-t')}=\sum_i\bar{n}_if^*_i(-t)f_i(-t')$, here, $f_i(-t)$ are the orthogonal modes and $\bar{n}_i$ are the corresponding mean photon number contained in each mode. We note that the photon field annihilation and creation operators in $G^{(1)}(t,t')$ are evaluated just after the TLS. 

To assess the performance of the dynamic stimulated emission and absorption processes, we analyse the mode with the highest mean photon number of each particular process. For the simulations without loss, we solve the closed-system dynamics using a matrix-product state approach \cite{Pichler2016PRL,Luo2024,ITensor2022}. When considering loss, we solve an open-system Lindblad Master Equation in QuTip \cite{qutip5}. The schematics of the simulated addition and subtraction processes, colour-coded by red and blue respectively, are presented in Fig.~\hyperref[fig:TLS_sub_add]{2(a)}. For up to five excitations, the incident states are photon Fock states in a Gaussian mode and the initial states of the TLS are prepared according to Fig.~\ref{fig:TLS_schematic}. We chose the input temporal mode to be a Gaussian ($f_{\rm in}(-t)=\frac{1}{\sqrt{\sigma\sqrt{\pi}}}e^{-(t-t_0)^2/2\sigma^2}$) of width $\sigma g^2=2/\sqrt{\pi}$, where $g^2=|g_{{\rm sub},1}(t_0)|^2$ is a reference coupling rate that is used to define the time-scale of the dynamics, and it is defined as the coupling strength of the single-photon subtraction process (Eq.~\eqref{one photon sub g}) evaluated at the center of a Gaussian input temporal mode. Most importantly, we chose the coupling strength for subtraction and addition for the TLS according to Eq.~\eqref{n photon sub g} and Eq.~\eqref{n photon add g} respectively.

In analysing the performance, we define the fidelity of the $n$-excitation subtraction process by $\mathcal{F}=\Tr[\ket{n-1}_{\rm out}\bra{n-1}\hat{\rho}]$. Here, $\hat{\rho}$ is the asymptotic final state of the TLS and the photon field; $\ket{n-1}_{\rm out}$ is an $(n-1)$-photon Fock state of the photon field with the corresponding temporal mode determined from factorising the first-order correlation function. We highlight that this fidelity captures both the absorption efficiency and the single-mode character of the output photon state. For the special single-excitation subtraction process, the entire output field is required to be in vacuum. Equivalently, in the lossless case, single-excitation subtraction fidelity is the excited state population at the final time i.e., $\mathcal{F}=\Tr[\hat{\sigma}_{ee}\hat{\rho}]$. Similarly, the fidelity of the $n$-excitation addition process is $\mathcal{F}=\Tr[\ket{n}_{\rm out}\bra{n}\hat{\rho}]$. In the lossless scenario, the fidelities of the subtraction (blue) and addition (red) processes are presented in Fig.~\hyperref[fig:TLS_sub_add]{2(b)}. Unit fidelities are expected for the single-excitation processes, as their solutions are exact; however, remarkably, with slight numerical optimisation on the variational parameters (Table~\ref{tab:TLS variational parameters}), the fidelities are monotonically increasing with excitation number larger than one, from $\mathcal{F}\approx0.998$ to $\mathcal{F}\approx0.999$ for the subtraction processes and $\mathcal{F}\approx0.996$ to $\mathcal{F}\approx0.998$ for the addition processes. The slightly lower fidelities of the addition processes is likely caused by the mode mismatch before applying the reversal transformations. 

In Fig.~\hyperref[fig:TLS_sub_add]{2(c)}, we have considered the impact of experimental imperfections on the fidelities of the two- and four-excitation processes. Here, the TLS can decay into non-guided modes, and we model these loss mechanisms by including an extra Lindblad operator of the form $\hat{L}_1=\sqrt{\gamma}\hat{\sigma}_-$ in the Lindblad Master equation. The subtraction (blue) processes are tolerant to losses with fidelities $\mathcal{F}\approx0.97$ at 10\% loss ($\gamma/g^2=0.1$). Because in order for the TLS to lose an excitation to the non-guided mode, it needs to absorb a photon from the guided mode, hence it is still a valid subtraction. On the other hand, the fidelities of the addition (red) processes suffer greatly from losses, with $\mathcal{F}\approx0.68$ at 10\% loss. This is because the TLS is initially prepared in the excited state and it has to idle while waiting for the input temporal mode to arrive. During the idling time, the excitation in the TLS could spontaneously decay into non-guided modes, thereby no excitation is left to be emitted into the guided mode. To minimise the effects of losses on the addition processes in experiments, one could prepare an exponential input temporal mode and excite the TLS just as the mode is arriving such that the idling time is minimised.

Figure~\hyperref[fig:TLS_sub_add]{2(e)} depicts the input temporal mode, $f_{\rm in}(-t)$, and the output temporal modes, $f_{\rm out}(-t)$, for the two- and four-excitation subtraction processes. Qualitatively, the overlap between these modes increases with the number of excitations, which justifies the approximation, $f_{\rm in}(-t)=f_{\rm out}(-t)$ that we have used to derive the time-dependent coupling strengths for the $n$-excitation addition process. In Fig.~\hyperref[fig:TLS_sub_add]{2(f)}, we showcase both the ground- (solid line) and excited-state (dashed line) population dynamics for the two-excitation subtraction (blue) and addition (red) processes.

\subsection{Connection to the Jaynes-Cummings model}
\label{connection to the JC model}

Here we provide an intuitive argument of how this excitation-number-dependent dynamic coupling can arise by approximating the TLS--photon interaction by a single-mode Jaynes-Cummings model. We also argue that no ``universal'' coupling exists to facilitate deterministic photon addition and subtraction across different excitation-number subspaces. Following that, we point out some of the similarities and differences of the results between the waveguide QED and that of the Jaynes-Cummings model.

The subtraction ansatz given by Eq.~\eqref{n photon sub g} can be intuitively understood by approximating the TLS-photon interaction as being Jaynes-Cummings-like, but with a time dependence \cite{Lund2024PRL}. One then approximates the photon-field operators of the interaction Hamiltonian in Eq.~\eqref{hamiltonian} by a product of the temporal mode and a single-mode bosonic operator i.e., $\hat{a}(-t)\rightarrow f_{\rm in}(-t)\hat{a}$, thereby Eq.~\eqref{hamiltonian} becomes the Jaynes-Cummings Hamiltonian, $\hat{H}(t)=g(t)f_{\rm in}(-t)\hat{\sigma}_+\hat{a}+{\rm h.c.}$. In this simplified model, the $n$-excitation subtraction process occurs when the system undergoes Rabi oscillations with the asymptotic transition, $\ket{n,g}\rightarrow\ket{n-1,e}$. A solution of the following form satisfies this transition by having the Rabi dynamics executing a $\pi$-pulse, that is $\sqrt{n}\int_0^\infty dt~g(t)f_{\rm in}(-t)=\pi/2$;
\begin{equation}
    g^*(t)=\frac{\pi f_{\rm in}(-t)}{4\sqrt{n\int_0^tdt'~|f_{\rm in}(-t')|^2}}.
\end{equation}
Although the $\pi$-pulse model fails to reproduce the exact solution of the single-excitation subtraction process when $n=1$, the only difference is a multiplicative factor. As an artifact of the Jaynes-Cummings model, the $\pi$-pulse model accounts for the variations in the Rabi frequency with the number of excitations, $n$. In the form of Eq.~\eqref{n photon sub g}, the $\pi$-pulse model produces a qualitative prediction for the optimised variational parameter, $s^{(\pi)}_n=\frac{\pi}{4}\sqrt{\frac{2n-1}{n}}=\{0.785,0.962,1.014,1.039,1.054\}$ for $n=\{1,2,\dots,5\}$. 

Importantly, we see that both the waveguide QED model and the Jaynes-Cummings model predict an excitation-number-dependent coupling is required for deterministic photon addition and subtraction. One naturally asks whether a ``universal'' coupling exists which accommodates for all excitation subspaces. Here we argue that no such coupling exists. Notice that the Hamiltonian in Eq.~\eqref{hamiltonian} is excitation-conserving which renders each distinct excitation-number subspace independent, i.e., evolution of $n$ excitations is independent from $m$, when $n\neq m$. Therefore, the zero-excitation state of the Hamiltonian, i.e., the vacuum and the ground state of the two-level system (TLS) $\ket{0,g}$ would never evolve to the excited state. Hence, if we wish to perform photon subtraction by starting with the TLS in $\ket{g}$ and the incoming state of light has nonzero support in vacuum (e.g., a squeezed vacuum) then the TLS cannot be excited deterministically. Furthermore, in the case of single-excitation photon subtraction, Eq.~\eqref{one photon sub g} suggests that there is a one-to-one correspondence between the coupling and the temporal mode function of the input single photon. The coupling is a solution of the deterministic single-excitation subtraction boundary condition ($B(x)=0$ in Eq.~\eqref{output relation}). Recall that Eq.~\eqref{output relation} is the deterministic two-excitation subtraction boundary condition. To simultaneously subtract a photon from a single- and two-photon superposition state, we must have a coupling that satisfies both $B(x)=0$ and Eq.~\eqref{output relation}. It is clear that these boundary conditions do not share the same solution. Therefore, no coupling exists such that a TLS can deterministically subtract a photon from an arbitrary superposition of one and two photons. We believe that this argument holds for all excitation numbers. The Jaynes-Cummings model shares the same feature with the evidence being the $\sqrt{n}$ factor for the Rabi dynamics to execute a $\pi$-pulse. Since photon addition is the time-reverse process of photon subtraction therefore, no coupling exists for a deterministic photon addition process on a superposition input state.

The single-mode Jaynes-Cummings model also predicts that, if the Rabi dynamics executes a $\pi$-pulse, then multiplying the coupling strength by odd integers would also satisfy the asymptotic transition of the subtraction process. Thereby completing multiple Rabi cycles. In the waveguide QED model, this would imply that when sweeping $s_n$, we should observe Rabi oscillations. From Fig.~\hyperref[fig:TLS_sub_add]{2(d)}, this is not the case. Apart from the first peak, we do not observe unit excited state population in the asymptotic limit as we sweep the variational parameter in the range, $s_n\in[0,10]$. This is a major discrepancy from the dynamics of the Jaynes-Cummings model, and qualitatively, it stems from the distinction between a TLS interacting with a single mode versus multiple propagating modes.

\subsection{Cascaded addition and subtraction}
\label{cascaded addition and subtraction}

\begin{figure}[t]
    \centering
    \begin{overpic}[width=0.48\textwidth]{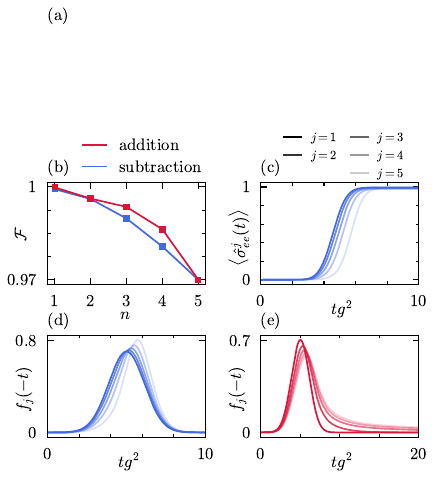}
        \put(9,73){\includegraphics[width=0.42\textwidth]{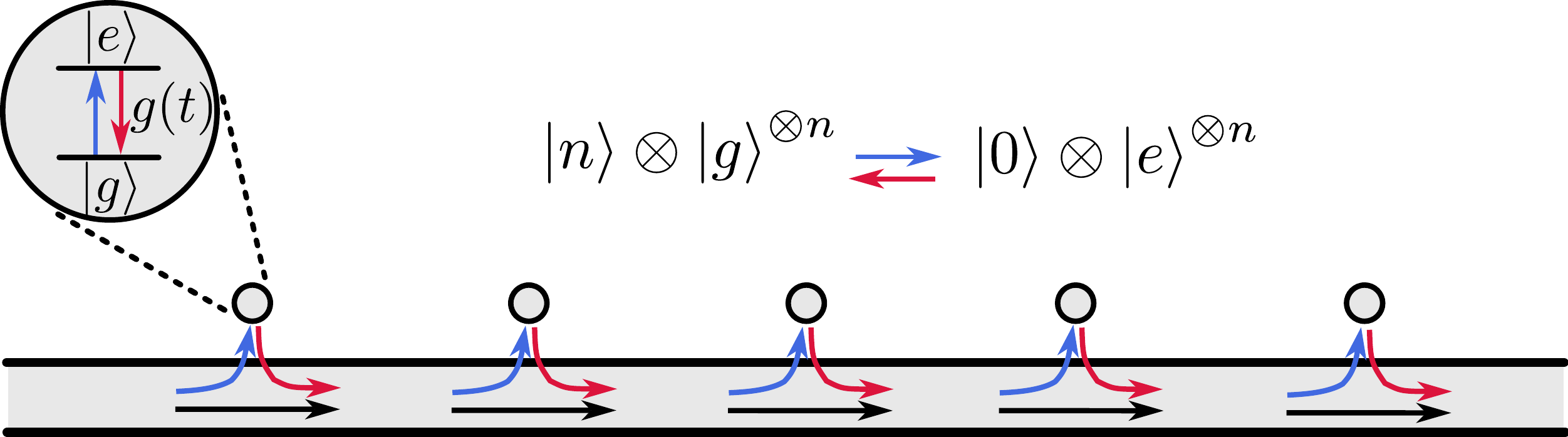}}
    \end{overpic}
    \caption{Simulation results for cascading up to five TLSs each with unique dynamic couplings. The subtraction and addition processes throughout this figure are labeled by colours blue and red respectively. (a) The schematic for cascading five TLSs for dynamic stimulated emission (red arrows) and absorption (blue arrows) with the corresponding ideal transitions, $\ket{5}\ket{g}^{\otimes 5}\rightarrow \ket{0}\ket{e}^{\otimes 5}$ and $\ket{0}\ket{e}^{\otimes 5}\rightarrow\ket{5}\ket{g}^{\otimes 5}$, respectively. (b) The unconditional fidelities from single- to five-excitation cascaded addition and subtraction processes; $n$ denotes the total number of excitations or equivalently the number of TLSs. The subtraction processes are all initialised with a Gaussian input temporal mode of width $\sigma g^2=2/\sqrt{\pi}$, where $g^2$ is a reference coupling rate (see main text). On the other hand, for the cascaded addition processes, the coupling strength of the first TLS is chosen such that it emits a photon with a Gaussian temporal mode of width $\sigma g^2=2/\sqrt{\pi}$. Sub-figures (c-e) show features of the five-excitation cascaded subtraction and addition. (c) The dynamics of the excited state populations for each cascaded TLS labeled by $j$. The temporal mode $f_j(-t)$ for subtraction (d) and addition (e) occupying the highest mean photon number before and after interacting with the $j$-th cascaded TLS respectively.}
    \label{fig:TLS_cascade}
\end{figure}

Subtracting and adding more than one single photon can be achieved by cascading multiple TLSs. Here, we cascade up to five TLSs and the schematics for these processes are depicted in Fig.~\hyperref[fig:TLS_cascade]{3(a)}. Ideal $n$-excitation cascade subtraction process occurs with the transition, $\ket{n}\ket{g}^{\otimes n}\rightarrow \ket{0}\ket{e}^{\otimes n}$, here $\ket{n}$ is an $n$-photon Fock state with a Gaussian input temporal mode and $\ket{0}$ is the vacuum, in other words, an $n$-photon Fock state excites all $n$ TLSs. Conversely, ideal $n$-excitation cascaded addition process occurs when all $n$ initially excited TLSs emit photons to produce an $n$-photon Fock state i.e., $\ket{0}\ket{e}^{\otimes n}\rightarrow\ket{n}\ket{g}^{\otimes n}$.

We define the fidelity of the $n$-excitation cascaded subtraction process to be the overlap between the output photon field and the vacuum state, in the lossless case, this is equivalent to the excited state population of all the TLSs at the final time, i.e., $\mathcal{F}=\Tr[\prod_{j=1}^n\hat{\sigma}^j_{ee}\hat{\rho}]$. We define the fidelity of the $n$-excitation cascaded addition process to be $\mathcal{F}=\Tr[\ket{n}_{\rm out}\bra{n}\hat{\rho}]$. The fidelities for the cascaded subtraction and addition processes with up to five excitations are presented in Fig.~\hyperref[fig:TLS_cascade]{3(b)}. We emphasize that we reused the variational parameters in Table~\ref{tab:TLS variational parameters} to ensure that the couplings are optimised for a single photon addition/subtraction event. Without backscattering, we do not believe that reoptimising them for cascaded addition and subtraction processes would bring substantial improvements to the fidelities. Most importantly, to extract the maximum performance from our protocol, we must account for the ever slightly changing mode function after each photon addition/subtraction event. Although the mode mismatches are small, in a deep cascade they compound, see Fig.~\hyperref[fig:TLS_cascade]{3(d-e)} for the evolution of the most dominant mode. To account for the mode mismatches, we diagonalise the first coherence function $G^{(1)}(t,t')=\expval{\hat{a}^\dagger(-t)\hat{a}(-t')}$ to determine the most dominant mode after each photon addition and subtraction, then tune the coupling of the subsequent TLS according to the newly updated most dominant mode. However, without conditioning on specific measurement outcomes of the TLSs, the imperfections from each individual subtraction and addition events (Fig.~\hyperref[fig:TLS_sub_add]{2(b)}) compound and ultimately reduce the total fidelity of the cascaded processes. For example, the fidelities are $\mathcal{F}\approx0.97$ for both five-excitation cascaded subtraction and addition processes. Although we have neglected the time taken for the photons to travel between subsequent TLSs, there still exist delays of the temporal mode due to interactions with the TLSs. Therefore, extra time is needed for subsequent TLSs to reach their excited states, as shown in Fig.~\hyperref[fig:TLS_cascade]{3(c)}.

\section{Dynamic stimulated emission and absorption using three-level systems}
\label{three-level system}

The processes of using TLSs to add and subtract photons introduced so far rely on modulating the coupling strength with time. This feature can be realised in circuit QED systems \cite{Korotkov2011PRB,Yin2013PRL} but would be infeasible to implement in optical systems. Additionally, the $n$-excitation addition process suffers greatly from experimental imperfections due to the idling time of the TLS as it needs to wait for the temporal mode to arrive. Hence, we are motivated to swap the TLS with a system that would be feasible to be considered in optical experiments and is more loss-tolerant. 

In the spirit of Ref.~\cite{Cirac1997PRL}, we introduce a $\Lambda$-type three-level system (3LS) that is driven by a coherent laser with a time-dependent Rabi frequency and phase to effectively describe a TLS and its dynamic coupling. The milestone protocol of Ref.~\cite{Cirac1997PRL} has been experimentally demonstrated with microwave \cite{Pechal2014PRX,Kurpiers2018Nature,Axline2018NPhy} and optical \cite{Chou2007Sci,Ritter2012Nature} photons and even phonons \cite{Bienfait2019Sci}. In this section, we start with the 3LS Hamiltonian, where under large detunings, the excited state can be adiabatically eliminated, leading to an effective TLS with time-dependent couplings controlled by the coherent drive. Here, we present our $n$-excitation subtraction and addition solutions of the required time-dependent Rabi frequency and phase of the coherent drive (see Appendix~\ref{appendix three-level} for detailed derivations of the single-, two- and $n$-excitation processes). Finally, we show that the performance of the 3LS is on par with that of the TLS.

\subsection{Model}

\begin{figure}[!t]
    \centering
    \includegraphics[width=0.9\linewidth]{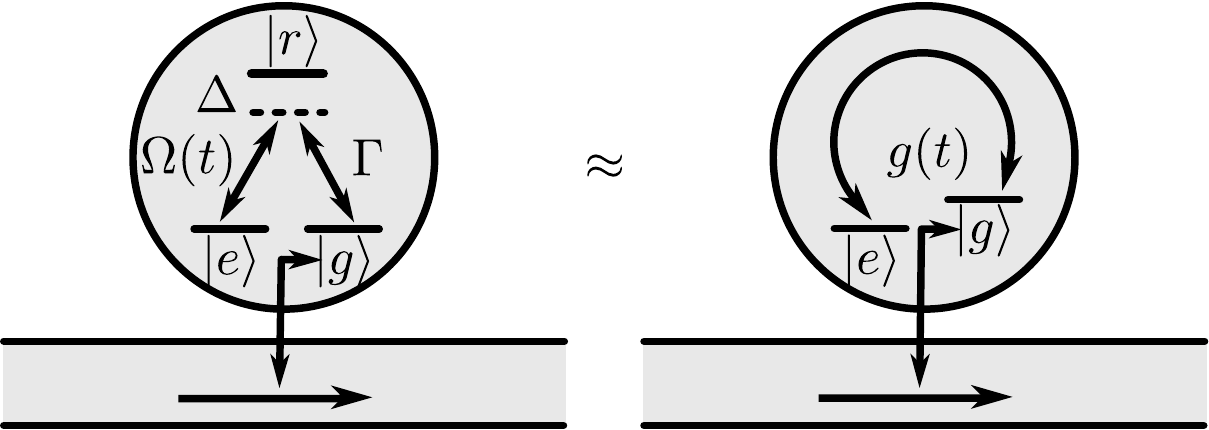}
    \caption{$\Lambda$-type three-level system (3LS) and the effective two-level system (TLS). On the left-hand side, a full 3LS with the excited state $\ket{r}$ and degenerate ground states $\ket{e}$, $\ket{g}$; that is under a large detuned coherent drive with time-dependent Rabi frequency and phase. Adiabatically eliminating the excited state, $\ket{r}$, yields the right-hand side, an effective TLS with AC Stark shifts to the energy of the ground states.}
    \label{fig:three-level_schematic}
\end{figure}

Here, we consider a $\Lambda$-type (3LS) consisting of an excited state $\ket{r}$ with excitation frequency $\omega_0$, and two degenerate ground states $\ket{e}$ and $\ket{g}$ (see Fig.~\ref{fig:three-level_schematic}). On the transition from $\ket{e}$ to $\ket{r}$, the 3LS is driven by a coherent laser with frequency, $\omega_L$, time-dependent Rabi frequency, $\Omega(t)$, and phase, $\phi(t)$. Meanwhile, the transition between $\ket{g}$ and $\ket{r}$ is coupled to the right-propagating photon field in a one-dimensional waveguide with decay rate, $\Gamma$. Under the dipole, rotating-wave and Markov approximations, the Hamiltonian to model this system reads, \cite{Cirac1997PRL},
\begin{align}
\label{3LS hamiltonian}
    \hat{H}(t)&=\int_{\mathbb{R}} dx~\hat{a}^\dagger(x)(\omega_L-i\partial_x)\hat{a}(x)+\omega_0\hat{\sigma}_{rr}\nonumber\\
    &+\frac{\Omega(t)}{2}(e^{-i(\omega_Lt+\phi(t))}\hat{\sigma}_{re}+{\rm h.c.})+\sqrt{\Gamma}(\hat{\sigma}_{rg}\hat{a}(0)+{\rm h.c.}),
\end{align}
with $\hat{\sigma}_{ij}=\ket{i}\bra{j}$. The first term is the free energy and the generator for propagation for the photon field. The second term is the excited state energy of the 3LS. The third and fourth terms facilitate the transition $\ket{e}\leftrightarrow\ket{g}$. We move to an interaction picture with respect to both the excited state and the photon field at the laser frequency, $\omega_L$, and the co-moving frame of the free-propagation term in the Hamiltonian. Furthermore, under large detunings ($\Delta=\omega_L-\omega_0$), $|\Delta|\gg \Omega(t),\Gamma$, the dynamics of the excited state are suppressed, hence the excited state can be adiabatically eliminated, which produces an effective TLS with AC Stark shifts on the ground state energies that depend on the Rabi frequency of the coherent drive and the photon field in the waveguide. Importantly, the $\ket{e}\leftrightarrow\ket{g}$ transition is now modulated by the time-dependent Rabi frequency and phase. All together, the dynamics of the 3LS and the photon field can be described by an effective Hamiltonian with only the degenerate ground states, $\ket{g}$ and $\ket{e}$, and an effective time-dependent coupling strength to the photon field (see Appendix~\ref{appendix three-level}),
\begin{align}
\label{effective hamiltonian}
    \hat{H}_{\rm eff}(t)=&\frac{\Gamma}{\Delta}\hat{\sigma}_{gg}\hat{a}^\dagger(-t)\hat{a}(-t)\nonumber\\
    &+\frac{\sqrt{\Gamma}\Omega(t)}{2\Delta}\left[e^{i\phi_g(t)}\hat{\sigma}_{eg}\hat{a}(-t)+{\rm h.c.}\right].
\end{align}
Note that we have eliminated the AC Stark shift of the energy of the $\ket{e}$ state by moving to an interaction picture and defining $\phi(t)=\phi_e(t)+\phi_g(t)$ and $\partial_t\phi_e(t)=-\Omega^2(t)/(4\Delta)$. The other AC Stark shift of the $\ket{g}$ state remains, and it depends on the local state of the photon field. The effective time-dependent coupling is $g(t)=\sqrt{\Gamma}\Omega(t)e^{i\phi_g(t)}/(2\Delta)$ and the Rabi frequency and phase are $\Omega(t)e^{i\phi(t)}/2=\Delta g(t)e^{-i\frac{\Delta}{\Gamma} G(0,t)}/\sqrt{\Gamma}$, where $G(0,t)=\int_0^tdt'|g(t')|^2$. We note that the Hamiltonian is very much like that of the TLS but has included the AC Stark shift term. If we use parameters such that $\Delta \gg \Omega(t) \gg \Gamma$, then one can likely neglect the AC Stark shift term and use our TLS ansatz (Eq.~\eqref{n photon add g} and Eq.~\eqref{n photon sub g}) to solve for the required time-dependent couplings for addition and subtraction. This approximation becomes worse for larger input photon states as $\hat{a}^\dagger(-t)\hat{a}(-t)$ increases the importance of the AC Stark shift term. Therefore, in the following sections we solve the processes with Eq.~\eqref{effective hamiltonian} in its entirety.

\subsection{$n$ excitations}

Under the effective model of the 3LS, ideal $n$-excitation subtraction occurs when the 3LS is initially prepared in $\ket{g}$, an incoming $n$-photon Fock state generates the transition to $\ket{e}$, and the outgoing state is an $(n-1)$-photon Fock state. In solving for the effective dynamic coupling strengths of the subtraction process, we adapt a similar treatment to the processes with a TLS, but the AC Stark shift in the effective Hamiltonian in Eq.~\eqref{effective hamiltonian} brings a level of complexity to the solutions. Nevertheless, the output boundary condition is (see Appendix~\ref{appendix three-level}),
\begin{equation}
\label{3 level bc}
    f_{\rm in}(-t)=\frac{1-(\xi^*/\xi)^n}{1-\xi^*/\xi}|\xi|^2g^*(t)e^{-\frac{\xi}{2} G(0,t)}S_{\xi/2}(0,t),
\end{equation}
Here, $\xi=[1+i\Gamma/(2\Delta)]^{-1}$ is a feature of the AC Stark effect. Importantly, this boundary condition is in the form of Eq.~\eqref{general bc} and its solution is given by Eq.~\eqref{general g(t)}. When using 3LSs for subtractions, we modify the coupling strength with two variational parameters for the real and imaginary parts of $u-v$ in Eq.~\eqref{general g(t)}. Despite including the extra parameters, the coupling strength is a solution to some boundary condition in the form of Eq.~\eqref{general bc}. Explicitly, the effective time-dependent coupling strength for the $n$-excitation subtraction using a 3LS is
\begin{equation}
    g^*(t)=\frac{s_{r,n}f_{\rm in}(-t)}{\sqrt{2\Re(\zeta)\left[\int_0^tdt'~|f_{\rm in}(-t')|^2\right]^{1+is_{i,n}\frac{\Im(\zeta)}{\Re(\zeta)}}}},
\end{equation}
here, $\zeta=\frac{1-(\xi^*/\xi)^n}{1-\xi^*/\xi}|\xi|^2-\xi/2$, $\Im(\cdot)$ denotes the imaginary part of a complex number, and the optimised variational parameters, $s_{r,n},s_{i,n}\in\mathbb{R}$, are given by Table~\ref{tab:3LS varitional parameters}. Notice that, if we had neglected the AC Stark shift by setting $\xi=1$, the coupling strength reduces to the solution for the TLS in Eq.~\eqref{n photon sub g}. Moreover, the additional variational parameter, $s_{i,n}$ would be unnecessary, since $\Im(\zeta)=0$. Nevertheless, the time-dependent Rabi frequency and phase for the $n$-excitation subtraction process are
\begin{align}
\label{3LS sub rabi}
    &\frac{\Omega_{{\rm sub},n}(t)}{2}e^{i\phi_{{\rm sub},n}(t)}\nonumber\\
    &=\frac{\Delta}{\sqrt{\Gamma}}\frac{s_{r,n}f^*_{\rm in}(-t)}{\sqrt{2\Re(\zeta)F(t)^{1+i\left[s^2_{r,n}\frac{\Delta}{\Gamma}-s_{i,n}\Im(\zeta)\right]/\Re(\zeta)}}},
\end{align}
here, $F(t)=\int_0^tdt'~|f_{\rm in}(-t')|^2$.

Conversely, the ideal $n$-excitation addition process is the reverse of the subtraction process. It occurs when an incoming $(n-1)$-photon Fock state gains a photon and leaves in an $n$-photon Fock state, accompanied by the $\ket{e}\rightarrow\ket{g}$ transition of the 3LS. To obtain the time-dependent Rabi frequency and phase for the addition processes, we apply the time-reversal, translation and parity inversion transformations to the subtraction process (see Appendix~\ref{appendix reversal}). Explicitly, the time-dependent Rabi frequency and phase for the $n$-excitation addition process reads
\begin{align}
\label{3LS add rabi}
    &\frac{\Omega_{{\rm add},n}(t)}{2}e^{i\phi_{{\rm add},n}(t)}\nonumber\\
    &=\frac{\Delta}{\sqrt{\Gamma}}\frac{a_{r,n}f^*_{\rm in}(-t)}{\sqrt{2\Re(\zeta)\Tilde{F}(t)^{1-i\left[a^2_{r,n}\frac{\Delta}{\Gamma}-a_{i,n}\Im(\zeta)\right]/\Re(\zeta)}}},
\end{align}
where $\Tilde{F}(t)=\int_t^\infty dt'|f_{\rm in}(-t')|^2$. Similarly, $a_{r,n},a_{i,n}\in \mathbb{R}$ are the variational parameters and the optimised value are given in Table~\ref{tab:3LS varitional parameters}. We emphasize that in the single-excitation scenario, the initial state is fixed with the 3LS in state $\ket{e}$ and the photon field in vacuum; therefore, one should make the substitution $f_{\rm in}(-t)\rightarrow if_{\rm out}(-t)$ in Eq.~\eqref{3LS add rabi} such that, the final state of the photon field is a single-photon state with temporal mode $f_{\rm out}(-t)$ (see Appendix~\ref{appendix three-level}). Notice that from Table~\ref{tab:3LS varitional parameters}, the variational parameters for the single-excitation processes deviate from 1; this is caused by the optimisation using the full 3LS, but the solutions are derived under the effective Hamiltonian given by Eq.~\eqref{effective hamiltonian}.

\begin{table}[!t]
    \centering
    \begin{tabular}{|c|c|c|c|c|}
        \hline
        $n$ & $s_{r,n}$ & $s_{i,n}$ & $a_{r,n}$ & $a_{i,n}$ \\
        \hline
         1 & 1.026 &  1.057 & 1.025 &  1.083 \\
         \hline
         2 & 1.080 & 0.858 & 0.819 &  0.921\\
         \hline
         3 & 1.087 & 0.795 & 0.874 &  0.875\\
         \hline
         4 & 1.082 & 0.750 & 0.897 &  0.839\\
         \hline
         5 & 1.066 & 0.702 & 0.902 & 0.805\\ 
         \hline
    \end{tabular}
    \caption{Numerically optimised parameters for subtraction ($s_{r,n}$, $s_{i,n}$) and addition ($a_{r,n}$, $a_{i,n}$) processes using 3LSs for up to five excitations.}
    \label{tab:3LS varitional parameters}
\end{table}

\begin{figure*}[ht!]
\centering
\begin{overpic}[width=0.85\textwidth]{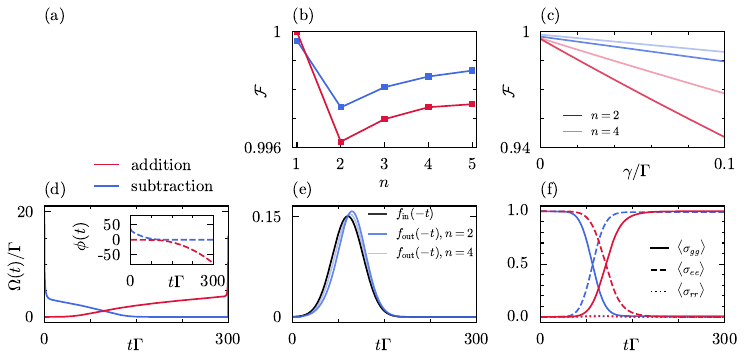}
  \put(5, 30){\includegraphics[width=0.2\textwidth]{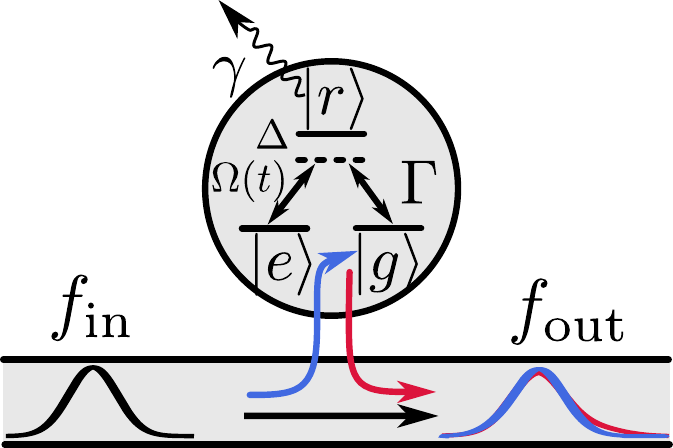}}
\end{overpic}
    \caption{Simulation results with the full 3LS. The subtraction and addition processes throughout this figure are labeled by colours blue and red, respectively. (a) The schematic of the 3LS dynamic stimulated emission (red arrows) and absorption (blue arrows) with input and output temporal modes labeled by $f_{\rm in}$ and $f_{\rm out}$ respectively. Results in all sub-figures are simulated with detuning $\Delta/\Gamma=5$ and with a Gaussian input temporal mode of width $\sigma\Gamma=25$. The lossless unconditional fidelities for the subtraction and addition processes for up to five excitations are presented in (b) and with loss for two- and four-excitations in (c). In (d), the time-dependent Rabi frequency (solid) and phase (inset, dashed) for the two-excitation subtraction  and addition  processes. (e) The input Gaussian temporal mode (black) and the output temporal modes for the two- and four-excitation subtraction processes. (f) The ground states, $\ket{g}$ (solid), $\ket{e}$ (dashed) and the excited state (dotted) population dynamics for the two-excitation subtraction and addition processes.}
    \label{fig:three-level_sub_add}
\end{figure*}

\subsection{Results and discussions}
\label{3LS results}

In this subsection, we present and discuss the performance of the $n$-excitation addition and subtraction processes using 3LSs. We briefly describe the numerical setup of the full 3LS simulation. The results are presented by Fig.~\ref{fig:three-level_sub_add} which includes the fidelities with and without loss for up to five excitations, the time-dependent Rabi frequency and phase, the state populations and the output temporal modes for subtraction. Remarkably, both the addition and subtraction processes are tolerant to losses.

As described in Sec.~\ref{TLS results}, we use the quantum pulses formalism where single-mode states are injected and absorbed into the waveguide using virtual cavities. In this subsection, we simulated the \textit{full} 3LS without adiabatically eliminating the excited state. We include coupling to non-guided modes of the environment by including an additional Lindblad operator in the master equation, $\hat{L}_1=\sqrt{\gamma}\hat{\sigma}_{gr}$.

The schematics of the $n$-excitation subtraction and addition processes are depicted in Fig.~\hyperref[fig:three-level_sub_add]{5(a)} by blue and red respectively. We chose the time-dependent Rabi frequency and phase for each process according to Eq.~\eqref{3LS sub rabi} and Eq.~\eqref{3LS add rabi}.  Since the analytic calculations are performed under the approximation of  adiabatically eliminating the excited state, to simulate the dynamics of the full 3LS, we must be in a regime where the approximation is justified, i.e., $|\Delta|\gg\Omega(t),\Gamma$. From Eq.~\eqref{3LS sub rabi} and Eq.~\eqref{3LS add rabi}, the Rabi frequency is proportional to the input temporal mode, hence choosing a wide Gaussian reduces its magnitude. In all simulations, we have set the detuning to be $\Delta/\Gamma=5$ and the input Gaussian mode with width $\sigma\Gamma=25$. We define the fidelities of each process identically to that of the TLS described in Sec.~\ref{TLS results}. The fidelity for the $n$-excitation subtraction process is $\mathcal{F}=\Tr[\ket{n-1}_{\rm out}\bra{n-1}\hat{\rho}]$, where $\ket{n-1}_{\rm out}$ is an $(n-1)$-photon Fock state. On the other hand, the fidelity for the $n$-excitation addition process is $\mathcal{F}=\Tr[\ket{n}_{\rm out}\bra{n}\hat{\rho}]$. 

In Fig.~\hyperref[fig:three-level_sub_add]{5(b)}, we present the fidelities of the subtraction (blue) and addition (red) processes for up to five excitations. The only approximation in the single-excitation processes is adiabatically eliminating the excited state, otherwise the expressions for the Rabi frequencies and phases are exact. Hence, both the single-excitation subtraction and addition processes have near-unit fidelity, with the addition process slightly higher because a single-photon state is always single-moded. For processes with excitation number greater than one, the fidelities are monotonically increasing; for subtraction, the fidelity with two excitations is $\mathcal{F}\approx0.997$ and increases to $\mathcal{F}\approx0.999$ with five excitations, for addition, the fidelity with two excitations is $\mathcal{F}\approx0.996$ and increases to $\mathcal{F}\approx0.997$ with five excitations. Furthermore, the optimised parameters in Table~\ref{tab:3LS varitional parameters} are responsible for correcting two approximations for the subtraction processes, namely, the adiabatic elimination of the excited state and the linearised analytic solutions. The addition processes have slightly lower fidelities because we have approximated the input temporal mode to be the same as the output before applying the reversal transformations to the subtraction process, Fig.~\hyperref[fig:three-level_sub_add]{5(e)} shows that this is a valid approximation. Nevertheless, this is another approximation that the optimised parameters have to correct for.

We show that the adiabatic condition, $|\Delta|\gg \Omega(t)$ is satisfied by plotting the time-dependent Rabi frequency and phase for the two-excitation subtraction (blue) and addition (red) processes in Fig.~\hyperref[fig:three-level_sub_add]{5(d)}. In Fig.~\hyperref[fig:three-level_sub_add]{5(e)}, we show that the input temporal mode is approximately equal to the output temporal mode for the two- and four-excitation subtraction processes; thus, it is appropriate to equate them for the reversal transformations. Figure~\hyperref[fig:three-level_sub_add]{5(f)} shows that the 3LS indeed acts like a TLS with most of the state population shared between the two ground states while the $\ket{r}$ state remains largely unpopulated.

Figure~\hyperref[fig:three-level_sub_add]{5(c)} presents the fidelities of the two- and four-excitation processes with loss. The loss mechanism considered here is the 3LS decaying into non-guided modes under the $\ket{r}\rightarrow\ket{g}$ transition. Remarkably, the four-excitation subtraction process achieves $\mathcal{F}\approx0.99$ with 10\% ($\gamma/\Gamma=0.1$) loss corresponding to a cooperativity of 10. As explained in Sec.~\ref{TLS results}, the 3LS decaying into non-guided modes is still a valid subtraction on the guided mode. Most importantly, the addition processes are also tolerant to losses, unlike the TLS where the excited state can spontaneously decay into non-guided modes whilst idling before the input temporal mode arrives. Here, the 3LS only gets excited by the Rabi drive as the temporal mode arrives, hence greatly reducing the impact of the spontaneous decay on the fidelities. 

In experimental settings, there are other imperfections to be considered, such as an additional decay channel between the excited state $\ket{r}$ and $\ket{e}$ or the presence of an initially mixed state $p_g\ket{g}\bra{g}+g_e\ket{e}\bra{e}$ \cite{LopesdaSilva24JOptB}. We qualitatively discuss their impacts on our results. The inclusion of another decay channel from $\ket{r}$ to $\ket{e}$ will introduce imperfections in our protocol. Let us look at the photon addition and subtraction cases separately. For addition, the goal is to perform the mapping $\ket{n}\ket{e}\mapsto\ket{n+1}\ket{g}$. The presence of the decay from $\ket{r}$ to $\ket{e}$ will introduce a dephasing channel to the interaction, and we expect some of the excitation to be trapped in $\ket{e}$ and unable to reach the guided mode. On the other hand, the idealised subtraction process is $\ket{n}\ket{g}\mapsto\ket{n-1}\ket{e}$, hence losing an excitation via the $\ket{r}$ to $\ket{e}$ decay channel still constitutes a valid subtraction. Therefore we expect the fidelity of the photon subtraction process to be more robust to losses compared to the addition process similar to the results presented by Fig.~\hyperref[fig:three-level_sub_add]{5(c)}.

The presence of an initially mixed state: due to the linearity of quantum mechanics, the ground and excited state components in the mixture $p_g \ket{g}\bra{g} + p_e \ket{e}\bra{e}$ will evolve independently. Suppose we wish to complete photon addition with an initially mixed state; the density matrix of the entire system evolves from $\ket{n}\bra{n}(p_g \ket{g}\bra{g} + p_e \ket{e}\bra{e})$ to $p_g \ket{\psi_n}\bra{\psi_n} + p_e\ket{n+1}\bra{n+1}\otimes\ket{g}\bra{g}$, here $\ket{\overline{n}}$ denotes an $n$-photon multimode state, while $\ket{\psi_n}=a\ket{\overline{n}}\ket{g}+b\ket{\overline{n-1}}\ket{e}$ and $a$, $b$ are complex coefficients. Here, we can deduce the photon number in the evolution because the dynamics of the 3LS mimics that of the TLS, whose dynamics are governed by the waveguide QED Hamiltonian (Eq.~\eqref{hamiltonian}), which conserves the excitation number. From the above equation, the initially mixed state leads to states with incorrect photon number ($n$ and $n-1$), which are orthogonal to the target state ($\ket{n+1}$). Therefore, the fidelity is bounded from above by $1-p_g$. A similar argument can be made for the photon subtraction process.

\section{Generation of Non-Gaussian Superposition States}
\label{non-gaussian states}

In previous sections, we used TLSs that are dynamically coupled to the propagating photon field to prepare photon Fock states with high fidelities. While photon Fock states have a wide range of applications in photonic quantum technologies \cite{Ourjoumtsev2007Nature,Winnel2024PRL,Yu2026NatPho}, non-Gaussian superpositions of Fock states, such as the Schr\"{o}dinger cat states and GKP states have enjoyed unparalleled attention in encoding quantum information \cite{Ralph2003PRA,Grimsmo2020PRX,Gottesman2001PRA,Grimsmo2021PRXQ}. Typical state preparation protocols rely on conditional measurements on the outputs of linear optical networks, which have low success probabilities. In this section, we explore the possibilities of using dynamic stimulated emission and absorption to prepare these non-Gaussian superposition states. However, since our solutions are strictly excitation-number dependent and as we argued in Sec.~\ref{connection to the JC model}, photon addition and subtraction with a superposition state input is fundamentally probabilistic. Therefore, we must herald on successful subtraction or addition of photons by measuring the TLS after the scattering events. Despite that, we show high success probabilities can be achieved if the input superposition state contains high-photon-number supports. This is because the variance in the coupling required is low for high-photon-number states. First, we demonstrate that successive subtraction \textit{or} addition of photons with a squeezed vacuum input state can prepare cat states with high fidelity. Next, we show that dynamic stimulated emission can be used to add a photon to an incoming general Gaussian state with high fidelity and success probability. 

\subsection{Cat states}
\label{cat states}

Here, we prepare cat states using successive subtractions. Cat states are defined as a superposition of opposite phase coherent states, $\ket{{\rm cat}_{\alpha,\pm}}\propto \ket{\alpha}\pm\ket{-\alpha}$. Large-scale cat states are key resources in quantum technologies. However, the efficient preparation of optical cat states has remained a challenge. Conventionally, optical cat states are prepared by the successive action of photon subtractions on a squeezed vacuum \cite{Dakna1997PRA,Ourjoumtsev2006science,Takase2021PRA}. In linear optical protocols, photon subtractions are typically realised by inserting beamsplitters and looking for photon-number-resolving detection patterns. Recently, Lund \textit{et al.} \cite{Lund2024PRL} showed that optical cat states can be prepared with higher success probabilities using a TLS with multiplexing. 

\begin{figure}[t!]
    \centering
    \begin{overpic}[width=0.48\textwidth]{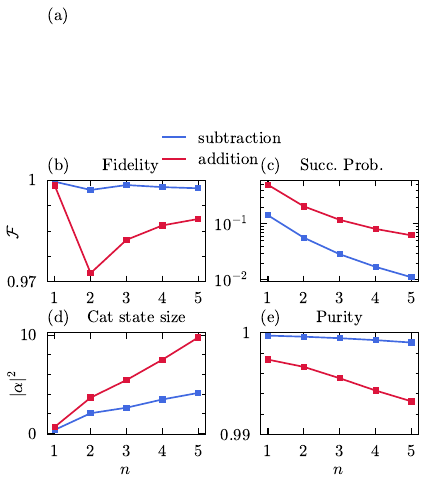}
        \put(9,74){\includegraphics[width=0.42\textwidth]{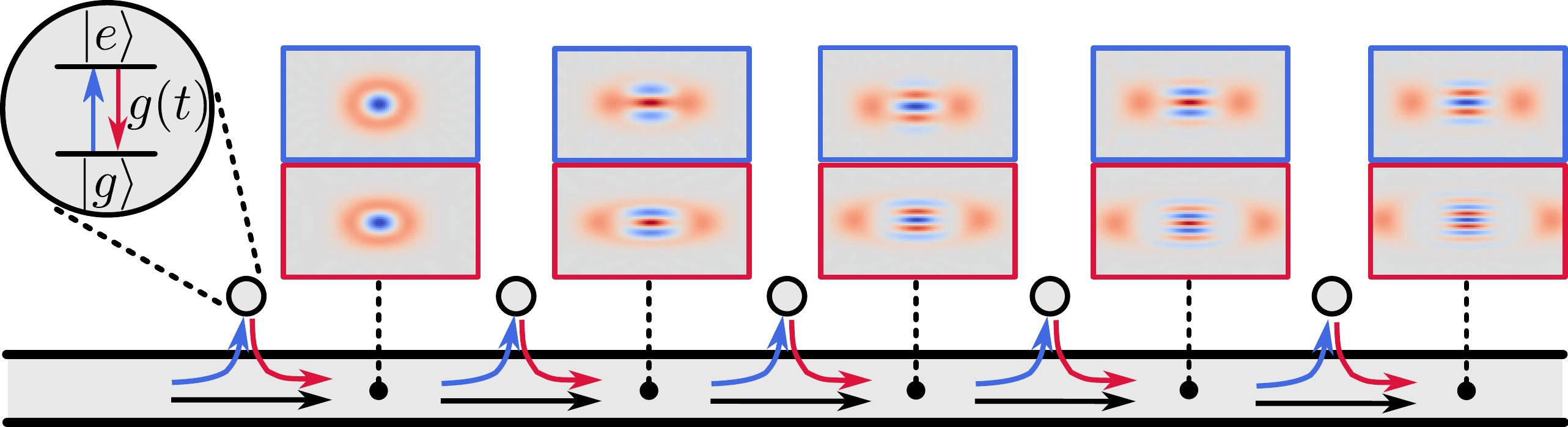}}
    \end{overpic}
    \caption{Simulation results of cat state generation using successive dynamic stimulated emissions or absorptions. The emission and absorption processes throughout this figure are labeled by colours red and blue, respectively. In (a), the schematic for successive subtractions or additions of single photons by cascading five TLSs (grey circles) with a 10~dB squeezed vacuum input state. Here, successive subtractions and additions are distinct processes, and the Wigner functions presented in the highlight boxes are the optimally unsqueezed single-mode state after each interaction. In the colour map of the Wigner functions, red (blue) indicates positive (negative) values. In sub-figures (b-c), the horizontal axis labels the number of successive subtractions or additions, from one up to five. In (b), the fidelity after each interaction, the fidelity is calculated against a cat state after the output state is optimally unsqueezed. In (c), the success probability after $n$ interactions. (d) The output cat state size, $|\alpha|^2$. (e) The purity, and it reflects if the output state is single-moded.}
    \label{fig:cat_state}
\end{figure}

We show that cat states can be prepared via dynamic absorption. In Fig.~\hyperref[fig:cat_state]{6(a)}, we cascade up to five TLSs to herald five subtractions (blue) from a 10~dB squeezed vacuum input. The Wigner functions after each subtraction are highlighted by blue boxes. Since our solutions for high-fidelity $n$-excitation subtraction, given by Eq.~\eqref{n photon sub g}, vary depending on the number of excitations, the process of dynamic absorption is not expected to be deterministic when the input is a superposition state. Moreover, it is unclear which dynamic coupling would produce an optimal result. Inspired by Ref.~\cite{Lund2024PRL}, we require that the dynamic absorption process act similarly to the annihilation operator on the photon field, i.e., we wish to execute the transition $\ket{\alpha}\ket{g}\rightarrow\ket{\alpha}\ket{e}$, where $\ket{\alpha}$ is a coherent state. But this transition violates the conservation of the total excitation therefore, the transition can only be satisfied by a nondeterministic process. Here, we emphasize that there is a trade-off between the fidelity and success probability; in the extreme case, a trivial process where no photon is absorbed would produce unit fidelity but also zero success probability. We found that by using the dynamic coupling $g(t)=0.15g_{{\rm sub},1}(t)$, the transition with $\alpha=\sqrt{10}$ can be approximately satisfied with fidelity $\mathcal{F}\approx0.997$ and success probability approximately equal to $0.635$. Recall that $g_{{\rm sub},1}(t)$ is the single-excitation subtraction solution, given by Eq.~\eqref{one photon sub g}. Therefore, we have chosen to use this dynamic coupling for each of the cascaded TLS depicted in Fig.~\hyperref[fig:cat_state]{6(a)}. We emphasize that, similar to the cascaded photon subtraction process discussed in Sec.~\ref{cascaded addition and subtraction}, we have tuned the couplings of each TLS to accommodate the slightly changing mode shape after each scattering event. The final state should ideally be a single-mode state of the photon field after other systems and the bath are traced out. The fidelity, success probability and the cat state size after each subtraction are presented by blue boxes in Fig.~\hyperref[fig:cat_state]{6(b-d)}.

Preparing large-scale cat states using successive subtractions is ultimately not efficient, Fig.~\hyperref[fig:cat_state]{6(c)} shows that at five subtractions, the cumulative success probability is approximately 1\%; this high degree of nondeterminism can be reasoned back to having a squeezed-vacuum input state. Using non-Gaussian input states, characterised by their non-zero stellar rank, can greatly improve the success probabilities of preparing large-scale cat states \cite{Ourjoumtsev2007Nature,Winnel2024PRL,Luo2024}. However, efficiently preparing optical non-Gaussian input states itself is also a challenge.

\label{photon-added gaussian state}
\begin{figure*}[ht!]
\centering
\begin{overpic}[width=0.9\textwidth]{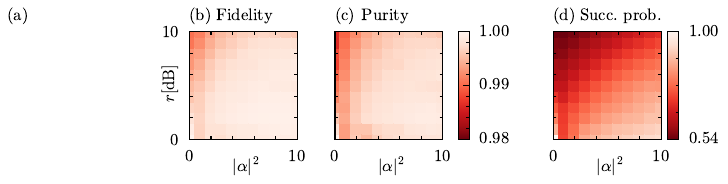}
  \put(0, 4){\includegraphics[width=0.2\textwidth]{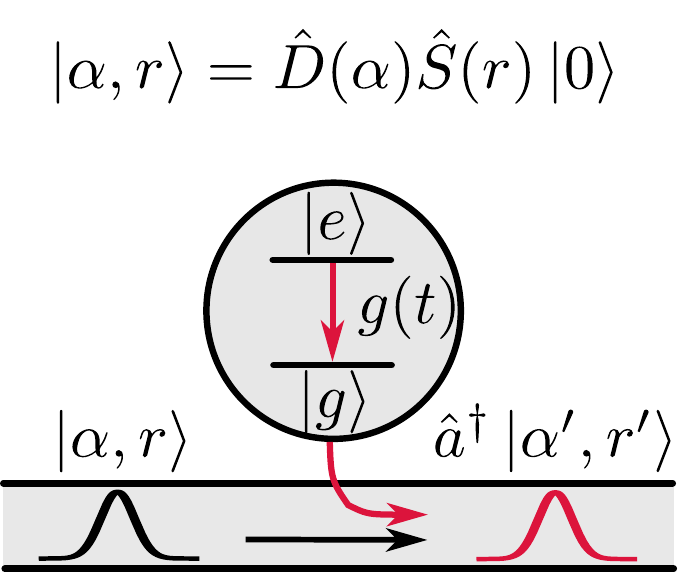}}
\end{overpic}
    \caption{Preparation of photon-added Gaussian states. In (a), the schematic where we use dynamic stimulated emission of the TLS to add a photon to a displaced squeezed vacuum state with displacement, $\alpha$ and squeezing strength, $r$. The target state is the normalised state after applying the creation operator on a different displaced squeezed vacuum. We considered displacement in the range $\alpha\in[0,\sqrt{10}]$ and squeezing up to 10~dB. In (b-d), the fidelities, purities and success probabilities for a combination of displacements and squeezing strengths.}
    \label{fig:TLS_stellar_1}
\end{figure*}

Here, we instead prepare cat states by successive \textit{additions} of photons with a 10~dB squeezed vacuum input, see Fig.~\hyperref[fig:cat_state]{6(a)} for a schematic. Throughout this work, we have relied on the deterministic feature of the subtraction processes on Fock states such that under reversal transformations we obtain the dynamic couplings which facilitate deterministic addition processes. However, as Fig.~\ref{fig:cat_state} suggests and argued in Sec.~\ref{connection to the JC model}, the subtraction processes with a superposition input state are not deterministic. There are two factors that contribute to the nondeterminism, namely; the input squeezed vacuum state has a significant population of vacuum which cannot excite the TLS, and the nonlinear response of the TLS produces distinct interactions between different Fock states, meaning that a superposition cannot excite a TLS deterministically. Because of the nondeterminism, we cannot rely on reversal transformations of the subtraction process to produce a coupling for the single-mode-preserving addition process. Hence, to produce a single-mode output state, we must introduce another time-dependent coupling strength for the addition of photons on superposition states. The following expression is our ansatz for the coupling and it is motivated by solving the Schr\"{o}dinger equation for the addition process on an $n$-excitation Fock state (see Appendix~\ref{appendix addition ansatz});
\begin{equation}
\label{superposition add g}
    g^*_{\rm add}(t)=\frac{ q_1f_{\rm in}(-t)}{\sqrt{1-q_2\int_0^tdt'|f_{\rm in}(-t')|^2}}.
\end{equation}
Here, $f_{\rm in}(-t)$ is the temporal mode just before the photon field interacts with the TLS, and $q_1,q_2\in\mathbb{R}$ are variational parameters with $q_2\leq1$. This ansatz offers great flexibility, and notice that, for $q_2=1$ and $q_1=a_n/\sqrt{2n-1}$, we recover the $n$-excitation addition coupling strength given by Eq.~\eqref{n photon add g}. We cascade up to five TLSs with time-dependent coupling strength according to Eq.~\eqref{superposition add g}, and we have diagonalised the first coherence function to feed the most dominant mode functions into the couplings of the TLSs. We optimised the parameters for the first TLS for success probability without sacrificing too much purity (or preserving a single mode); the values of the optimised parameters are $\{q_1,q_2\}=\{0.509, 0.425\}$. Notice that a creation operator or an annihilation operator acting on a squeezed vacuum produces the same state; hence, in Fig.~\hyperref[fig:cat_state]{6(b)} we observe overlapping fidelities for a single-photon subtracted or added squeezed vacuum state. However, successive action of the creation operator on a squeezed vacuum state produces a different state from that of the annihilation operator. Therefore, without compromising on the fidelities, we found that the variational parameters, $\{q_1,q_2\}=\{1/\sqrt{10},1/10\}$ for the subsequent TLSs, are a middle ground for generating large-scale cat states with increasing fidelities and high success probabilities. Although Fig.~\hyperref[fig:cat_state]{6(b)} shows that the cat states generated after each round of addition have lower fidelities than that with the subtractions, the successive addition protocol is however far more efficient, at nearly an order-of-magnitude higher in success probability (Fig.~\hyperref[fig:cat_state]{6(c)}) and the size of the generated cat state (Fig.~\hyperref[fig:cat_state]{6(d)}) is much larger. The higher success probabilities come at a cost of lowered purities i.e., the state of the photon field is less contained in a single-mode, but nonetheless the purities are above 0.99 after five successive additions of single photons (Fig.~\hyperref[fig:cat_state]{6(e)}).

\subsection{Photon-added Gaussian state}
\label{photon-added gaussian state}
A general Gaussian pure state can be described as a displaced and squeezed vacuum, $\hat{D}(\alpha)\hat{S}(r)\ket{0}$; here, $\hat{D}(\alpha)=e^{\alpha\hat{a}^\dagger-\alpha^*\hat{a}}$ and $\hat{S}(r)=e^{-\frac{r}{2}(\hat{a}^2-\hat{a}^{\dagger 2})}$ are the single-mode displacement operator and squeezing operator, respectively. In this subsection, we consider $\{\alpha,r\} \in \mathbb R$. We refer to a photon-added Gaussian state as the normalised state after applying a photon creation operator on a displaced squeezed vacuum. In this subsection, we use dynamic stimulated emission of a TLS to produce a photon-added Gaussian state. These states are key resource states, as they serve as starting points for generating non-Gaussian states of higher stellar ranks \cite{Chabaud2020PRL}.

Figure~\hyperref[fig:TLS_stellar_1]{7(a)} shows a schematic of dynamic stimulated emission of a photon to a displaced squeezed vacuum in a temporal mode. The target state is a photon-added displaced squeezed vacuum. Note that the output state does not necessarily have the same displacement and squeezing values. This means that for a Gaussian state, the TLS mimics the action of the creation operator up to a Gaussian operation. In the absence of loss, dynamic stimulated emission preserves excitation number. The differences in displacement and squeezing between the initial state and the target state can be qualitatively explained by the difference between an excitation-conserving unitary evolution and the action of the creation operator. Suppose the process shown in Fig.~\hyperref[fig:TLS_stellar_1]{7(a)} is deterministic, by the excitation-conserving Hamiltonian in Eq.~\eqref{hamiltonian}, the excitation number of the photon field after the TLS has emitted a photon is $\sinh^2(r)+|\alpha|^2+1$; to match the excitation number, the displacement and squeezing on the target state must be different except for the case with vacuum. For conditional processes, the excitation number is not conserved; hence, we allow the target state to have different amounts of displacement and squeezing. 

In this subsection, we used the ansatz given by Eq.~\eqref{superposition add g} for the addition of a single photon with the two variational parameters, $\{q_1,q_2\}$. Since a particular dynamic coupling cannot deterministically carry out photon addition for a superposition of Fock states, we must condition on measuring the TLS in the ground state to herald a successful addition. We optimised the variational parameters for a balance between success probability and purity for the input displacements in the range, $\alpha\in[0,\sqrt{10}]$ and squeezing strength from 0~dB to 10~dB.

The fidelities against photon-added Gaussian states and the purities are present in Fig.~\hyperref[fig:TLS_stellar_1]{7(b-c)}; the success probabilities of addition are presented in Fig.~\hyperref[fig:TLS_stellar_1]{7(d)}. The fidelities, purities and success probabilities decrease from unity with increased values of squeezing on the vacuum; at 10~dB of squeezing, the fidelity and the success probability reduce to approximately 0.98 and 0.55, respectively. The reductions in the fidelity and success probability for the photon-added squeezed vacuum can be explained by the increased variance in the Fock-state distribution with increased values of squeezing on the vacuum. Nonetheless, this feature can be compensated by introducing displacements, since it can reduce the relative difference in the excitation number of the dominant probability amplitudes. Moreover, the photon-added coherent state with $|\alpha|^2=10$, can achieve approximately 0.999 fidelity and 0.94 success probability, as shown in Fig.~\ref{fig:TLS_stellar_1}. On average, most photon-added Gaussian states have fidelities above 0.99 and success probabilities above 0.6. We note that the TLS can introduce anti-displacement and anti-squeezing; the discrepancies between the initial and the target displacements and squeezing strengths are presented by Fig.~\ref{fig:appendix_stellar_1_diff} in Appendix~\ref{appendix addition ansatz}.

\section{Conclusions and outlook}
\label{conclusion}

In summary, we have introduced the concept of dynamic stimulated emission and absorption of two-level systems by providing the time-dependent coupling strengths to the $n$-excitation addition and subtraction processes; we showed that our solutions can achieve fidelities greater than 0.996 for arbitrary number of excitations. We cascade up to five TLSs for five successive photon additions and subtractions and the unconditional fidelities are above 0.97, demonstrating the possibility of preparing $n$-excitation Fock states using quantum emitters. Furthermore, we showed that the same $n$-excitation addition and subtraction processes can occur if the TLS and its time-dependent coupling strength are replaced with a 3LS that is driven by a coherent laser with a time-dependent Rabi frequency and phase, which can be realised in optical experiments. As a proof-of-concept, we demonstrated high success probabilities of preparing Schr\"{o}dinger cat states from a squeezed vacuum input by cascading up to five TLSs for successive photon subtractions and additions. However, a more practical use of quantum emitters may be preparing photon-added Gaussian states. While Gaussian states such as the squeezed vacuum state has been widely considered as resource states for the preparation of useful non-Gaussian states \cite{Dakna1997PRA,Ourjoumtsev2007PRL,Xanadu2025Nat,Takase2021PRA,Takase2024PRA,Hanamura2025}, the lack of non-Gaussianity reflected by their zero stellar rank and the lack of mean photon number ultimately hinder the efficiency of preparing exotic non-Gaussian states such as the large-scale cat states or the GKP states. On the other hand, using a photon-added Gaussian state with stellar rank one \cite{Chabaud2020PRL} or other few-photon non-Gaussian states can greatly improve the feasibility of preparing exotic non-Gaussian states using linear optical protocols \cite{Hastrup2020OL, Luo2024,Winnel2024PRL,Yu2026NatPho}.

Future directions include improvement on the success probabilities of photon subtraction and addition when the input is a superposition state, thus directly improving the efficiency of preparing a photon-added Gaussian state. Moreover, one could cascade quantum emitters and prepare multi-photon-added Gaussian states, which have higher stellar ranks and could potentially improve the preparation of exotic non-Gaussian states. On the other hand, by entangling the cascaded quantum emitters, one can also prepare superposition states in a temporal mode. The cubic phase state is another key non-Gaussian state for universal bosonic quantum computation \cite{Lloyd1999PRL,Gu2009PRA,Miyata2016PRA,Yukawa2013OE}, incorporating quantum emitters into the preparation schemes could potentially improve the efficiency. Ultimately, our work provides a path for integrating quantum emitters into optical non-Gaussian state preparation.

\section*{Acknowledgments}
H.L. thanks Yangming Wang for useful discussions and acknowledges support from the Sydney Quantum Academy. S.M. acknowledges support from
the Australian Research Council (ARC) via the Future Fellowship, `Emergent many-body phenomena in engineered quantum optical systems', project no. FT200100844. M.M. acknowledges support from the Office of Naval Research (Award No. N00014-24-1-2052). P.S.S. gratefully acknowledges
support from the S2I-Gupta Fellowship. F.Y. gratefully acknowledges support from the NSF Graduate Research Fellowship. 
\bibliography{main}

@article{Cirac1997PRL,
  title = {Quantum State Transfer and Entanglement Distribution among Distant Nodes in a Quantum Network},
  author = {Cirac, J. I. and Zoller, P. and Kimble, H. J. and Mabuchi, H.},
  journal = {Phys. Rev. Lett.},
  volume = {78},
  issue = {16},
  pages = {3221--3224},
  numpages = {0},
  year = {1997},
  month = {Apr},
  publisher = {American Physical Society},
  doi = {10.1103/PhysRevLett.78.3221},
  url = {https://link.aps.org/doi/10.1103/PhysRevLett.78.3221}
}

@article{Gough2015EPJ,
    title= {Generating nonclassical quantum input field states with modulating filters},
	author = {Gough, John E. and Zhang, Guofeng},
	date = {2015/06/19},
	doi = {10.1140/epjqt/s40507-015-0027-z},
	id = {Gough2015},
	isbn = {2196-0763},
	journal = {EPJ Quantum Technology},
	number = {1},
	pages = {15},
	url = {https://doi.org/10.1140/epjqt/s40507-015-0027-z},
	volume = {2},
	year = {2015}
}

@INPROCEEDINGS{Nurdin2016IEEE,
  author={Nurdin, Hendra I. and James, Matthew R. and Yamamoto, N.},
  booktitle={2016 IEEE 55th Conference on Decision and Control (CDC)}, 
  title={Perfect single device absorber of arbitrary traveling single photon fields with a tunable coupling parameter: A QSDE approach}, 
  year={2016},
  volume={},
  number={},
  pages={2513-2518},
  doi={10.1109/CDC.2016.7798639}}

@article{Kiilerich2019PRL,
  title = {Input-Output Theory with Quantum Pulses},
  author = {Kiilerich, Alexander Holm and M\o{}lmer, Klaus},
  journal = {Phys. Rev. Lett.},
  volume = {123},
  issue = {12},
  pages = {123604},
  numpages = {6},
  year = {2019},
  month = {Sep},
  publisher = {American Physical Society},
  doi = {10.1103/PhysRevLett.123.123604},
  url = {https://link.aps.org/doi/10.1103/PhysRevLett.123.123604}
}

@article{Dicke1954PR,
  title = {Coherence in Spontaneous Radiation Processes},
  author = {Dicke, R. H.},
  journal = {Phys. Rev.},
  volume = {93},
  issue = {1},
  pages = {99--110},
  numpages = {0},
  year = {1954},
  month = {Jan},
  publisher = {American Physical Society},
  doi = {10.1103/PhysRev.93.99},
  url = {https://link.aps.org/doi/10.1103/PhysRev.93.99}
}

@article{Fischer2018OSA,
author = {Kevin A. Fischer},
journal = {OSA Continuum},
number = {2},
pages = {772--782},
publisher = {Optica Publishing Group},
title = {Exact calculation of stimulated emission driven by pulsed light},
volume = {1},
month = {Oct},
year = {2018},
url = {https://opg.optica.org/osac/abstract.cfm?URI=osac-1-2-772},
doi = {10.1364/OSAC.1.000772},
}

@article{Chabaud2020PRL,
  title = {Stellar Representation of Non-Gaussian Quantum States},
  author = {Chabaud, Ulysse and Markham, Damian and Grosshans, Fr\'ed\'eric},
  journal = {Phys. Rev. Lett.},
  volume = {124},
  issue = {6},
  pages = {063605},
  numpages = {6},
  year = {2020},
  month = {Feb},
  publisher = {American Physical Society},
  doi = {10.1103/PhysRevLett.124.063605},
  url = {https://link.aps.org/doi/10.1103/PhysRevLett.124.063605}
}

@article{Lund2024PRL,
  title = {Subtraction and Addition of Propagating Photons by Two-Level Emitters},
  author = {Lund, Mads M. and Yang, Fan and Christiansen, Victor Rueskov and Kornovan, Danil and M\o{}lmer, Klaus},
  journal = {Phys. Rev. Lett.},
  volume = {133},
  issue = {10},
  pages = {103601},
  numpages = {7},
  year = {2024},
  month = {Sep},
  publisher = {American Physical Society},
  doi = {10.1103/PhysRevLett.133.103601},
  url = {https://link.aps.org/doi/10.1103/PhysRevLett.133.103601}
}

@article{Ourjoumtsev2007PRL,
  title = {Increasing Entanglement between Gaussian States by Coherent Photon Subtraction},
  author = {Ourjoumtsev, Alexei and Dantan, Aur\'elien and Tualle-Brouri, Rosa and Grangier, Philippe},
  journal = {Phys. Rev. Lett.},
  volume = {98},
  issue = {3},
  pages = {030502},
  numpages = {4},
  year = {2007},
  month = {Jan},
  publisher = {American Physical Society},
  doi = {10.1103/PhysRevLett.98.030502},
  url = {https://link.aps.org/doi/10.1103/PhysRevLett.98.030502}
}

@article{Serino2023PRXQ,
  title = {Realization of a Multi-Output Quantum Pulse Gate for Decoding High-Dimensional Temporal Modes of Single-Photon States},
  author = {Serino, Laura and Gil-Lopez, Jano and Stefszky, Michael and Ricken, Raimund and Eigner, Christof and Brecht, Benjamin and Silberhorn, Christine},
  journal = {PRX Quantum},
  volume = {4},
  issue = {2},
  pages = {020306},
  numpages = {12},
  year = {2023},
  month = {Apr},
  publisher = {American Physical Society},
  doi = {10.1103/PRXQuantum.4.020306},
  url = {https://link.aps.org/doi/10.1103/PRXQuantum.4.020306}
}

@article{
Ourjoumtsev2006science,
author = {Alexei Ourjoumtsev  and Rosa Tualle-Brouri  and Julien Laurat  and Philippe Grangier },
title = {Generating Optical Schrödinger Kittens for Quantum Information Processing},
journal = {Science},
volume = {312},
number = {5770},
pages = {83-86},
year = {2006},
doi = {10.1126/science.1122858},
URL = {https://www.science.org/doi/abs/10.1126/science.1122858},
eprint = {https://www.science.org/doi/pdf/10.1126/science.1122858}}

@article{Ourjoumtsev2007Nature,
	author = {Ourjoumtsev, Alexei and Jeong, Hyunseok and Tualle-Brouri, Rosa and Grangier, Philippe},
	date = {2007/08/01},
	doi = {10.1038/nature06054},
	id = {Ourjoumtsev2007},
	isbn = {1476-4687},
	journal = {Nature},
	number = {7155},
	pages = {784--786},
	title = {Generation of optical `Schr{\"o}dinger cats'from photon number states},
	url = {https://doi.org/10.1038/nature06054},
	volume = {448},
	year = {2007}}

@article{Dakna1997PRA,
  title = {Generating Schr\"odinger-cat-like states by means of conditional measurements on a beam splitter},
  author = {Dakna, M. and Anhut, T. and Opatrn\'y, T. and Kn\"oll, L. and Welsch, D.-G.},
  journal = {Phys. Rev. A},
  volume = {55},
  issue = {4},
  pages = {3184--3194},
  numpages = {0},
  year = {1997},
  month = {Apr},
  publisher = {American Physical Society},
  doi = {10.1103/PhysRevA.55.3184},
  url = {https://link.aps.org/doi/10.1103/PhysRevA.55.3184}
}

@article{Combes2017APX,
author = {Joshua Combes and Joseph Kerckhoff and Mohan Sarovar},
title = {The SLH framework for modeling quantum input-output networks},
journal = {Advances in Physics: X},
volume = {2},
number = {3},
pages = {784--888},
year = {2017},
publisher = {Taylor \& Francis},
doi = {10.1080/23746149.2017.1343097},
URL = {https://doi.org/10.1080/23746149.2017.1343097},
eprint = {https://doi.org/10.1080/23746149.2017.1343097
}}

@article{Gheri1998,
author = {Gheri, Klaus M. and Ellinger, Klaus and Pellizzari, Thomas and Zoller, Peter},
title = {Photon-Wavepackets as Flying Quantum Bits},
journal = {Fortschritte der Physik},
volume = {46},
number = {4-5},
pages = {401-415},
doi = {https://doi.org/10.1002/(SICI)1521-3978(199806)46:4/5<401::AID-PROP401>3.0.CO;2-W},
url = {https://onlinelibrary.wiley.com/doi/abs/10.1002/%28SICI%291521-3978%28199806%2946%3A4/5%3C401%3A%3AAID-PROP401%3E3.0.CO%3B2-W},
year = {1998}
}

@Article{Lodahl2017Nature,
author={Lodahl, Peter
and Mahmoodian, Sahand
and Stobbe, S{\o}ren
and Rauschenbeutel, Arno
and Schneeweiss, Philipp
and Volz, J{\"u}rgen
and Pichler, Hannes
and Zoller, Peter},
title={Chiral quantum optics},
journal={Nature},
year={2017},
month={Jan},
day={01},
volume={541},
number={7638},
pages={473-480},
issn={1476-4687},
doi={10.1038/nature21037},
url={https://doi.org/10.1038/nature21037}
}

@article{Pichler2016PRL,
  title = {Photonic Circuits with Time Delays and Quantum Feedback},
  author = {Pichler, Hannes and Zoller, Peter},
  journal = {Phys. Rev. Lett.},
  volume = {116},
  issue = {9},
  pages = {093601},
  numpages = {6},
  year = {2016},
  month = {Mar},
  publisher = {American Physical Society},
  doi = {10.1103/PhysRevLett.116.093601},
  url = {https://link.aps.org/doi/10.1103/PhysRevLett.116.093601}
}

@misc{Luo2024,
      title={Efficient optical cat state generation using squeezed few-photon superposition states}, 
      author={Haoyuan Luo and Sahand Mahmoodian},
      year={2024},
      eprint={2412.14798},
      archivePrefix={arXiv},
      primaryClass={quant-ph},
      url={https://arxiv.org/abs/2412.14798}, 
}

@article{Winnel2024PRL,
  title = {Deterministic Preparation of Optical Squeezed Cat and Gottesman-Kitaev-Preskill States},
  author = {Winnel, Matthew S. and Guanzon, Joshua J. and Singh, Deepesh and Ralph, Timothy C.},
  journal = {Phys. Rev. Lett.},
  volume = {132},
  issue = {23},
  pages = {230602},
  numpages = {6},
  year = {2024},
  month = {Jun},
  publisher = {American Physical Society},
  doi = {10.1103/PhysRevLett.132.230602},
  url = {https://link.aps.org/doi/10.1103/PhysRevLett.132.230602}
}

@article{Paulisch2019PRA,
  title = {Quantum metrology with one-dimensional superradiant photonic states},
  author = {Paulisch, V. and Perarnau-Llobet, M. and Gonz\'alez-Tudela, A. and Cirac, J. I.},
  journal = {Phys. Rev. A},
  volume = {99},
  issue = {4},
  pages = {043807},
  numpages = {15},
  year = {2019},
  month = {Apr},
  publisher = {American Physical Society},
  doi = {10.1103/PhysRevA.99.043807},
  url = {https://link.aps.org/doi/10.1103/PhysRevA.99.043807}
}

@article{Thomas2022Nature,
	author = {Thomas, Philip and Ruscio, Leonardo and Morin, Olivier and Rempe, Gerhard},
	date = {2022/08/01},
	doi = {10.1038/s41586-022-04987-5},
	id = {Thomas2022},
	isbn = {1476-4687},
	journal = {Nature},
	number = {7924},
	pages = {677--681},
	title = {Efficient generation of entangled multiphoton graph states from a single atom},
	url = {https://doi.org/10.1038/s41586-022-04987-5},
	volume = {608},
	year = {2022}}

@article{Fluhmann2019Nature,
	author = {Fl{\"u}hmann, C. and Nguyen, T. L. and Marinelli, M. and Negnevitsky, V. and Mehta, K. and Home, J. P.},
	date = {2019/02/01},
	doi = {10.1038/s41586-019-0960-6},
	id = {Fl{\"u}hmann2019},
	isbn = {1476-4687},
	journal = {Nature},
	number = {7745},
	pages = {513--517},
	title = {Encoding a qubit in a trapped-ion mechanical oscillator},
	url = {https://doi.org/10.1038/s41586-019-0960-6},
	volume = {566},
	year = {2019}}

@Article{Tomm2021NNANO,
author={Tomm, Natasha
and Javadi, Alisa
and Antoniadis, Nadia Olympia
and Najer, Daniel
and L{\"o}bl, Matthias Christian
and Korsch, Alexander Rolf
and Schott, R{\"u}diger
and Valentin, Sascha Ren{\'e}
and Wieck, Andreas Dirk
and Ludwig, Arne
and Warburton, Richard John},
title={A bright and fast source of coherent single photons},
journal={Nature Nanotechnology},
year={2021},
month={Apr},
day={01},
volume={16},
number={4},
pages={399-403},
issn={1748-3395},
doi={10.1038/s41565-020-00831-x},
url={https://doi.org/10.1038/s41565-020-00831-x}
}

@article{Valente2012NJP,
doi = {10.1088/1367-2630/14/8/083029},
url = {https://doi.org/10.1088/1367-2630/14/8/083029},
year = {2012},
month = {aug},
publisher = {IOP Publishing},
volume = {14},
number = {8},
pages = {083029},
author = {Valente, D and Li, Y and Poizat, J P and Gérard, J M and Kwek, L C and Santos, M F and Auffèves, A},
title = {Optimal irreversible stimulated emission},
journal = {New Journal of Physics}
}

@article{Lodahl2015RMP,
  title = {Interfacing single photons and single quantum dots with photonic nanostructures},
  author = {Lodahl, Peter and Mahmoodian, Sahand and Stobbe, S\o{}ren},
  journal = {Rev. Mod. Phys.},
  volume = {87},
  issue = {2},
  pages = {347--400},
  numpages = {54},
  year = {2015},
  month = {May},
  publisher = {American Physical Society},
  doi = {10.1103/RevModPhys.87.347},
  url = {https://link.aps.org/doi/10.1103/RevModPhys.87.347}
}

@article{Yudson1985JETP,
  title={Dynamics of integrable quantum systems},
  author={Yudson, V. I.},
  journal={Sov. Phys. JETP},
  volume={88},
  year = {1985},
  pages={1043}
}

@ARTICLE{Shen2007PRA,
  author = {Jung-Tsung Shen and Shanhui Fan},
  title = {Strongly correlated multiparticle transport in one dimension through a quantum impurity},
  journal = PRA,
  year = {2007},
  volume = {76},
  pages = {062709},
  doi={10.1103/PhysRevA.76.062709}
}

@article{Hastrup2020OL,
author = {Jacob Hastrup and Jonas Schou Neergaard-Nielsen and Ulrik Lund Andersen},
journal = {Opt. Lett.},
number = {3},
pages = {640--643},
publisher = {Optica Publishing Group},
title = {Deterministic generation of a four-component optical cat state},
volume = {45},
month = {Feb},
year = {2020},
url = {https://opg.optica.org/ol/abstract.cfm?URI=ol-45-3-640},
doi = {10.1364/OL.383194},
}

@article{Grimsmo2020PRX,
  title = {Quantum Computing with Rotation-Symmetric Bosonic Codes},
  author = {Grimsmo, Arne L. and Combes, Joshua and Baragiola, Ben Q.},
  journal = {Phys. Rev. X},
  volume = {10},
  issue = {1},
  pages = {011058},
  numpages = {32},
  year = {2020},
  month = {Mar},
  publisher = {American Physical Society},
  doi = {10.1103/PhysRevX.10.011058},
  url = {https://link.aps.org/doi/10.1103/PhysRevX.10.011058}
}

@article{Gottesman2001PRA,
  title = {Encoding a qubit in an oscillator},
  author = {Gottesman, Daniel and Kitaev, Alexei and Preskill, John},
  journal = {Phys. Rev. A},
  volume = {64},
  issue = {1},
  pages = {012310},
  numpages = {21},
  year = {2001},
  month = {Jun},
  publisher = {American Physical Society},
  doi = {10.1103/PhysRevA.64.012310},
  url = {https://link.aps.org/doi/10.1103/PhysRevA.64.012310}
}

@article{Xanadu2025Nat,
	author = {Larsen, M. V. and Bourassa, J. E. and Kocsis, S. and Tasker, J. F. and Chadwick, R. S. and Gonz{\'a}lez-Arciniegas, C. and Hastrup, J. and Lopetegui-Gonz{\'a}lez, C. E. and Miatto, F. M. and Motamedi, A. and Noro, R. and Roeland, G. and Baby, R. and Chen, H. and Contu, P. and Di Luch, I. and Drago, C. and Giesbrecht, M. and Grainge, T. and Krasnokutska, I. and Menotti, M. and Morrison, B. and Puviraj, C. and Rezaei Shad, K. and Hussain, B. and McMahon, J. and Ortmann, J. E. and Collins, M. J. and Ma, C. and Phillips, D. S. and Seymour, M. and Tang, Q. Y. and Yang, B. and Vernon, Z. and Alexander, R. N. and Mahler, D. H.},
	date = {2025/06/01},
	doi = {10.1038/s41586-025-09044-5},
	id = {Larsen2025},
	isbn = {1476-4687},
	journal = {Nature},
	number = {8068},
	pages = {587--591},
	title = {Integrated photonic source of Gottesman--Kitaev--Preskill qubits},
	url = {https://doi.org/10.1038/s41586-025-09044-5},
	volume = {642},
	year = {2025}
}

@article{Sollner2015NNA,
	author = {S{\"o}llner, Immo and Mahmoodian, Sahand and Hansen, Sofie Lindskov and Midolo, Leonardo and Javadi, Alisa and Kir{\v s}ansk{\.e}, Gabija and Pregnolato, Tommaso and El-Ella, Haitham and Lee, Eun Hye and Song, Jin Dong and Stobbe, S{\o}ren and Lodahl, Peter},
	date = {2015/09/01},
	doi = {10.1038/nnano.2015.159},
	id = {S{\"o}llner2015},
	isbn = {1748-3395},
	journal = {Nature Nanotechnology},
	number = {9},
	pages = {775--778},
	title = {Deterministic photon--emitter coupling in chiral photonic circuits},
	url = {https://doi.org/10.1038/nnano.2015.159},
	volume = {10},
	year = {2015}
}

@article{Mahmoodian2020PRX,
  title = {Dynamics of Many-Body Photon Bound States in Chiral Waveguide QED},
  author = {Mahmoodian, Sahand and Calaj\'o, Giuseppe and Chang, Darrick E. and Hammerer, Klemens and S\o{}rensen, Anders S.},
  journal = {Phys. Rev. X},
  volume = {10},
  issue = {3},
  pages = {031011},
  numpages = {14},
  year = {2020},
  month = {Jul},
  publisher = {American Physical Society},
  doi = {10.1103/PhysRevX.10.031011},
  url = {https://link.aps.org/doi/10.1103/PhysRevX.10.031011}
}

@article{Ralph2015PRL,
  title = {Photon Sorting, Efficient Bell Measurements, and a Deterministic Controlled-$Z$ Gate Using a Passive Two-Level Nonlinearity},
  author = {Ralph, T. C. and S\"ollner, I. and Mahmoodian, S. and White, A. G. and Lodahl, P.},
  journal = {Phys. Rev. Lett.},
  volume = {114},
  issue = {17},
  pages = {173603},
  numpages = {5},
  year = {2015},
  month = {Apr},
  publisher = {American Physical Society},
  doi = {10.1103/PhysRevLett.114.173603},
  url = {https://link.aps.org/doi/10.1103/PhysRevLett.114.173603}
}

@article{Yang2022PRL,
  title = {Deterministic Photon Sorting in Waveguide QED Systems},
  author = {Yang, Fan and Lund, Mads M. and Pohl, Thomas and Lodahl, Peter and M\o{}lmer, Klaus},
  journal = {Phys. Rev. Lett.},
  volume = {128},
  issue = {21},
  pages = {213603},
  numpages = {6},
  year = {2022},
  month = {May},
  publisher = {American Physical Society},
  doi = {10.1103/PhysRevLett.128.213603},
  url = {https://link.aps.org/doi/10.1103/PhysRevLett.128.213603}
}

@article{Takase2021PRA,
  title = {Generation of optical Schr\"odinger cat states by generalized photon subtraction},
  author = {Takase, Kan and Yoshikawa, Jun-ichi and Asavanant, Warit and Endo, Mamoru and Furusawa, Akira},
  journal = {Phys. Rev. A},
  volume = {103},
  issue = {1},
  pages = {013710},
  numpages = {8},
  year = {2021},
  month = {Jan},
  publisher = {American Physical Society},
  doi = {10.1103/PhysRevA.103.013710},
  url = {https://link.aps.org/doi/10.1103/PhysRevA.103.013710}
}

@misc{Hanamura2025,
      title={Beyond Stellar Rank: Control Parameters for Scalable Optical Non-Gaussian State Generation}, 
      author={Fumiya Hanamura and Kan Takase and Hironari Nagayoshi and Ryuhoh Ide and Warit Asavanant and Kosuke Fukui and Petr Marek and Radim Filip and Akira Furusawa},
      year={2025},
      eprint={2509.06255},
      archivePrefix={arXiv},
      primaryClass={quant-ph},
      url={https://arxiv.org/abs/2509.06255}, 
}

@article{Takase2024PRA,
  title = {Generation of flying logical qubits using generalized photon subtraction with adaptive Gaussian operations},
  author = {Takase, Kan and Hanamura, Fumiya and Nagayoshi, Hironari and Bourassa, J. Eli and Alexander, Rafael N. and Kawasaki, Akito and Asavanant, Warit and Endo, Mamoru and Furusawa, Akira},
  journal = {Phys. Rev. A},
  volume = {110},
  issue = {1},
  pages = {012436},
  numpages = {10},
  year = {2024},
  month = {Jul},
  publisher = {American Physical Society},
  doi = {10.1103/PhysRevA.110.012436},
  url = {https://link.aps.org/doi/10.1103/PhysRevA.110.012436}
}

@article{Bartlett2002PRL,
  title = {Efficient Classical Simulation of Continuous Variable Quantum Information Processes},
  author = {Bartlett, Stephen D. and Sanders, Barry C. and Braunstein, Samuel L. and Nemoto, Kae},
  journal = {Phys. Rev. Lett.},
  volume = {88},
  issue = {9},
  pages = {097904},
  numpages = {4},
  year = {2002},
  month = {Feb},
  publisher = {American Physical Society},
  doi = {10.1103/PhysRevLett.88.097904},
  url = {https://link.aps.org/doi/10.1103/PhysRevLett.88.097904}
}

@article{Mari2012PRL,
  title = {Positive Wigner Functions Render Classical Simulation of Quantum Computation Efficient},
  author = {Mari, A. and Eisert, J.},
  journal = {Phys. Rev. Lett.},
  volume = {109},
  issue = {23},
  pages = {230503},
  numpages = {5},
  year = {2012},
  month = {Dec},
  publisher = {American Physical Society},
  doi = {10.1103/PhysRevLett.109.230503},
  url = {https://link.aps.org/doi/10.1103/PhysRevLett.109.230503}
}

@article{Lloyd1999PRL,
  title = {Quantum Computation over Continuous Variables},
  author = {Lloyd, Seth and Braunstein, Samuel L.},
  journal = {Phys. Rev. Lett.},
  volume = {82},
  issue = {8},
  pages = {1784--1787},
  numpages = {0},
  year = {1999},
  month = {Feb},
  publisher = {American Physical Society},
  doi = {10.1103/PhysRevLett.82.1784},
  url = {https://link.aps.org/doi/10.1103/PhysRevLett.82.1784}
}

@article{Mirrahimi2014NJP,
doi = {10.1088/1367-2630/16/4/045014},
url = {https://doi.org/10.1088/1367-2630/16/4/045014},
year = {2014},
month = {apr},
publisher = {IOP Publishing},
volume = {16},
number = {4},
pages = {045014},
author = {Mirrahimi, Mazyar and Leghtas, Zaki and Albert, Victor V and Touzard, Steven and Schoelkopf, Robert J and Jiang, Liang and Devoret, Michel H},
title = {Dynamically protected cat-qubits: a new paradigm for universal quantum computation},
journal = {New Journal of Physics}
}

@article{Ralph2003PRA,
  title = {Quantum computation with optical coherent states},
  author = {Ralph, T. C. and Gilchrist, A. and Milburn, G. J. and Munro, W. J. and Glancy, S.},
  journal = {Phys. Rev. A},
  volume = {68},
  issue = {4},
  pages = {042319},
  numpages = {11},
  year = {2003},
  month = {Oct},
  publisher = {American Physical Society},
  doi = {10.1103/PhysRevA.68.042319},
  url = {https://link.aps.org/doi/10.1103/PhysRevA.68.042319}
}

@article{Gilchrist2004JOB,
doi = {10.1088/1464-4266/6/8/032},
url = {https://doi.org/10.1088/1464-4266/6/8/032},
year = {2004},
month = {jul},
publisher = {},
volume = {6},
number = {8},
pages = {S828},
author = {A Gilchrist and Kae Nemoto and W J Munro and T C Ralph and S Glancy and Samuel L Braunstein and G J Milburn},
title = {Schrödinger cats and their power for quantum information processing},
journal = {Journal of Optics B: Quantum and Semiclassical Optics}
}

@article{Munro2002PRA,
  title = {Weak-force detection with superposed coherent states},
  author = {Munro, W. J. and Nemoto, K. and Milburn, G. J. and Braunstein, S. L.},
  journal = {Phys. Rev. A},
  volume = {66},
  issue = {2},
  pages = {023819},
  numpages = {6},
  year = {2002},
  month = {Aug},
  publisher = {American Physical Society},
  doi = {10.1103/PhysRevA.66.023819},
  url = {https://link.aps.org/doi/10.1103/PhysRevA.66.023819}
}

@article{Knill2001Nature,
	author = {Knill, E. and Laflamme, R. and Milburn, G. J.},
	date = {2001/01/01},
	doi = {10.1038/35051009},
	id = {Knill2001},
	isbn = {1476-4687},
	journal = {Nature},
	number = {6816},
	pages = {46--52},
	title = {A scheme for efficient quantum computation with linear optics},
	url = {https://doi.org/10.1038/35051009},
	volume = {409},
	year = {2001}}

@article{Kok2007RevMP,
  title = {Linear optical quantum computing with photonic qubits},
  author = {Kok, Pieter and Munro, W. J. and Nemoto, Kae and Ralph, T. C. and Dowling, Jonathan P. and Milburn, G. J.},
  journal = {Rev. Mod. Phys.},
  volume = {79},
  issue = {1},
  pages = {135--174},
  numpages = {0},
  year = {2007},
  month = {Jan},
  publisher = {American Physical Society},
  doi = {10.1103/RevModPhys.79.135},
  url = {https://link.aps.org/doi/10.1103/RevModPhys.79.135}
}

@article{Brask2010PRL,
  title = {Hybrid Long-Distance Entanglement Distribution Protocol},
  author = {Brask, J. B. and Rigas, I. and Polzik, E. S. and Andersen, U. L. and S\o{}rensen, A. S.},
  journal = {Phys. Rev. Lett.},
  volume = {105},
  issue = {16},
  pages = {160501},
  numpages = {4},
  year = {2010},
  month = {Oct},
  publisher = {American Physical Society},
  doi = {10.1103/PhysRevLett.105.160501},
  url = {https://link.aps.org/doi/10.1103/PhysRevLett.105.160501}
}

@article{Braunstein2005RevMP,
  title = {Quantum information with continuous variables},
  author = {Braunstein, Samuel L. and van Loock, Peter},
  journal = {Rev. Mod. Phys.},
  volume = {77},
  issue = {2},
  pages = {513--577},
  numpages = {0},
  year = {2005},
  month = {Jun},
  publisher = {American Physical Society},
  doi = {10.1103/RevModPhys.77.513},
  url = {https://link.aps.org/doi/10.1103/RevModPhys.77.513}
}

@article{Grimsmo2021PRXQ,
  title = {Quantum Error Correction with the Gottesman-Kitaev-Preskill Code},
  author = {Grimsmo, Arne L. and Puri, Shruti},
  journal = {PRX Quantum},
  volume = {2},
  issue = {2},
  pages = {020101},
  numpages = {20},
  year = {2021},
  month = {Jun},
  publisher = {American Physical Society},
  doi = {10.1103/PRXQuantum.2.020101},
  url = {https://link.aps.org/doi/10.1103/PRXQuantum.2.020101}
}

@article{Matsos2024PRL,
  title = {Robust and Deterministic Preparation of Bosonic Logical States in a Trapped Ion},
  author = {Matsos, V. G. and Valahu, C. H. and Navickas, T. and Rao, A. D. and Millican, M. J. and Kolesnikow, X. C. and Biercuk, M. J. and Tan, T. R.},
  journal = {Phys. Rev. Lett.},
  volume = {133},
  issue = {5},
  pages = {050602},
  numpages = {7},
  year = {2024},
  month = {Jul},
  publisher = {American Physical Society},
  doi = {10.1103/PhysRevLett.133.050602},
  url = {https://link.aps.org/doi/10.1103/PhysRevLett.133.050602}
}

@article{Ofek2016Nature,
	author = {Ofek, Nissim and Petrenko, Andrei and Heeres, Reinier and Reinhold, Philip and Leghtas, Zaki and Vlastakis, Brian and Liu, Yehan and Frunzio, Luigi and Girvin, S. M. and Jiang, L. and Mirrahimi, Mazyar and Devoret, M. H. and Schoelkopf, R. J.},
	date = {2016/08/01},
	doi = {10.1038/nature18949},
	id = {Ofek2016},
	isbn = {1476-4687},
	journal = {Nature},
	number = {7617},
	pages = {441--445},
	title = {Extending the lifetime of a quantum bit with error correction in superconducting circuits},
	url = {https://doi.org/10.1038/nature18949},
	volume = {536},
	year = {2016}}

@article{Aharonovich2016NPho,
	author = {Aharonovich, Igor and Englund, Dirk and Toth, Milos},
	date = {2016/10/01},
	doi = {10.1038/nphoton.2016.186},
	id = {Aharonovich2016},
	isbn = {1749-4893},
	journal = {Nature Photonics},
	number = {10},
	pages = {631--641},
	title = {Solid-state single-photon emitters},
	url = {https://doi.org/10.1038/nphoton.2016.186},
	volume = {10},
	year = {2016}}

@article{Senellart2017NNano,
	author = {Senellart, Pascale and Solomon, Glenn and White, Andrew},
	date = {2017/11/01},
	doi = {10.1038/nnano.2017.218},
	id = {Senellart2017},
	isbn = {1748-3395},
	journal = {Nature Nanotechnology},
	number = {11},
	pages = {1026--1039},
	title = {High-performance semiconductor quantum-dot single-photon sources},
	url = {https://doi.org/10.1038/nnano.2017.218},
	volume = {12},
	year = {2017}}

@article{Suarez2025PRXQ,
  title = {Chiral Quantum Optics: Recent Developments and Future Directions},
  author = {Su\'arez-Forero, D.G. and Jalali Mehrabad, M. and Vega, C. and Gonz\'alez-Tudela, A. and Hafezi, M.},
  journal = {PRX Quantum},
  volume = {6},
  issue = {2},
  pages = {020101},
  numpages = {18},
  year = {2025},
  month = {Apr},
  publisher = {American Physical Society},
  doi = {10.1103/PRXQuantum.6.020101},
  url = {https://link.aps.org/doi/10.1103/PRXQuantum.6.020101}
}

@article{Yin2013PRL,
  title = {Catch and Release of Microwave Photon States},
  author = {Yin, Yi and Chen, Yu and Sank, Daniel and O'Malley, P. J. J. and White, T. C. and Barends, R. and Kelly, J. and Lucero, Erik and Mariantoni, Matteo and Megrant, A. and Neill, C. and Vainsencher, A. and Wenner, J. and Korotkov, Alexander N. and Cleland, A. N. and Martinis, John M.},
  journal = {Phys. Rev. Lett.},
  volume = {110},
  issue = {10},
  pages = {107001},
  numpages = {5},
  year = {2013},
  month = {Mar},
  publisher = {American Physical Society},
  doi = {10.1103/PhysRevLett.110.107001},
  url = {https://link.aps.org/doi/10.1103/PhysRevLett.110.107001}
}

@article{Pechal2014PRX,
  title = {Microwave-Controlled Generation of Shaped Single Photons in Circuit Quantum Electrodynamics},
  author = {Pechal, M. and Huthmacher, L. and Eichler, C. and Zeytino\ifmmode \breve{g}\else \u{g}\fi{}lu, S. and Abdumalikov, A. A. and Berger, S. and Wallraff, A. and Filipp, S.},
  journal = {Phys. Rev. X},
  volume = {4},
  issue = {4},
  pages = {041010},
  numpages = {9},
  year = {2014},
  month = {Oct},
  publisher = {American Physical Society},
  doi = {10.1103/PhysRevX.4.041010},
  url = {https://link.aps.org/doi/10.1103/PhysRevX.4.041010}
}

@article{Ritter2012Nature,
	author = {Ritter, Stephan and N{\"o}lleke, Christian and Hahn, Carolin and Reiserer, Andreas and Neuzner, Andreas and Uphoff, Manuel and M{\"u}cke, Martin and Figueroa, Eden and Bochmann, Joerg and Rempe, Gerhard},
	date = {2012/04/01},
	doi = {10.1038/nature11023},
	id = {Ritter2012},
	isbn = {1476-4687},
	journal = {Nature},
	number = {7393},
	pages = {195--200},
	title = {An elementary quantum network of single atoms in optical cavities},
	url = {https://doi.org/10.1038/nature11023},
	volume = {484},
	year = {2012}}

@article{
Chou2007Sci,
author = {Chin-Wen Chou  and Julien Laurat  and Hui Deng  and Kyung Soo Choi  and Hugues de Riedmatten  and Daniel Felinto  and H. Jeff Kimble },
title = {Functional Quantum Nodes for Entanglement Distribution over Scalable Quantum Networks},
journal = {Science},
volume = {316},
number = {5829},
pages = {1316-1320},
year = {2007},
doi = {10.1126/science.1140300},
URL = {https://www.science.org/doi/abs/10.1126/science.1140300},
eprint = {https://www.science.org/doi/pdf/10.1126/science.1140300}}

@article{Joshi2023PRX,
  title = {Resonance Fluorescence of a Chiral Artificial Atom},
  author = {Joshi, Chaitali and Yang, Frank and Mirhosseini, Mohammad},
  journal = {Phys. Rev. X},
  volume = {13},
  issue = {2},
  pages = {021039},
  numpages = {27},
  year = {2023},
  month = {Jun},
  publisher = {American Physical Society},
  doi = {10.1103/PhysRevX.13.021039},
  url = {https://link.aps.org/doi/10.1103/PhysRevX.13.021039}
}

@article{Korotkov2011PRB,
  title = {Flying microwave qubits with nearly perfect transfer efficiency},
  author = {Korotkov, Alexander N.},
  journal = {Phys. Rev. B},
  volume = {84},
  issue = {1},
  pages = {014510},
  numpages = {10},
  year = {2011},
  month = {Jul},
  publisher = {American Physical Society},
  doi = {10.1103/PhysRevB.84.014510},
  url = {https://link.aps.org/doi/10.1103/PhysRevB.84.014510}
}

@article{Axline2018NPhy,
	author = {Axline, Christopher J. and Burkhart, Luke D. and Pfaff, Wolfgang and Zhang, Mengzhen and Chou, Kevin and Campagne-Ibarcq, Philippe and Reinhold, Philip and Frunzio, Luigi and Girvin, S. M. and Jiang, Liang and Devoret, M. H. and Schoelkopf, R. J.},
	date = {2018/07/01},
	doi = {10.1038/s41567-018-0115-y},
	id = {Axline2018},
	isbn = {1745-2481},
	journal = {Nature Physics},
	number = {7},
	pages = {705--710},
	title = {On-demand quantum state transfer and entanglement between remote microwave cavity memories},
	url = {https://doi.org/10.1038/s41567-018-0115-y},
	volume = {14},
	year = {2018}}

@article{Kurpiers2018Nature,
	author = {Kurpiers, P. and Magnard, P. and Walter, T. and Royer, B. and Pechal, M. and Heinsoo, J. and Salath{\'e}, Y. and Akin, A. and Storz, S. and Besse, J. -C. and Gasparinetti, S. and Blais, A. and Wallraff, A.},
	date = {2018/06/01},
	doi = {10.1038/s41586-018-0195-y},
	id = {Kurpiers2018},
	isbn = {1476-4687},
	journal = {Nature},
	number = {7709},
	pages = {264--267},
	title = {Deterministic quantum state transfer and remote entanglement using microwave photons},
	url = {https://doi.org/10.1038/s41586-018-0195-y},
	volume = {558},
	year = {2018}}

@article{Bienfait2019Sci,
author = {A. Bienfait  and K. J. Satzinger  and Y. P. Zhong  and H.-S. Chang  and M.-H. Chou  and C. R. Conner  and {\'E}. Dumur  and J. Grebel  and G. A. Peairs  and R. G. Povey  and A. N. Cleland },
title = {Phonon-mediated quantum state transfer and remote qubit entanglement},
journal = {Science},
volume = {364},
number = {6438},
pages = {368-371},
year = {2019},
doi = {10.1126/science.aaw8415},
URL = {https://www.science.org/doi/abs/10.1126/science.aaw8415}}

@article{qutip5,
  title = {QuTiP 5: The Quantum Toolbox in {Python}},
  author = {
    Lambert, Neill and Gigu{`e}re, Eric and Menczel, Paul and Li, Boxi and
    Hopf, Patrick and Su{'a}rez, Gerardo and Gali, Marc and Lishman, Jake and
    Gadhvi, Rushiraj and Agarwal, Rochisha and Galicia, Asier and Shammah, Nathan and
    Nation, Paul and Johansson, J. R. and Ahmed, Shahnawaz and Cross, Simon and
    Pitchford, Alexander and Nori, Franco
  },
  journal = {Physics Reports},
  volume = {1153},
  pages = {1-62},
  year = {2026},
  issn = {0370-1573},
  doi = {10.1016/j.physrep.2025.10.001},
  url = {https://www.sciencedirect.com/science/article/pii/S0370157325002704},
}

@Article{ITensor2022,
	title={{The ITensor Software Library for Tensor Network Calculations}},
	author={Matthew Fishman and Steven R. White and E. Miles Stoudenmire},
	journal={SciPost Phys. Codebases},
	pages={4},
	year={2022},
	publisher={SciPost},
	doi={10.21468/SciPostPhysCodeb.4},
	url={https://scipost.org/10.21468/SciPostPhysCodeb.4},
}

@article{Ding2025NPho,
	author = {Ding, Xing and Guo, Yong-Peng and Xu, Mo-Chi and Liu, Run-Ze and Zou, Geng-Yan and Zhao, Jun-Yi and Ge, Zhen-Xuan and Zhang, Qi-Hang and Liu, Hua-Liang and Wang, Lin-Jun and Chen, Ming-Cheng and Wang, Hui and He, Yu-Ming and Huo, Yong-Heng and Lu, Chao-Yang and Pan, Jian-Wei},
	date = {2025/04/01},
	doi = {10.1038/s41566-025-01639-8},
	id = {Ding2025},
	isbn = {1749-4893},
	journal = {Nature Photonics},
	number = {4},
	pages = {387--391},
	title = {High-efficiency single-photon source above the loss-tolerant threshold for efficient linear optical quantum computing},
	url = {https://doi.org/10.1038/s41566-025-01639-8},
	volume = {19},
	year = {2025}}

@article{Gu2009PRA,
  title = {Quantum computing with continuous-variable clusters},
  author = {Gu, Mile and Weedbrook, Christian and Menicucci, Nicolas C. and Ralph, Timothy C. and van Loock, Peter},
  journal = {Phys. Rev. A},
  volume = {79},
  issue = {6},
  pages = {062318},
  numpages = {16},
  year = {2009},
  month = {Jun},
  publisher = {American Physical Society},
  doi = {10.1103/PhysRevA.79.062318},
  url = {https://link.aps.org/doi/10.1103/PhysRevA.79.062318}
}

@article{Miyata2016PRA,
  title = {Implementation of a quantum cubic gate by an adaptive non-Gaussian measurement},
  author = {Miyata, Kazunori and Ogawa, Hisashi and Marek, Petr and Filip, Radim and Yonezawa, Hidehiro and Yoshikawa, Jun-ichi and Furusawa, Akira},
  journal = {Phys. Rev. A},
  volume = {93},
  issue = {2},
  pages = {022301},
  numpages = {10},
  year = {2016},
  month = {Feb},
  publisher = {American Physical Society},
  doi = {10.1103/PhysRevA.93.022301},
  url = {https://link.aps.org/doi/10.1103/PhysRevA.93.022301}
}

@article{Yukawa2013OE,
author = {Mitsuyoshi Yukawa and Kazunori Miyata and Takahiro Mizuta and Hidehiro Yonezawa and Petr Marek and Radim Filip and Akira Furusawa},
journal = {Opt. Express},
number = {5},
pages = {5529--5535},
publisher = {Optica Publishing Group},
title = {Generating superposition of up-to three photons for continuous variable quantum information processing},
volume = {21},
month = {Mar},
year = {2013},
url = {https://opg.optica.org/oe/abstract.cfm?URI=oe-21-5-5529},
doi = {10.1364/OE.21.005529},
}

@article{Gonzalez2017PRL,
  title = {Efficient Multiphoton Generation in Waveguide Quantum Electrodynamics},
  author = {Gonz\'alez-Tudela, A. and Paulisch, V. and Kimble, H. J. and Cirac, J. I.},
  journal = {Phys. Rev. Lett.},
  volume = {118},
  issue = {21},
  pages = {213601},
  numpages = {6},
  year = {2017},
  month = {May},
  publisher = {American Physical Society},
  doi = {10.1103/PhysRevLett.118.213601},
  url = {https://link.aps.org/doi/10.1103/PhysRevLett.118.213601}
}

@article{Gonzalez2015PRL,
  title = {Deterministic Generation of Arbitrary Photonic States Assisted by Dissipation},
  author = {Gonz\'alez-Tudela, A. and Paulisch, V. and Chang, D. E. and Kimble, H. J. and Cirac, J. I.},
  journal = {Phys. Rev. Lett.},
  volume = {115},
  issue = {16},
  pages = {163603},
  numpages = {6},
  year = {2015},
  month = {Oct},
  publisher = {American Physical Society},
  doi = {10.1103/PhysRevLett.115.163603},
  url = {https://link.aps.org/doi/10.1103/PhysRevLett.115.163603}
}

@article{LopesdaSilva24JOptB,
author = {Wendel Lopes da Silva and Daniel Valente},
journal = {J. Opt. Soc. Am. B},
number = {8},
pages = {C1--C8},
publisher = {Optica Publishing Group},
title = {Exact solution of a lambda quantum system driven by a two-photon wave packet},
volume = {41},
month = {Aug},
year = {2024},
url = {https://opg.optica.org/josab/abstract.cfm?URI=josab-41-8-C1},
doi = {10.1364/JOSAB.515618},
}

@article{Yu2026NatPho,
	author = {Yu, Shang and Sun, Jinzhao and Chen, Kuan-Cheng and Yang, Zhi-Huai and Li, Zhenghao and Mer, Ewan and Alwehaibi, Yazeed K. and Winston, Shana H. and Lopena, Dayne Marcus and Zhang, Zi-Cheng and Yang, Guang and Tao, Runxia and Zhou, Mingti and Machado, Gerard J. and Dong, Ying and Bondesan, Roberto and Vedral, Vlatko and Kim, M. S. and Walmsley, Ian A. and Patel, Raj B.},
	date = {2026/07/31},
	doi = {10.1038/s41566-026-01962-8},
	id = {Yu2026},
	isbn = {1749-4893},
	journal = {Nature Photonics},
	title = {Extensible universal photonic quantum computing with nonlinearity},
	url = {https://doi.org/10.1038/s41566-026-01962-8},
	year = {2026}}

\newpage
\appendix
\onecolumngrid

\section{The general boundary condition and its solution}
\label{appendix inversion}
In this Appendix, we derive the solution of the general boundary condition presented in Eq.~\eqref{general bc} in the main text.
In the following, we find an explicit expression for $g(t)$ from the equation
\begin{align}
\label{A general bc}
    f(-t)=ug^*(t)\int_0^t dt'~g(t')e^{v G(t,t')}f(-t'),
\end{align}
where $u,v\in\mathbb{C}$ and $f(-t)$ is a square-integrable function and $G(0,t)=\int_0^t dt'~|g(t)|^2$. We start with an ansatz of the form,
\begin{align}
    g^*(t)=\frac{f(-t)}{aF(t)^b},
\end{align}
where $F(t)=\int_0^tdt'~|f(-t')|^2$, $a\in\mathbb{R}$ and $b\in\mathbb{C}$ with $\Re(b)=1/2$.
Substituting the ansatz into Eq.~\eqref{A general bc}, we have
\begin{equation}
    1=\frac{u}{a^2}F(t)^{-\frac{v}{a^2}-b}\int_0^tdt'~F(t')^{\frac{v}{a^2}-b^*}\partial_{t'}F(t').
\end{equation}
Performing the integral with $F(0)=0$ we have,
\begin{equation}
    1=\frac{u}{a^2}\frac{F(t)^{-\frac{v}{a^2}-b}}{\frac{v}{a^2}-b^*+1}F(t)^{\frac{v}{a^2}-b^*+1}.
\end{equation}
To satisfy the above equation, we arrive at two constraints on the parameters in our ansatz,
\begin{align}
    v+a^2(1-b^*)=u,\\
    b+b^*=1.
\end{align}
The second constraint is automatically satisfied by setting $\Re(b)=1/2$. Separating the first constraint into its real and imaginary parts, we can solve for $a$ by taking the real part. Then substituting $a^2$ into the imaginary part of the constraint allows us to solve for $\Im(b)$. Together, the solution for $a$ and $b$ are 
\begin{align}
    &a=\sqrt{2\Re(u-v)},\\
    &b=\frac{1}{2}+i\frac{\Im(u)-\Im(v)}{2\Re(u-v)}=\frac{u-v}{2\Re(u-v)}.
\end{align}
Then we have the explicit expression for $g(t)$, and it is presented in the main text as Eq.~\eqref{general g(t)}
\begin{align}
    g^*(t)=\frac{f(-t)}{\sqrt{2\Re(u-v)F(t)^{\frac{u-v}{\Re(u-v)}}}}.
\end{align}
Notice that $\Re(u-v)\neq0$ in the above expression.

\section{Time and space reversal}
\label{appendix reversal}
Here we present the derivation of the dynamic coupling for the photon addition process using a TLS and a 3LS. This Appendix is relevant for obtaining Eq.~\eqref{two photon add g}, Eq.~\eqref{n photon add g} and Eq.~\eqref{3LS add rabi} from Eq.~\eqref{two photon sub g}, Eq.~\eqref{n photon sub g} and Eq.~\eqref{3LS sub rabi} respectively.

Photon subtraction via TLS is near deterministic, hence the reversed process is the photon addition process. Suppose the Hamiltonian that performs the photon subtraction process is $H_s(t)=g_s(t)\hat{\sigma}_+\hat{a}(-t)+g_s^*(t)\hat{a}^\dagger(-t)\hat{\sigma}_-$. To find the Hamiltonian, $H_a(t)=g_a(t)\hat{\sigma}_+\hat{a}(-t)+g_a^*(t)\hat{a}^\dagger(-t)\hat{\sigma}_-$, that performs the photon addition process, we need to apply the time-reversal $\hat{\Theta}$, translation $\hat{T}(r)$, and parity-reversal $\hat{\mathcal{P}}$ transformations to the subtraction Hamiltonian, $H_s(t)$. In this appendix, the subscripts of relevant quantities label the corresponding processes. The intuitive explanation behind each transformation is that: the time-reversal generates the reverse process, but in the interaction picture, the TLS has traveled to position $x=-t_f$, therefore we must return it to the origin by translation, lastly, the parity un-flips the direction of travel caused by the time-reversal transformation. 

Let us introduce the operators by defining their actions on the photon field annihilation operator,
\begin{align}
    &\hat{\Theta}\hat{a}(x)\hat{\Theta}^{-1}=\hat{a}(x),\quad \hat{\Theta}i\hat{\Theta}^{-1}=-i, \\
    &\hat{T}(r)\hat{a}(x)\hat{T}^{-1}(r)=\hat{a}(x+r),\\
    &\hat{\mathcal{P}}\hat{a}(x)\hat{\mathcal{P}}^{-1}=\hat{a}(-x).
\end{align}
These operators act trivially on the TLS operators. Given that for ideal photon subtraction we have $\ket{\psi_s(t_f)}=\hat{U}_s(t_f,0)\ket{\psi_s(0)}$, applying the three transformations we have,
\begin{equation}
    \hat{U}_a(t_f,0)=\hat{\mathcal{P}}\hat{T}(t_f)\hat{\Theta}\hat{U}^\dagger_s(t_f,0)(\hat{\mathcal{P}}\hat{T}(t_f)\hat{\Theta})^{-1}=\mathcal{T}\exp{-i\int_0^{t_f} dt~\left(g^*_s(t_f-t)\hat{\sigma}_+\hat{a}(-t)+g_s(t_f-t)\hat{a}^\dagger(-t)\hat{\sigma}_-\right)},
\end{equation}
where $\mathcal{T\exp{}}$ denotes the time-ordered exponential. We define the initial and final states of the addition by
\begin{equation}
    \ket{\psi_a(0)}=\hat{\mathcal{P}}\hat{T}(t_f)\hat{\Theta}\ket{\psi_s(t_f)}, \quad \ket{\psi_a(t_f)}=\hat{\mathcal{P}}\hat{T}(t_f)\hat{\Theta}\ket{\psi_s(0)}.
\end{equation}
Then we have $\ket{\psi_a(t_f)}=\hat{U}_a(t_f,0)\ket{\psi_a(0)}$ with 
\begin{equation}
    g_a(t)=g^*_s(t_f-t).
\end{equation}
Taking the infinite-time limit ($t_f\rightarrow\infty$) produces Eq.~\eqref{two photon sub g} and Eq.~\eqref{n photon sub g} of the main text. Recall that $g_s(t)$ depends on the input temporal mode of the initial state, $\ket{\psi_s(0)}$. Under these transformations, the final state of the subtraction process is mapped to the initial state of the addition process and vice versa. The dynamic coupling strength $g_a(t)$ would have to depend on the final state of the addition process, which is unwanted. Ideally the dynamic coupling should only depend on the initial state. Hence, to use the dynamic coupling in the ways described in the main text, we must make another approximation such that the initial and final states of the process share the same temporal mode (see Sec.~\ref{two-excitation} and Sec.~\ref{TLS results} of the main text for justifications of making this approximation).

The full 3LS is described by a different Hamiltonian (see Eq.~\eqref{3LS hamiltonian}), but the key time dependence that facilitate both the dynamic stimulated emission and absorption processes are the dynamic Rabi frequencies and phases. Applying the same argument as above, we obtain the Rabi frequency and phase for the photon addition process from the subtraction process, i.e., Eq.~\eqref{3LS add rabi} from Eq.~\eqref{3LS sub rabi} of the main text. Explicitly, the relation is
\begin{equation}
    \frac{\Omega_a(t)}{2}e^{i\phi_a(t)}=\frac{\Omega_s(t_f-t)}{2}e^{-i\phi_s(t_f-t)}.
\end{equation}

\section{Linearised $n$-excitation subtraction}
\label{appendix n-excitation}

In this Appendix, we derive the ansatz of time-dependent coupling for the $n$-excitation subtraction process. The strategy is to split the unsymmetrised wavefunction into different regions and solve the Schr\"{o}dinger equation sequentially from region to region. To keep track of the rapidly growing number of terms, we introduce the linear approximation at appropriate steps. Finally, we arrive at a general boundary that has the form of Eq.~\eqref{general bc} of the main text and the explicit expression for the time-dependent coupling strength is given by Eq.~\eqref{general g(t)} and derived in Appendix~\ref{appendix inversion}.

Initialising with an $n$-excitation Fock state, and by the conservation of excitation, the ansatz of the $n$-excitation subspace that captures all of the dynamics is
\begin{equation}
    \ket{\psi(t)}=\frac{1}{\sqrt{n!}}\int_{\mathbb{R}^n} d^n\mathbf{x}~\alpha(\mathbf{x},t)\hat{\mathbf{a}}^\dagger(\mathbf{x})\ket{0,g}
    +\frac{1}{\sqrt{(n-1)!}}\int_{\mathbb{R}^{n-1}} d^{n-1}\mathbf{x}~\beta(\mathbf{x},t)\hat{\mathbf{a}}^\dagger(\mathbf{x})\ket{0,e},\label{A n excitation ansatz}
\end{equation}
where $d^j\mathbf{x}=dx_1dx_2\dots dx_j$, $\hat{\mathbf{a}}^\dagger(\mathbf{x})=\hat{a}^\dagger(x_1)\hat{a}^\dagger(x_2)\dots\hat{a}^\dagger(x_j)$ and $j$ is the dimension of $\mathbf{x}$. In the $n$-excitation subspace, $\alpha(\mathbf{x},t)$ is the symmetric wavefunction corresponds to $n$-photon with the TLS in the ground state. Similarly, $\beta(\mathbf{x},t)$ is the symmetric wavefunction corresponds to $(n-1)$-photon with an excited TLS. Moreover, $\alpha(\mathbf{x},t)$ and $\beta(\mathbf{x},t)$ are understood to have $n$ and $n-1$ position arguments, respectively. From the Schr\"{o}dinger equation, the equations of motion relate the wavefunctions by,
\begin{align}
    &i\partial_t\alpha(\mathbf{x},t)=\frac{g^*(t)}{\sqrt{n}}\sum_{j=1}^n\beta(\mathbf{x},t)\delta(x_j+t),\label{A n photon d alpha}\\
    &i\partial_t\beta(\mathbf{x},t)=\sqrt{n}g(t)\alpha(\mathbf{x},t)\rvert_{x_m=-t}.\label{A n photon d beta}
\end{align}
Abusing the notation, the position argument, $\mathbf{x}$, of the $n$-photon wavefunction has $n$ dimensions, while the position argument of the $(n-1)$-photon wavefunction on the right-hand side of Eq.~\eqref{A n photon d alpha} has $n-1$ dimensions, the missing position variable, $x_j$, appears in the $\delta$-function and each term in the sum exhausts all $n$ position variables, such that the right-hand side is also symmetric; e.g., see Eq.~\eqref{two photon d alpha} for the two-excitation case. Likewise in Eq.~\eqref{A n photon d beta}, the missing position variable of the $(n-1)$-photon wavefunction is the one evaluated at $x_m=-t$ on the right-hand side, e.g., Eq.~\eqref{two photon d beta}. Restricting to $x_1\geq x_2\geq\dots\geq x_n$, we work with the unsymmetrised $n$-photon and $(n-1)$-photon wavefunctions; we label them in regions with respect to the position orderings of the TLS and the photons in a manner similar to the two-excitation scenario of Sec.~\ref{two-excitation}. For example, $\alpha_m(\mathbf{x},t)$ is the $n$-photon wavefunction in the region $x_1\geq\dots\geq x_m\geq-t\geq x_{m+1}\geq\dots\geq x_n$, again, the subscript denotes the number of photons to the right of the TLS, in this case, $m$ photons. Similarly, $\beta_m(\mathbf{x},t)$ is the $(n-1)$-photon wavefunction in the region $x_1\geq\dots\geq x_{m-1}\geq -t \geq x_{m+1}\geq \dots\geq x_n$ with the subscript denoting the $m$-th photon is absorbed; additionally, it signals the position variable, $x_m$, is missing from the position argument, $\mathbf{x}$, i.e., $\beta_m(\mathbf{x},t)=\beta_m(x_1,\dots,x_{m-1},x_{m+1},\dots,x_n,t)$. 

Integrating Eq.~\eqref{A n photon d alpha} away from the position of the photons produces the free-propagation condition, and integrating over the position of each of the photons reveal the effects of the interactions in the form of $n$ boundary conditions; particularly, the boundary condition that is a consequence of the interaction between the TLS and the $m$-th photon is
\begin{align}
    \alpha_{m}(\mathbf{x},t)\bigg|_{x_{m}=-t}&-\alpha_{m-1}(\mathbf{x},t)\bigg|_{x_{m}=-t}=-i\frac{g^*(t)}{\sqrt{n}}\beta_{m}(\mathbf{x},t),\label{A n photon bc}
\end{align}
with $m=1,2,\dots,n$. Together with the regularisation condition---when any one of the photons is evaluated at the position of the TLS, the unsymmetrical $n$-photon wavefunction is the average of the wavefunction in neighbouring regions, i.e., $\alpha_{\rm unsym}(\mathbf{x},t)\rvert_{x_{m}=-t}=[\alpha_{m-1}(\mathbf{x},t)\rvert_{x_{m}=-t}+\alpha_m(\mathbf{x},t)\rvert_{x_{m}=-t}]/2$---and the above boundary condition, Eq.~\eqref{A n photon d beta} produces $n$ differential equations, 
\begin{equation}
\label{A n photon diff equ}
    \partial_t\beta_m(\mathbf{x},t)=-|g(t)|^2\beta_m(\mathbf{x},t)/2-i\sqrt{n}g(t)\alpha_{m-1}(\mathbf{x},t)\bigg|_{x_m=-t}
\end{equation}
where the corresponding domains are $t\in[-x_{m-1},-x_{m}]$ with $x_0=0$. 

We now solve the $n$-excitation subtraction process with the an initial $n$-photon Fock state in the waveguide and the TLS in the ground state, 
\begin{equation}
    \ket{\psi(t=0)}=\frac{1}{\sqrt{n!}}\left[\int_\mathbb{R} dxf_{\rm in}(x)\hat{a}^\dagger(x)\right]^n\ket{0,g}.
\end{equation}

By the free propagation property, the $n$-photon unsymmetrised wavefunction before any interactions is $\alpha_0(\mathbf{x},t)=\alpha_0(\mathbf{x},0)=\prod_{j=1}^n f_{\rm in}(x_j)$. The solution given by Eq.~\eqref{A n photon diff equ} for $\beta_1(\mathbf{x},t)$ is
\begin{equation}
    \beta_1(\mathbf{x},t)=-i\sqrt{n}\prod_{j=2}^nf_{\rm in}(x_j)e^{-G(0,t)/2}S_{1/2}(0,t),
\end{equation}
recall that $S_v(a,b)=\int_a^b dt'~g(t')e^{vG(a,t')}f_{\rm in}(-t')$ from Eq.~\eqref{source term}. From the boundary condition in Eq.~\eqref{A n photon bc}, we can relate $n$-photon unsymmetrised wavefunction in the following region, $\alpha_1(\mathbf{x},t)$, to the initial conditions by
\begin{equation}
    \alpha_1(\mathbf{x},t)\bigg|_{x_1=-t}=\prod_{j=2}^nf_{\rm in}(x_j)\left[f_{\rm in}(-t)-g^*(t)e^{-G(0,t)/2}S_{1/2}(0,t)\right].
\end{equation}
Instead of solving for $\beta_2(\mathbf{x},t)$ and $\alpha_2(\mathbf{x},t)$, we introduce the linear approximation step by step, $x_2=-t$, such that the TLS does not interact with subsequent photons. In other words, $\beta_2(\mathbf{x},t)$ and $\alpha_2(\mathbf{x},t)$ only have contributions from free propagation. 

The same argument can be applied to both unsymmetriesd wavefunctions in any region, provided we introduce the linear approximation appropriately, i.e., we can relate the $(n-1)$-photon wavefunctions in arbitrary regions to the first region, i.e., $\beta_m(\mathbf{x},t)\rvert_{x_{1:m-1}=-t}=\beta_1(\mathbf{x},t)\rvert_{x_{2:m}=-t}$ with $x_{1:m}=-t$ denoting $x_j=-t$ for $j=1,2,\dots,m$. This is because Eq.~\eqref{A n photon d beta} does not contain any $\delta$-functions, hence the relation is simply a result of matching the wavefunctions at the boundary of each region. To see the matching boundary condition in general, let us introduce the position argument, $\mathbf{y}$, with $n-1$ dimensions, for the $(n-1)$-photon wavefunction. We label the unsymmetrical $(n-1)$-photon wavefunction by $\beta_m(\mathbf{y},t)$ if it is in the region with position ordering $y_1\geq \dots \geq y_{m-1}\geq-t\geq y_m\geq\dots\geq y_{n-1}$. Then integrating Eq.~\eqref{A n photon d beta} with a vanishing interval centered at $t=-y_{m-1}$ produces the matching boundary condition that relates the wavefunction in neighbouring regions, $\beta_m(\mathbf{y},-y_{m-1})=\beta_{m-1}(\mathbf{y},-y_{m-1})$, which is the same as 
\begin{equation}
    \beta_m(\mathbf{y},t)\bigg|_{y_{m-1}=-t}=\beta_{m-1}(\mathbf{y},t)\bigg|_{y_{m-1}=-t},
\end{equation}
up to a change of variable. Reiterating this expression until the label of the right-hand side reaches $1$, we relate the wavefunction with label $m$ to label $1$,
\begin{equation}
    \beta_m(\mathbf{y},t)\bigg|_{y_{1:m-1}=-t}=\beta_1(\mathbf{y},t)\bigg|_{y_{1:m-1}=-t}.
\end{equation}
Relating the position argument, $\mathbf{y}$, to the position argument of the $n$-photon wavefunction, $\mathbf{x}$, we must pay careful attention to the meaning of each label. Specifically, $\beta_m(\mathbf{x},t)$ is the wavefunction corresponding to the $m$-th photon is absorbed, and hence it is understood that $x_m$ should be missing from its position argument. Therefore, on the left-hand side, we evaluate $y_{1:m-1}=x_{1:m-1}=-t$ and $y_{m:n-1}=x_{m+1:n}$. Likewise, on the right-hand side, we evaluate $y_{1:m-1}=x_{2:m}=-t$ and $y_{m:n-1}=x_{m+1:n}$. All together we have the matching boundary condition that relates the $(n-1)$-photon wavefunction in region labeled by $m$ to 1, using the position arguments of the $n$-photon wavefunction,
\begin{equation}
    \beta_m(\mathbf{x},t)\bigg|_{x_{1:m-1}=-t}=\beta_1(\mathbf{x},t)\bigg|_{x_{2:m}=-t}.
\end{equation}
Combining this result with the boundary condition in Eq.~\eqref{A n photon bc} we have the linearised boundary condition relating the $n$-photon wavefunctions in neighbouring regions to the initial conditions, 
\begin{equation}
    \alpha_{m}(\mathbf{x},t)\bigg|_{x_{1:m}=-t}-\alpha_{m-1}(\mathbf{x},t)\bigg|_{x_{1:m}=-t}=-i\frac{g^*(t)}{\sqrt{n}}\beta_m(\mathbf{x},t)\bigg|_{x_{1:m-1}=-t}=-i\frac{g^*(t)}{\sqrt{n}}\beta_1(\mathbf{x},t)\bigg|_{x_{2:m}=-t}.\label{A n photon linear bc}
\end{equation}
By induction, the closed form of the $n$-photon wavefunction with label $m$ in terms of the initial conditions with the appropriate linearisations is
\begin{equation}
    \alpha_m(\mathbf{x},t)\bigg|_{x_{1:m}=-t}=\prod_{j=m+1}^n f_{\rm in}(x_j)f_{\rm in}^{m-1}(-t)\left[f_{\rm in}(-t)-mg^*(t)e^{-G(0,t)/2}S_{1/2}(0,t)\right].\label{A n photon final bc}
\end{equation}
We prove the above closed form by induction. Since $\alpha_0(\mathbf{x},t)$ and $\alpha_1(\mathbf{x},t)\rvert_{x_1=-t}$ are already in the closed form. From the linearised boundary condition in Eq.~\eqref{A n photon linear bc} and applying the induction step, we have,
\begin{align}
    \alpha_{m+1}(\mathbf{x},t)\bigg|_{x_{1:m+1=-t}}&=\alpha_m(\mathbf{x},t)\bigg|_{x_{1:m+1}=-t}-i\frac{g^*(t)}{\sqrt{n}}\beta_1(\mathbf{x},t)\bigg|_{x_{2:m+1}=-t}\\
    &=\prod_{j=m+2}^n f_{\rm in}(x_j)f_{\rm in}^{m}(-t)\left[f_{\rm in}(-t)-mg^*(t)e^{-G(0,t)/2}S_{1/2}(0,t)\right]\nonumber\\
    &\quad-\prod_{j=m+2}^nf_{\rm in}(x_j)f_{\rm in}^m(-t)g^*(t)e^{-G(0,t)/2}S_{1/2}(0,t)\\
    &=\prod_{j=m+2}^n f_{\rm in}(x_j)f_{\rm in}^{m}(-t)\left[f_{\rm in}(-t)-(m+1)g^*(t)e^{-G(0,t)/2}S_{1/2}(0,t)\right],
\end{align}
as required.

Lastly, the necessary output boundary condition for the ideal $n$-excitation subtraction process is $\alpha_n(\mathbf{x},t)=0$; together with the linear approximation, Eq.~\eqref{A n photon bc} becomes the linearised output boundary condition, 
\begin{equation}
    \sqrt{n}\alpha_{n-1}(\mathbf{x},t)\bigg|_{x_{1:n}=-t}=ig^*(t)\beta_{n}(\mathbf{x},t)\bigg|_{x_{1:n-1=-t}}=ig^*(t)\beta_{1}(\mathbf{x},t)\bigg|_{x_{2:n=-t}}.
\end{equation}
We emphasize that the solution obtained from the above boundary condition will not guarantee the ideal $n$-excitation subtraction process, but it provides a high-performance ansatz. In terms of the initial conditions, the linearised output boundary condition, and according to Appendix~\ref{appendix inversion}, the time-dependent coupling strength are
\begin{equation}
    f_{\rm in}(-t)=ng^*(t)e^{-G(0,t)/2}S_{1/2}(0,t),\quad g^*(t)=\frac{f_{\rm in}(-t)}{\sqrt{(2n-1)\int_0^tdt'~|f_{\rm in}(-t')|^2}}.
\end{equation}

\section{Photon addition ansatz for superposition states}
\label{appendix addition ansatz}

Here we provide a motivation for the ansatz given by Eq.~\eqref{superposition add g} which facilitates the photon addition process when the input is a superposition state. We also present the differences in displacements and squeezing strengths between the prepared state and an ideal photon-added Gaussian state. 

As mentioned in the main text, since the subtraction process with a superposition is nondeterministic, then using the reversed dynamic coupling for the addition process would lead to a multimode output state. Therefore, we must choose an alternate coupling strength. In this Appendix, we solve the Schr\"{o}dinger equation for the $n$-excitation photon addition process directly. In order to obtain an explicit expression for the dynamic coupling, we still need to make the linear approximation described in Appendix~\ref{appendix n-excitation}. The procedure is largely the same as Appendix~\ref{appendix n-excitation}, except for the initial state has changed to an $(n-1)$-photon Fock state with an excited TLS,
\begin{equation}
    \ket{\psi(t=0)}=\frac{1}{\sqrt{(n-1)!}}\left[\int_\mathbb{R}dx~f_{\rm in}(x)\hat{a}^\dagger(x)\right]^{n-1}\ket{0,e}.
\end{equation}
The corresponding unsymmetrised wavefunctions before any interactions are $\alpha_0(\mathbf{x},t)=0$ and $\beta_1(\mathbf{x},0)=\prod_{j=2}^nf_{\rm in}(x_j)$. Solving Eq.~\eqref{A n photon diff equ} for $\beta_1(\mathbf{x},t)$, we obtain
\begin{equation}
    \beta_1(\mathbf{x},t)=e^{-G(0,t)/2}\prod_{j=2}^nf_{\rm in}(x_j).
\end{equation}
From Eq.~\eqref{A n photon bc}, the unsymmetrised $n$-photon wavefunction in the region with label 1 reads
\begin{equation}
    \alpha_1(\mathbf{x},t)\rvert_{x_1=-t}=-i\frac{g^*(t)}{\sqrt{n}}e^{-G(0,t)/2}\prod_{j=2}^nf_{\rm in}(x_j).
\end{equation}
This corresponds to the TLS emitting a photon before any photons from the $(n-1)$-photon Fock state have arrived. Akin to Appendix~\ref{appendix n-excitation}, we introduce the linear approximation in steps. The closed form of the unsymmetrised $n$-photon wavefunction in the region with label $m$ is 
\begin{equation}
    \alpha_m(\mathbf{x},t)\bigg|_{x_{1:m}=-t}=-im\frac{g^*(t)}{\sqrt{n}}e^{-G(0,t)/2}f^{m-1}_{\rm in}(-t)\prod_{j=m+1}^nf_{\rm in}(x_j).
\end{equation}
We prove the above closed form by induction. Since $\alpha_0(\mathbf{x},t)$ and $\alpha_1(\mathbf{x},t)$ are already in the closed form, we apply the induction step with Eq.~\eqref{A n photon linear bc};
\begin{align}
    \alpha_{m+1}(\mathbf{x},t)\bigg|_{x_{1:m+1=-t}}&=\alpha_m(\mathbf{x},t)\bigg|_{x_{1:m+1}=-t}-i\frac{g^*(t)}{\sqrt{n}}\beta_1(\mathbf{x},t)\bigg|_{x_{2:m+1}=-t}\\
    &=-im\frac{g^*(t)}{\sqrt{n}}e^{-G(0,t)/2}f^{m}_{\rm in}(-t)\prod_{j=m+2}^nf_{\rm in}(x_j)-i\frac{g^*(t)}{\sqrt{n}}e^{-G(0,t)/2}f^{m}_{\rm in}(-t)\prod_{j=m+2}^nf_{\rm in}(x_j)\nonumber\\
    &=-i(m+1)\frac{g^*(t)}{\sqrt{n}}e^{-G(0,t)/2}f^{m}_{\rm in}(-t)\prod_{j=m+2}^nf_{\rm in}(x_j),
\end{align}
as required.

We require the output $n$-photon wavefunction to be a Fock state, i.e., the unsymmetrised wavefunction in the region $n$ must be $\alpha_n(\mathbf{x},t)=\prod_{j=1}^nf_{\rm out}(x_j)$, where $f_{\rm out}(x)$ is a normalised mode function. Using the closed form of the unsymmetrised $n$-photon wavefunction we can relate the desired output wavefunction to the initial condition,
\begin{equation}
    f^n_{\rm out}(-t)=-i\sqrt{n}g^*(-t)e^{-G(-0,t)/2}f_{\rm in}^{n-1}(-t).
\end{equation}
The explicit expression for the dynamic coupling can be obtained by taking the squared modulus and integrating $|f^n_{\rm out}(-t)/f_{\rm in}^{n-1}(-t)|^2=-n\partial_te^{-G(0,t)}$ from 0 to $t$. Finally,
\begin{equation}
\label{A add g}
    g^*(t)=\frac{if^n_{\rm out}(-t)}{f_{\rm in}^{n-1}(-t)\sqrt{n-\int_0^tdt'~|f^n_{\rm out}(-t)/f_{\rm in}^{n-1}(-t)|^2}}.
\end{equation}
Notice that in the single-excitation case ($n=1$), the expression is identical to Eq.~\eqref{one photon add g}. However, this ansatz would not produce the desired addition processes when the excitation number is greater than one. From here we make modifications to this coupling strength to improve its flexibility. First, we approximate the output mode with the input mode $f_{\rm out}(-t)=f_{\rm in}(-t)$ such that it only depends on the initial conditions. Notice that when the input state is a Fock state, we are allowed to omit the global phase of the mode function, whereas for a superposition input state, we must set the global phase to 1 such that we do not introduce any unwanted relative phases within the superposition of Fock states. Furthermore, we include additional variational parameters, $\{q_1,q_2\}\in\mathbb{R}$, such that our ansatz reads
\begin{equation}
\label{A superposition add g}
    g^*(t)=\frac{q_1f_{\rm in}(-t)}{\sqrt{1-q_2\int_0^tdt'~|f_{\rm in}(-t')|^2}},
\end{equation}
where $q_2\leq1$. The above ansatz offers a wide range of flexibility; specifically, $q_1$ controls the overall magnitude, while $q_2$ can dramatically change the time-dependence of the ansatz. By choosing $\{q_1,q_2\}=\{1/\sqrt{n},1/n\}$, we recover Eq.~\eqref{A add g}. Moreover, if $q_2=0$, the coupling is proportional to the mode function, and if we set $\{q_1,q_2\}=\{a_n/\sqrt{2n-1},1\}$, we would recover the coupling for the $n$-excitation addition process given by Eq.~\eqref{n photon add g}.

As mentioned in Sec.~\ref{photon-added gaussian state}, we use Eq.~\eqref{A superposition add g} as the dynamic coupling for a TLS to add a photon to a displaced squeezed vacuum state, $\ket{\alpha,r}$. We optimised the variational parameters in Eq.~\eqref{A superposition add g} such that the process has a high probability of success without compromising on the fidelity against a target photon-added displaced squeezed vacuum, $\ket{\alpha',r'}$. In this process, a TLS can introduce additional displacement and squeezing; therefore, we allow different displacements and squeezings between the target state and the output state, the differences are presented in Fig.~\ref{fig:appendix_stellar_1_diff}. We found that the anti-displacement becomes significant when the input Gaussian state is highly squeezed. Conversely, the anti-squeezing is only noticeable when the initial Gaussian state is not displaced. 
\begin{figure}[t!]
    \centering
    \includegraphics[width=0.55\linewidth]{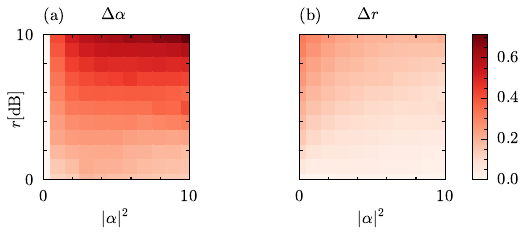}
    \caption{Differences in displacement (a) and squeezing (b) between the prepared state and the ideal photon-added Gaussian state. The differences are defined as $\Delta\alpha=\alpha-\alpha'$ in (a) and  $\Delta r=r-r'$ in (b). Here $\alpha$ is the displacement parameter on the input Gaussian state and $\alpha'$ is that of the target photon-added Gaussian state, and similarly for the squeezing strengths $r$ and $r'$.}
    \label{fig:appendix_stellar_1_diff}
\end{figure}

\section{Three-level system}
\label{appendix three-level}

In this appendix, we provide details of how we derive the time-dependent coupling for the three-level system (3LS) presented in Sec.~\ref{three-level system} of the main text. We derive the result \cite{Cirac1997PRL}, that an effective TLS with a time-dependent coupling is obtained using a $\Lambda$-type 3LS with a time-dependent drive. By adiabatically eliminating the excited state, we arrive at an effective Hamiltonian with time-dependent couplings between the ground states and the waveguide. We present this derivation to introduce the system using our notation, which then allows us to move on to our new result of solving for the time-dependence required for dynamic stimulated emission.

We start with a $\Lambda$-type 3LS with an excited state $\ket{r}$ and two degenerate ground states $\ket{e}$ and $\ket{g}$. The 3LS has excitation frequency, $\omega_0$ and the $\ket{e}$ to $\ket{r}$ transition is driven by a laser with frequency $\omega_L$, time-dependent Rabi frequency $\Omega(t)$, and phase $\phi(t)$. The transition between $\ket{g}$ and $\ket{r}$ is coupled to the waveguide with decay rate $\Gamma$. The waveguide is modeled by a right-propagating photon field whose dispersion is linearised about the frequency $\omega_L$. The Hamiltonian for this system is 
\begin{align}
    \hat{H}(t)=\int dx~\hat{a}^\dagger(x)(\omega_L-i\partial_x)\hat{a}(x)+\omega_0\hat{\sigma}_{rr}+\frac{\Omega(t)}{2}\left[e^{-i(\omega_Lt+\phi(t))}\hat{\sigma}_{re}+{\rm h.c.}\right]+\sqrt{\Gamma}\left[\hat{\sigma}_{rg}\hat{a}(0)+{\rm h.c.}\right],
\end{align}
where $\hat{\sigma}_{ij}=\ket{i}\bra{j}$ and $\hat{a}(x)$ is the photon field annihilation operator at position $x$. Moving to an interaction picture with respect to the excited state and the photon field at the laser frequency $\omega_L$, the Hamiltonian is
\begin{align}
    \hat{H}(t)=-\Delta\hat{\sigma}_{rr}+\frac{\Omega(t)}{2}\left[e^{-i\phi(t)}\hat{\sigma}_{re}+{\rm h.c.}\right]+\sqrt{\Gamma}\left[\hat{\sigma}_{rg}\hat{a}(-t)+{\rm h.c.}\right],
\end{align}
where $\Delta=\omega_L-\omega_0$.

\subsection{Adiabatic elimination}
Here we adiabatically eliminate the excited state, $\ket{r}$ and show that the 3LS can be effectively described by a TLS. From the Heisenberg picture, the equation of motion for transition operators are
\begin{align}
    i\partial_t\hat{\sigma}_{re}=\Delta\hat{\sigma}_{re}-\sqrt{\Gamma}\hat{a}^\dagger(-t)\hat{\sigma}_{ge}+\frac{\Omega(t)}{2}e^{i\phi(t)}(\hat{\sigma}_{rr}-\hat{\sigma}_{ee}),\\
     i\partial_t\hat{\sigma}_{rg}=\Delta\hat{\sigma}_{rg}+\sqrt{\Gamma}\hat{a}^\dagger(-t)(\hat{\sigma}_{rr}-\hat{{\sigma}}_{gg})-\frac{\Omega(t)}{2}e^{i\phi(t)}\hat{\sigma}_{eg}.
\end{align}
Assuming the adiabatic condition, $|\Delta|\gg\Omega(t),\Gamma$, the dynamics of the transition operators involving the excited state are much faster than the rest of the system; therefore, from the perspective of the slowly evolving time scale of the system, the fast-evolving operators reach their steady state; hence, we can adiabatically eliminate the excited state $\ket{r}$ by setting $\partial_t\hat{\sigma}_{re}\sim0$, $\partial_t\hat{\sigma}_{rg}\sim0$. We arrive at,
\begin{align}
    \hat{\sigma}_{re}\sim[\sqrt{\Gamma}\hat{a}^\dagger(-t)\hat{\sigma}_{ge}+\Omega(t)e^{i\phi(t)}\hat{\sigma}_{ee}/2]/\Delta,\\
    \hat{\sigma}_{rg}\sim[\sqrt{\Gamma}\hat{a}^\dagger(-t)\hat{\sigma}_{gg}+\Omega(t)e^{i\phi(t)}\hat{\sigma}_{eg}/2]/\Delta,
\end{align}
where we have used the approximation $\hat{\sigma}_{rr}=\hat{\sigma}_{re}\hat{\sigma}_{er}\sim1/\Delta^2\sim0$.
The equation of motion for a system operator $\hat{s}$ between the ground states $\ket{g}$ and $\ket{e}$ is,
\begin{align}
    i\partial_t\hat{s}=&\frac{\Omega(t)}{2}(e^{-i\phi(t)}[\hat{s},\hat{\sigma}_{re}]+e^{i\phi(t)}[\hat{s},\hat{\sigma}_{er}])+\sqrt{\Gamma}([\hat{s},\hat{\sigma}_{rg}]\hat{a}(-t)+\hat{a}^\dagger(-t)[\hat{s},\hat{\sigma}_{gr}])\\
    &=\frac{\Omega(t)}{2}(e^{i\phi(t)}\hat{s}\hat{\sigma}_{er}-e^{-i\phi(t)}\hat{\sigma}_{re}\hat{s})+\sqrt{\Gamma}(\hat{a}^\dagger(-t)\hat{s}\hat{\sigma}_{gr}-\hat{\sigma}_{rg}\hat{s}\hat{a}(-t))\\
    &=\frac{\Gamma}{\Delta}\hat{a}^\dagger(-t)\hat{a}(-t)[\hat{s},\hat{\sigma}_{gg}]+\frac{\Omega^2(t)}{4\Delta}[\hat{s},\hat{\sigma}_{ee}]+\frac{\Omega(t)\sqrt{\Gamma}}{2\Delta}e^{i\phi(t)}[\hat{s},\hat{\sigma}_{eg}]\hat{a}(-t)+\frac{\Omega(t)\sqrt{\Gamma}}{2\Delta}e^{-i\phi(t)}\hat{a}^\dagger(-t)[\hat{s},\hat{\sigma}_{ge}].
\end{align}
The equation of motion for the photon field operator is,
\begin{align}
    i\partial_t\hat{a}(x,t)=\sqrt{\Gamma}\hat{\sigma}_{gr}\delta(x+t)\sim\sqrt{\Gamma}\delta(x+t)[\sqrt{\Gamma}\hat{a}(-t)\hat{\sigma}_{gg}+\Omega(t)e^{-xi\phi(t)}\hat{\sigma}_{ge}/2]/\Delta.
\end{align}
We then have an effective Hamiltonian describing the dynamics between the ground states and the photon field,
\begin{align}
    \hat{H}(t)=\frac{\Gamma}{\Delta}\hat{\sigma}_{gg}\hat{a}^\dagger(-t)\hat{a}(-t)+\frac{\Omega^2(t)}{4\Delta}\hat{\sigma}_{ee}+\left[\frac{\Omega(t)\sqrt{\Gamma}}{2\Delta}e^{i\phi(t)}\hat{\sigma}_{eg}\hat{a}(-t)+{\rm h.c.}\right].
\end{align}
The first term is the detuning of the photon field followed by two AC Stark shifts of the ground states, the last term is the Jaynes-Cummings-interaction term with an effective time-dependent coupling. Let $\phi(t)=\phi_g(t)+\phi_e(t)$ and move into a rotating frame of the $\ket{e}$ state and choose $\partial_t\phi_e(t)=-\Omega^2(t)/4\Delta$, we recover the effective Hamiltonian given by Eq.~\eqref{effective hamiltonian},
\begin{align}
\label{A effective hamiltonian}
        \hat{H}(t)=\frac{\Gamma}{\Delta}\hat{\sigma}_{gg}\hat{a}^\dagger(-t)\hat{a}(-t)+\frac{\Omega(t)\sqrt{\Gamma}}{2\Delta}\left[e^{i\phi_g(t)}\hat{\sigma}_{eg}\hat{a}(-t)+{\rm h.c.}\right].
\end{align}
In the following subsections, we carry out the calculations using this effective Hamiltonian.

\subsection{Single excitation}
We now provide a derivation of the solution for the time-dependent coupling in the single-excitation limit. We start with the ansatz,
\begin{align}
    \ket{\psi(t)}=\int dx~\alpha(x,t)\hat{a}^\dagger(x)\ket{0,g}+\beta(t)\ket{0,e}.
\end{align}
From this, we obtain two equations of motion,
\begin{align}
    i\partial_t\beta(t)&=g(t)\alpha(-t,t),\\
    i\partial_t\alpha(x,t)&=\left[\frac{\Gamma}{\Delta}\alpha(-t,t)+g^*(t)\beta(t)\right]\delta(x+t),
\end{align}
where $g(t)=\frac{\Omega(t)\sqrt{\Gamma}}{2\Delta}e^{i\phi_g(t)}$. Integrating $t$ over the small interval $[x-\epsilon,x+\epsilon]$ on either side of the $\delta$-function to obtain the boundary condition (where we have used $\alpha(-t,t)=[\alpha_0(-t,t)+\alpha_1(-t,t)]/2$ and evaluated $x=-t$)
\begin{align}
    \bigg(1-\frac{i\Gamma}{2\Delta}\bigg)\alpha_0(-t,t)-\bigg(1+\frac{i\Gamma}{2\Delta}\bigg)\alpha_1(-t,t)=ig^*(t)\beta(t).
\end{align}
Substitute  $\alpha_1(-t,t)$ from the boundary condition into the $\beta(t)$ equation, we have a Bernoulli equation that governs the dynamics of the $\ket{e}$ state,
\begin{align}
    \partial_t\beta(t)=-\frac{\xi|g(t)|^2}{2}\beta(t)-i\xi g(t)\alpha_0(-t,t),
\end{align}
where we have defined $\xi=[1+i\Gamma/(2\Delta)]^{-1}$ as a complex factor that is caused by the AC Stark shift in the effective Hamiltonian of Eq.~\eqref{A effective hamiltonian}. Notice that from the definition of $\xi$ we have $\Re(\xi)=|\xi|^2$. The Bernoulli equation of $\beta(t)$ has solution
\begin{align}
    \beta(t)=e^{-\xi G(0,t)/2}\left[\beta(0)-i\xi\int_0^t dt' ~g(t')e^{\xi G(0,t')/2}\alpha_0(-t',t')\right],
\end{align}
where $G(0,t)=\int_0^tdt'~|g(t')|^2$.
Solving for the release of a single photon in the mode shape $f_{\rm out}(-t)=\alpha_1(-t,t)$ and $\beta(0)=1$, $\alpha_0(-t,t)=0$, from the boundary condition, we need to invert the following equation for $g(t)$
\begin{equation}
    f_{\rm out}(-t)=-i\xi g^*(t)e^{-\xi G(0,t)/2}.
\end{equation}
Notice that
\begin{align}
    &|f_{\rm out}(-t)|^2=|\xi|^2|g(t)|^2e^{-\Re(\xi)G(0,t)}=-|\xi|^2\partial_te^{-\Re(\xi)G(0,t)}/\Re(\xi)=-\partial_te^{-|\xi|^2G(0,t)},\\
    &e^{-\Re{\xi}G(0,t)}=1-\frac{\Re(\xi)}{|\xi|^2}\int_0^tdt'~|f_{\rm out}(-t')|^2=\int_t^\infty dt'|f_{\rm out}(-t')|^2.
\end{align}
We therefore obtain the effective coupling strength for emission,
\begin{equation}
\label{A 3LS one photon add g}
    g^*(t)=\frac{if_{\rm out}(-t)}{\xi \sqrt{\left[\int_t^\infty dt'|f_{\rm out}(-t')|^2\right]^{1/\xi^*}}}.
\end{equation}
Similarly, for absorption of a single photon in the mode shape $f_{\rm in}(-t)=\alpha_0(-t,t)$ and $\beta(0)=0$, $\alpha_1=(-t,t)=0$, the boundary condition is of the form of Eq.~\eqref{general bc},
\begin{equation}
    f_{\rm in}(-t)=|\xi|^2g^*(t)e^{-\xi G(0,t)/2}\int_0^t dt'~g(t')e^{\xi G(0,t')/2}f_{\rm in}(-t').
\end{equation}
Whose solution is given by Eq.~\eqref{general g(t)},
\begin{equation}
\label{A 3LS one photon sub g}
    g^*(t)=\frac{f_{\rm in}(-t)}{|\xi|\sqrt{\left[\int_0^t dt'~|f_{\rm in}(-t')|^2\right]^{1/\xi}}}.
\end{equation}
These effective couplings can be related to the time-dependent Rabi frequency and phase using the relation.
\begin{equation}
    \frac{\Omega(t)}{2}e^{i\phi(t)}=\frac{\Delta}{\sqrt{\Gamma}} g(t)e^{-i\frac{\Delta}{\Gamma} G(0,t)}
\end{equation}

\subsection{Two excitations}
We begin with the two-excitation ansatz,
\begin{equation}
    \ket{\psi(t)}_I=\int dx_1dx_2 \, \alpha(x_1,x_2,t)\frac{\hat{a}^\dagger(x_1)\hat{a}^\dagger(x_2)}{\sqrt{2}}\ket{0,g}+\int dx \, \beta(x,t)\hat{a}^\dagger(x)\ket{0,e}.
\end{equation}
The equations of motion are,
\begin{align}
    i\partial_t\beta(x,t)&=\sqrt{2}g(t)\alpha(x,-t,t)\label{A 3LS two photon d beta}\\
    i\partial_t\alpha(x_1,x_2,t)&=\left[\frac{\Gamma}{\Delta}\alpha(-t,x_2,t)+\frac{g^*(t)}{\sqrt{2}}\beta(x_2,t)\right]\delta(x_1+t)+\left[\frac{\Gamma}{\Delta}\alpha(-t,x_1,t)+\frac{g^*(t)}{\sqrt{2}}\beta(x_1,t)\right]\delta(x_2+t).\label{A 3LS two photon d alpha}
\end{align}
where $g(t)=\frac{\Omega(t)\sqrt{\Gamma}}{2\Delta}e^{i\phi_g(t)}$. Like in Sec,~\ref{two-excitation}, we now restrict the wavefunctions to $x_1\geq x_2$ and split them into the different regions. The boundary conditions are derived by integrating over the $\delta$-functions in Eq.~\eqref{A 3LS two photon d alpha}, but due to the AC Stark shift of the ground state, they contain the $\xi$ complex factor,
\begin{align}
    \alpha_0(-t,x_2,t)/\xi^*-\alpha_1(-t,x_2,t)/\xi=\frac{ig^*(t)}{\sqrt{2}}\beta_1(x_2,t),\\
    \alpha_1(x_1,-t,t)/\xi^*-\alpha_2(x_1,-t,t)/\xi=\frac{ig^*(t)}{\sqrt{2}}\beta_2(x_1,t).
\end{align}
Substituting $\alpha_1(-t,x_2,t)$ and $\alpha_2(x_1,-t,t)$ from the boundary conditions into the $\beta_1(x_2,t)$ and $\beta_2(x_1,t)$ equations we have the differential equations for the single-photon wavefunctions,
\begin{align}
    \partial_t\beta_1(x_2,t)=-\frac{\xi|g(t)|^2}{2}\beta_1(x_2,t)-i\xi\sqrt{2} g(t)\alpha_0(-t,x_2,t),\\
    \partial_t\beta_2(x_1,t)=-\frac{\xi|g(t)|^2}{2}\beta_2(x_1,t)-i\xi\sqrt{2} g(t)\alpha_1(x_1,-t,t).
\end{align} 
These Bernoulli equations have solution
\begin{align}
    \beta_1(x_2,t)=e^{-\xi G(0,t)/2}\left[\beta_1(x_2,0)-i\xi\sqrt{2}\int_0^t dt' ~g(t')e^{\xi G(0,t')/2}\alpha_0(-t',x_2,t')\right],\\
    \beta_2(x_1,t)=e^{-\xi G(-x_1,t)/2}\left[\beta_1(x_1,-x_1)-i\xi\sqrt{2}\int_{-x_1}^t dt' ~g(t')e^{\xi G(-x_1,t')/2}\alpha_1(x_1,-t',t')\right],
\end{align}
where $G(0,t)=\int_0^tdt'~|g(t')|^2$. Here, we can use $\beta_2(x_1,-x_1)=\beta_1(x_1,-x_1)$ and $\alpha_1(x_1,-t,t)=\alpha_1(x_1,-t,-x_1)$, to relate $\beta_2(x_1,t)$ to the initial conditions.

Solving for subtraction, we have the initial state
\begin{align}
    \ket{\psi(t=0)}=\frac{1}{\sqrt{2}}\left[\int dx~f_{\rm in}(x)\hat{a}^\dagger(x)\right]^2\ket{0,g}.
\end{align}
Ideal subtraction process would require  $\alpha_2(x_1,x_2,t)=0$, and thus the boundary condition becomes $\alpha_1(x_1,-t,t)=i\xi^*g^*(t)\beta_2(x_1,t)/\sqrt{2}$, in terms of the initial conditions, we have
\begin{align}
    &\frac{\xi}{\xi^*}f_{\rm in}(-t)\left[f_{\rm in}(x_1)-|\xi|^2g^*(-x_1)e^{-\xi G(0,-x_1)/2}S_{\xi/2}(0,-x_1)\right]\\
    =&|\xi|^2g^*(t)e^{-\xi G(-x_1,t)/2}\biggl\{e^{-\xi G(0,-x_1)/2}f_{\rm in}(x_1)S_{\xi/2}(0,-x_1)\\
    &+\frac{\xi}{\xi^*}S_{\xi/2}(-x_1,t)\left[f_{\rm in}(x_1)-|\xi|^2g^*(-x_1)e^{-\xi G(0,-x_1)/2}S_{\xi/2}(0,-x_1)\right]\biggr\}.
\end{align}
Notice that by setting $\xi=1$, we recover Eq.~\eqref{output relation}. Nonetheless, this boundary condition is difficult to solve; hence, we make the linear approximation by setting $-x_1=t$, the effect is that we are left with a boundary condition that is in the form of Eq.~\eqref{general bc},
\begin{align}
    f_{\rm in}(-t)=\frac{2|\xi|^4}{\xi}g^*(t)e^{-\xi G(0,t)/2}S_{\xi/2}(0,t).
\end{align}
Then the explicit expression for the effective coupling strength is given by Eq.~\eqref{general g(t)},
\begin{align}
    g^*(t)=\frac{f_{\rm in}(-t)}{\sqrt{2\Re(\zeta)\left[\int_0^tdt'|f_{\rm in}(-t')|^2\right]^{\zeta/\Re(\zeta)}}},
\end{align}
with $\zeta=\frac{2|\xi|^4}{\xi}-\frac{\xi}{2}$.

\subsection{Linearised $n$-excitation subtraction}

We now generalise the two-excitation ansatz to the $n$-excitation problem. The derivation in this subsection is similar to Appendix~\ref{appendix n-excitation}, we shall skip most of the detail and only present important steps. The ansatz for the $n$-excitation subspace is identical to that of Eq.~\eqref{A n excitation ansatz}. From the Schr\"{o}dinger equation, the equations of motion are
\begin{align}
    &i\partial_t\alpha(\mathbf{x},t)=\sum_{j=1}^n\delta(x_j+t)\left[\frac{g^*(t)}{\sqrt{n}}\beta(\mathbf{x},t)+\frac{\Gamma}{\Delta}\alpha(\mathbf{x},t)\bigg|_{x_j=-t}\,\right],\label{A 3 level n photon d alpha}\\
    &i\partial_t\beta(\mathbf{x},t)=\sqrt{n}g(t)\alpha(\mathbf{x},t)\big|_{x_m=-t}.\label{A 3 level n photon d beta}
\end{align}
Similar to Eq.~\eqref{A n photon bc}, the boundary conditions are,
\begin{equation}
    \frac{\alpha_{m}(\mathbf{x},t)}{\xi}\bigg|_{x_m=-t}-\frac{\alpha_{m-1}(\mathbf{x},t)}{\xi^*}\bigg|_{x_m=-t}=-i\frac{g^*(t)}{\sqrt{n}}\beta_m(\mathbf{x},t),\label{A 3 level n photon bc}
\end{equation}
where $m=0,1,\dots,n-1$. Together with the regularisation condition and the above boundary conditions, Eq.~\eqref{A 3 level n photon d beta} produces $n$ differential equations,
\begin{equation} 
\partial_t\beta_m(\mathbf{x},t)=-\frac{\xi|g(t)|^2}{2}\beta_m(\mathbf{x},t)-i\xi\sqrt{n}g(t)\alpha_{m-1}(\mathbf{x},t)\bigg|_{x_{m}=-t},\label{A 3 level n photon beta de}
\end{equation}
each with domain $-t\in[x_{m-1},x_{m}]$ with $x_0=0$.

We solve the $n$-excitation subtraction process with $\alpha_0(\mathbf{x},0)=\alpha_0(\mathbf{x},t)=\prod_{j=1}^n f_{\rm in}(x_j)$. Solving Eq.~\eqref{A 3 level n photon beta de} for $\beta_1(\mathbf{x},t)$ gives
\begin{equation}
    \beta_1(\mathbf{x},t)=-i\xi\sqrt{n}\prod_{j=2}^nf_{\rm in}(x_j)e^{-\xi G(0,t)/2}S_{\xi/2}(0,t),
\end{equation}
From the boundary condition in Eq.~\eqref{A 3 level n photon bc}, we can relate $\alpha_1(\mathbf{x},t)$ to the initial conditions by
\begin{equation}
    \alpha_1(\mathbf{x},t)\bigg|_{x_1=-t}=\frac{\xi}{\xi^
    *}\prod_{j=2}^nf_{\rm in}(x_j)\left[f_{\rm in}(-t)-|\xi|^2g^*(t)e^{-\xi G(0,t)/2}S_{\xi/2} (0,t)\right].
\end{equation}
Since Eq.~\eqref{A 3 level n photon d beta} does not contain any $\delta$-functions, we have the matching boundary condition for the $(n-1)$-photon wavefunctions, $\beta_m(\mathbf{x},t)\rvert_{x_{m-1}=-t}=\beta_{m-1}(\mathbf{x},t)\rvert_{x_m=-t}$. Combined with the boundary condition in Eq.~\eqref{A 3 level n photon bc}, this allows us to relate the $n$-photon in arbitrary regions to the initial conditions, provided we evaluate the position coordinates appropriately,
\begin{equation}
    \frac{1}{\xi}\alpha_{m}(\mathbf{x},t)\bigg|_{x_{1:m}=-t}-\frac{1}{\xi^*}\alpha_{m-1}(\mathbf{x},t)\bigg|_{x_{1:m}=-t}=-i\frac{g^*(t)}{\sqrt{n}}\beta_m(\mathbf{x},t)\bigg|_{x_{1:m-1}=-t}=-i\frac{g^*(t)}{\sqrt{n}}\beta_1(\mathbf{x},t)\bigg|_{x_{2:m}=-t}.\label{A 3 level n photon linear bc}
\end{equation}
By induction, one can relate the $n$-photon wavefunction with label $m$ to the initial conditions with the appropriate linearisations,
\begin{equation}
    \alpha_m(\mathbf{x},t)\rvert_{x_{1:m}=-t}=\prod_{j=m+1}^n f_{\rm in}(x_j)f_{\rm in}^{m-1}(-t)\left[\left(\frac{\xi}{\xi^*}\right)^mf_{\rm in}(-t)-\sum_{k=1}^m\left(\frac{\xi}{\xi^*}\right)^k|\xi|^2g^*(t)e^{-\xi G(0,t)/2}S_{\xi/2}(0,t)\right].\label{A 3 level final bc}
\end{equation}
We prove the above closed form by induction. Since $\alpha_0(\mathbf{x},t)$ and $\alpha_1(\mathbf{x},t)\rvert_{x_1=-t}$ are already in the closed form. From the linearised boundary condition given by Eq.~\eqref{A 3 level n photon linear bc} and applying the induction step, we have,
\begin{align}
    \alpha_{m+1}(\mathbf{x},t)\rvert_{x_{1:m+1=-t}}&=\frac{\xi}{\xi^*}\alpha_m(\mathbf{x},t)\bigg|_{x_{1:m+1}=-t}-\frac{i\xi g^*(t)}{\sqrt{n}}\beta_1(\mathbf{x},t)\bigg|_{x_{2:m+1}=-t}\\
    &=\frac{\xi}{\xi^*}\prod_{j=m+2}^n f_{\rm in}(x_j)f_{\rm in}^{m}(-t)\left[\left(\frac{\xi}{\xi^*}\right)^mf_{\rm in}(-t)-\sum_{k=1}^m\left(\frac{\xi}{\xi^*}\right)^k|\xi|^2g^*(t)e^{-\xi G(0,t)/2}S_{\xi/2}(0,t)\right]\nonumber\\
    &\quad-\frac{\xi}{\xi^*}|\xi|^2\prod_{j=m+2}^nf_{\rm in}(x_j)f_{\rm in}^m(-t)g^*(t)e^{-\xi G(0,t)/2}S_{\xi/2}(0,t)\\
    &=\prod_{j=m+2}^n f_{\rm in}(x_j)f_{\rm in}^{m}(-t)\left[\left(\frac{\xi}{\xi^*}\right)^{m+1}f_{\rm in}(-t)-\sum_{k=1}^{m+1}\left(\frac{\xi}{\xi^*}\right)^k|\xi|^2g^*(t)e^{-\xi G(0,t)/2}S_{\xi/2}(0,t)\right],
\end{align}
as required.

Finally, the output boundary condition for the ideal subtraction process requires $\alpha_n(\mathbf{x},t)=0$, together with the linear approximation, the output boundary condition is $\alpha_{n-1}(\mathbf{x},t)\rvert_{x_{1:n}=-t}/\xi^*=ig^*(t)\beta_{1}(\mathbf{x},t)\rvert_{x_{2:n=-t}}/\sqrt{n}$. In terms of the initial conditions,
\begin{align}
    &\left(\frac{\xi}{\xi^*}\right)f_{\rm in}^{n-1}(-t)\left[\left(\frac{\xi}{\xi^*}\right)^{n-1}f_{\rm in}(-t)-\sum_{k=1}^{n-1}\left(\frac{\xi}{\xi^*}\right)^k|\xi|^2g^*(t)e^{-\xi G(0,t)/2}\phi_\xi(0,t)\right]\nonumber\\
    &\quad =f^{n-1}_{\rm in}(-t)\frac{\xi}{\xi^*}|\xi|^2g^*(t)e^{-\xi G(0,t)/2}S_{\xi/2}(0,t)\\
    &f_{\rm in}(-t)=\frac{1-(\xi^*/\xi)^{n}}{1-\xi^*/\xi}|\xi|^2g^*(t)e^{-\xi G(0,t)/2}S_{\xi/2}(0,t).
\end{align}
Recognize that the above boundary condition is in the form of Eq.~\eqref{general bc}, then the dynamic coupling is
\begin{equation}
    g^*(t)=\frac{f_{\rm in}(-t)}{\sqrt{2\Re(\zeta)\left[\int_0^tdt'|f_{\rm in}(-t')|^2\right]^{\zeta/\Re(\zeta)}}},
\end{equation}
with $\zeta=\frac{1-(\xi^*/\xi)^{n}}{1-\xi^*/\xi}|\xi|^2-\frac{\xi}{2}$.
\end{document}